\documentclass[11pt]{article}
\makeatletter\renewcommand{\section}{\@startsection
{section}{1}{\z@}{-3.5ex plus -1ex minus
    -.2ex}{2.3ex plus .2ex}{\large\bf }}

\makeatletter\renewcommand{\subsection}{\@startsection{subsection}{2}{\z@}{-3.25ex
plus -1ex minus
   -.2ex}{1.5ex plus .2ex}{\bf }}
\makeatletter\renewcommand{\subsubsection}{\@startsection{subsubsection}{3}{-2.45ex}{-3.25ex
plus -1ex minus -.2ex}{1.5ex plus .2ex}{\it }}
\makeatletter\renewcommand{\paragraph}%
  {\@startsection{paragraph}{3}{\z@\hspace{0.06cm}}{.25ex \@plus1ex \@minus.2ex}{-0.5em}{\bf }}

\renewcommand{\thesection}{\arabic{section}.}
\renewcommand{\thesubsection}{\arabic{section}.\arabic{subsection}.}

\renewcommand{\theequation}{\thesection\arabic{equation}}

\renewcommand*\l@section{\@dottedtocline{1}{0em}{1.5em}}
\renewcommand*\l@subsubsection{\@dottedtocline{4}{3.8em}{3.2em}}

\renewcommand\tableofcontents{%
    \section*{\Large\contentsname
        \@mkboth{%
           \MakeUppercase\contentsname}{\MakeUppercase\contentsname}}%
       {\baselineskip=15pt plus 2pt minus 1pt
    \@starttoc{toc}}%
\vspace{-3mm}\centerline{{\vrule height 0.5pt width 15.5cm depth
0pt}} }

\renewenvironment{thebibliography}[1]
     {\newpage\baselineskip=16pt plus 2pt minus 1pt
      \section*{\Large\refname
        \@mkboth{\MakeUppercase\refname}{\MakeUppercase\refname}}%
     \list{\@biblabel{\@arabic\c@enumiv}}%
           {\settowidth\labelwidth{\@biblabel{#1}}%
            \leftmargin\labelwidth
            \advance\leftmargin\labelsep
            \@openbib@code
            \usecounter{enumiv}%
            \let\p@enumiv\@empty
            \renewcommand\theenumiv{\@arabic\c@enumiv}}%
      \sloppy
      \clubpenalty4000
      \@clubpenalty \clubpenalty
      \widowpenalty4000%
      \sfcode`\.\@m}

\let\fn\footnote
\renewcommand{\footnote}[1]{\linespread{1.1}\fn{#1}\linespread{1.29}}

\usepackage{amsfonts}
\usepackage{amssymb}

\usepackage{mathrsfs}

\usepackage{amsmath,amssymb}
\usepackage{bbm}

\def\slasha#1{\setbox0=\hbox{$#1$}#1\hskip-\wd0\hbox to\wd0{\hss\sl/\/\hss}}

\def\periodb#1{\setbox0=\hbox{$#1$}#1\hskip-\wd0\hbox to\wd0{-}}
\newcommand{\unit}{\mathbbm{1}}   
\newcommand{\od}{\ell}   
\newcommand{\rone}{\mathbf{1}}    
\newcommand{\rthree}{\mathbf{3}}    
\newcommand{\rfour}{\mathbf{4}}    
\newcommand{\CA}{\mathcal{A}}    
\newcommand{\CAt}{\tilde{\mathcal{A}}}    
\newcommand{\CB}{\mathcal{B}}    
\newcommand{\CD}{\mathcal{D}}    
\newcommand{\CCD}{\mathscr{D}}    
\newcommand{\CCDb}{\bar{\mathscr{D}}}    
\newcommand{\CF}{\mathcal{F}}    
\newcommand{\CCG}{\mathscr{G}}    
\newcommand{\CJ}{\mathcal{J}}    
\newcommand{\CL}{\mathcal{L}}    
\newcommand{\CM}{\mathcal{M}}    
\newcommand{\CN}{\mathcal{N}}    
\newcommand{\CO}{\mathcal{O}}    
\newcommand{\COb}{\bar{\mathcal{O}}}    
\newcommand{\CP}{\mathcal{P}}    

\newcommand{\CS}{\mathcal{S}}    
\newcommand{\CT}{\mathcal{T}}    
\newcommand{\CCT}{\mathscr{T}}    
\newcommand{\CU}{\mathcal{U}}    
\newcommand{\CV}{\mathcal{V}}    
\newcommand{\CCV}{\mathscr{V}}    
\newcommand{\CW}{\mathcal{W}}    
\newcommand{\CCX}{\mathscr{X}}    
\newcommand{\CCY}{\mathscr{Y}}    
\newcommand{\CZ}{\mathcal{Z}}    
\newcommand{\CCZ}{\mathscr{Z}}    
\newcommand{\CE}{\mathcal{E}}    
\newcommand{\FR}{\mathbbm{R}}     
\newcommand{\FC}{\mathbbm{C}}     
\newcommand{\CPP}{{\mathbbm{C}P}}    
\newcommand{\dd}{\mathrm{d}}     
\newcommand{\dpar}{\partial}     
\newcommand{\dparb}{{\bar{\partial}}}     
\newcommand{\etab}{{\bar{\eta}}}     
\newcommand{\etat}{{\tilde{\eta}}}     
\newcommand{\Thetab}{{\bar{\Theta}}}     
\newcommand{\embd}{{\hookrightarrow}}     
\newcommand{\diag}{{\mathrm{diag}}}     
\newcommand{\ed}{{\dot{1}}}    
\newcommand{\zd}{{\dot{2}}}    
\newcommand{\de}{\mathrm{e}}     
\newcommand{\di}{\mathrm{i}}     
\newcommand{\bz}{{\bar{z}}}     
\newcommand{\bl}{{\bar{\lambda}}}     
\newcommand{\hl}{{\hat{\lambda}}}     
\newcommand{\bV}{{\bar{V}}}     
\newcommand{\bw}{{\bar{w}}}     
\newcommand{\bW}{{\bar{W}}}     
\newcommand{\by}{{\bar{y}}}     
\newcommand{\ald}{{\dot{\alpha}}}     
\newcommand{\bed}{{\dot{\beta}}}     
\newcommand{\gad}{{\dot{\gamma}}}     
\newcommand{\ded}{{\dot{\delta}}}     
\newcommand{\eps}{{\varepsilon}}     
\newcommand{\tphi}{{\tilde{\phi}}}     
\newcommand{\eand}{{~~~\mbox{and}~~~}}     
\newcommand{\der}[1]{\frac{\dpar}{\dpar #1}}   
\newcommand{\derr}[2]{\frac{\dpar #1}{\dpar #2}}   
\newcommand{\ddpart}[1]{\dd #1 \der{#1}}   
\newcommand{\ci}[1]{\overset{\circ}{#1}{}}   
\newcommand{\tr}{\,\mathrm{tr}\,}     

\newcommand{\agl}{\mathrm{gl}}     
\newcommand{\asu}{\mathrm{su}}     
\newcommand{\ahu}{\mathrm{u}}     
\newcommand{\sU}{\mathrm{U}}     
\newcommand{\sSU}{\mathrm{SU}}     
\newcommand{\sGL}{\mathrm{GL}}     
\newcommand{\sSO}{\mathrm{SO}}     
\newcommand{\sSpin}{\mathrm{Spin}}     
\newcommand{\sEnd}{\mathrm{End}\,}     
\newcommand{\spn}{\mathrm{span}}     

\newcommand{\remark}[1]{}     

\makeatletter \@addtoreset{equation}{section} \makeatother
\renewcommand{\thesection}{\arabic{section}.}
\renewcommand{\theequation}{\thesection\arabic{equation}}

\begin{document}
\begin{titlepage}
\setcounter{page}{0}
\begin{flushright}
  hep-th/0505161\\
  ITP--UH--07/05
\end{flushright}
\vskip 1.0cm
\begin{center}
{\Large \bf The Topological B-model on a Mini-Supertwistor
Space\\[0.5cm] and Supersymmetric Bogomolny Monopole Equations}\\
\vskip 1.0cm
\renewcommand{\thefootnote}{\fnsymbol{footnote}}
{\Large Alexander D. Popov\footnote{On leave from Bogoliubov
Laboratory of Theoretical Physics, JINR, Dubna, Russia.},
Christian S\"{a}mann and Martin Wolf} \setcounter{footnote}{0}
\renewcommand{\thefootnote}{\arabic{thefootnote}}
\vspace{.8cm}

{\em Institut f\"{u}r Theoretische Physik\\
     Universit\"{a}t Hannover\\
     Appelstra{\ss}e 2, 30167 Hannover, Germany}
\vspace{.6cm}

 {E-mail: {\ttfamily popov, saemann,
wolf@itp.uni-hannover.de} } \vspace{0.8cm}
\end{center}
\begin{center}
{\bf Abstract}
\end{center}
\begin{quote}
In the recent paper hep-th/0502076, it was argued that the open
topological B-model whose target space is a complex
$(2|4)$-dimensional mini-supertwistor space with D3- and D1-branes
added corresponds to a super Yang-Mills theory in three
dimensions. Without the D1-branes, this topological B-model is
equivalent to a dimensionally reduced holomorphic Chern-Simons
theory. Identifying the latter with a holomorphic BF-type theory,
we describe a twistor correspondence between this theory and a
supersymmetric Bogomolny model on $\FR^3$. The connecting link in
this correspondence is a partially holomorphic Chern-Simons theory
on a Cauchy-Riemann supermanifold which is a real one-dimensional
fibration over the mini-supertwistor space. Along the way of
proving this twistor correspondence, we review the necessary basic
geometric notions and construct action functionals for the
involved theories. Furthermore, we discuss the geometric aspect of
a recently proposed deformation of the mini-supertwistor space,
which gives rise to mass terms in the supersymmetric Bogomolny
equations. Eventually, we present solution generating techniques
based on the developed twistorial description together with some
examples and comment briefly on a twistor correspondence for super
Yang-Mills theory in three dimensions.
\end{quote}
\end{titlepage}
\newpage

\tableofcontents

\section{Introduction and summary}

Twistor string theory \cite{Witten:2003nn} is built upon the
observation that the open topological B-model on the Calabi-Yau
supermanifold given by the open subset
$\CP^{3|4}=\CPP^{3|4}\backslash\CPP^{1|4}$ of the supertwistor
space $\CPP^{3|4}$ with a stack of $n$ D5-branes\footnote{These
D5-branes are not quite space-filling and defined by the condition
that all open string vertex operators do not depend on
antiholomorphic Gra{\ss}mann coordinates on $\CP^{3|4}$.} is
equivalent to holomorphic Chern-Simons (hCS) theory on the same
space. This theory describes holomorphic structures on a rank $n$
complex vector bundle $\CE$ over $\CP^{3|4}$ which are given by
the (0,1) part $\CA^{0,1}$ of a connection one-form $\CA$ on
$\CE$. The components of $\CA^{0,1}$ appear as the excitations of
open strings ending on the D5-branes (i.e., they are zero modes of
these strings). Furthermore, the spectrum of physical states
contained in $\CA^{0,1}$ is the same as that of $\CN=4$ super
Yang-Mills (SYM) theory, but the interactions of both theories
differ. In fact, by analyzing the linearized \cite{Witten:2003nn}
and the full \cite{Popov:2004rb} field equations, it was shown
that hCS theory on $\CP^{3|4}$ is equivalent to the $\CN=4$
supersymmetric self-dual Yang-Mills (SDYM) theory on $\FR^4$
introduced in \cite{Siegel:1992za}, which can be considered as a
truncation of the full $\CN=4$ SYM theory.

It was conjectured by Witten that the perturbative amplitudes of
the full $\CN=4$ SYM theory are recovered by including
D-instantons wrapping holomorphic curves in $\CP^{3|4}$ into the
topological B-model \cite{Witten:2003nn}. The presence of these
D1-branes leads to additional fermionic states from zero modes of
strings stretching between the D5- and the D1-branes. The
scattering amplitudes are then computed in terms of currents
constructed from these additional fields, which localize on the
D1-branes, by integrating certain correlation functions over the
moduli space of these D1-branes in $\CP^{3|4}$. This proposal
generalizes an earlier construction of maximally
helicity-violating amplitudes by Nair \cite{Nair:bq}. Thus by
incorporating D1-branes into the topological B-model, one can
complement the $\CN=4$ SDYM theory to the full $\CN=4$ SYM theory,
at least at tree-level. This conjecture has then been verified in
several cases and it has been used in a number of papers for
calculating field theory amplitudes by using methods inspired by
string theory and twistor geometry. For a good account of the
progress made in this area, see e.g.\
\cite{webpage,Cachazo:2005ga} and references therein. For other
aspects of twistor string theories discussed lately, see e.g.\
\cite{supergravity}-\cite{Burinskii:2004tt}.

In a recent paper \cite{Chiou:2005jn}, a dimensional reduction of
the above correspondence was considered: It was shown that
scattering amplitudes of $\CN=4$ SYM theory which are localized on
holomorphic curves in the supertwistor space $\CP^{3|4}$ can be
reduced to amplitudes of $\CN=8$ SYM theory in three dimensions
which are localized on holomorphic curves in the supersymmetric
extension $\CP^{2|4}$ of the mini-twistor space
$\CP^2:=T^{1,0}\CPP^1$. Note that the simplest of such curves in
the mini-twistor space $\CP^2$ is the Riemann sphere $\CPP^1$
which coincides with the {\em spectral curve} of the BPS $\sSU(2)$
monopole.\footnote{Every static $\sSU(2)$ monopole of charge $k$
may be constructed from an algebraic curve in $\CP^2$
\cite{Hitchin:1982gh}, and an $\sSU(n)$ monopole is defined by
$n-1$ such holomorphic curves \cite{Murray:1985ji}.} The
corresponding string theory after this reduction is the
topological B-model on the mini-supertwistor space $\CP^{2|4}$
with $n$ not quite space-filling D3-branes (defined analogously to
the D5-branes in the six-dimensional case) and additional
D1-branes wrapping holomorphic cycles in $\CP^{2|4}$. It is
reasonable to assume that the latter correspond to monopoles and
substitute the D-instantons in the case of the supertwistor space
$\CP^{3|4}$. In \cite{Chiou:2005jn}, also a twistor string theory
corresponding to a certain massive SYM theory in three dimensions
was described. The target space of the underlying topological
B-model is a Calabi-Yau supermanifold obtained from the
mini-supertwistor space $\CP^{2|4}$ by deforming its complex
structure along the fermionic directions.

The goal of this paper is to complement \cite{Chiou:2005jn} by
considering the open topological B-model on the Calabi-Yau
supermanifold $\CP^{2|4}$ in the presence of the D3-branes but
without additional D1-branes. This model corresponds to a field
theory on the mini-supertwistor space $\CP^{2|4}$ obtained by a
reduction of holomorphic Chern-Simons theory on the supertwistor
space $\CP^{3|4}$. We show that this field theory on $\CP^{2|4}$
is a {\em holomorphic BF-type} (hBF) {\em theory}\/\footnote{These
theories were introduced in \cite{Popov:1999cq} and considered
e.g.\ in \cite{Ivanova:2000xr,Baulieu:2004pv}.} which in turn is
equivalent to a supersymmetric Bogomolny model. This model can be
understood as a truncation of $\CN=8$ SYM theory in three
dimensions.

Recall that the open topological B-model on a Calabi-Yau
(super)manifold $Y$ with a stack of $n$ (not quite) space-filling
D-branes can be described in terms of $\sEnd\CE$-valued
$(0,q)$-forms from the Dolbeault cohomology group
$H_\dparb^{0,q}(Y,\sEnd\CE)$, where $\CE$ is a rank $n$ vector
bundle over $Y$. It was argued in
\cite{Witten:1992fb,Witten:2003nn} that in three complex
dimensions, the relevant field is the $(0,1)$ part $\CA^{0,1}$ of
a connection one-form $\CA$ on the complex vector bundle $\CE$.
All the remaining fields are unphysical and only needed when
quantizing the theory. This is also supported by an example
presented in this paper: the B-model on the space $\CP^{2|4}$
corresponds to a gauge theory, which contains besides $\CA^{0,1}$
also an unphysical scalar field from
$H_\dparb^{0,0}(\CP^{2|4},\sEnd\CE)$ as a Lagrange multiplier in
the action functional. This theory describes again holomorphic
structures on $\CE$.

The action functional of hBF theory on $\CP^{2|4}$ is not of
Chern-Simons (CS) type, but one can introduce a CS type action on
the correspondence space $\CF^{5|8}\cong \FR^{3|8}\times S^2$
which enters into the double fibration
\begin{equation}
\begin{aligned}
\begin{picture}(50,40)
\put(0.0,0.0){\makebox(0,0)[c]{$\CP^{2|4}$}}
\put(64.0,0.0){\makebox(0,0)[c]{$\FR^{3|8}$}}
\put(34.0,33.0){\makebox(0,0)[c]{$\CF^{5|8}$}}
\put(25.0,25.0){\vector(-1,-1){18}}
\put(37.0,25.0){\vector(1,-1){18}}
\end{picture}
\end{aligned}
\end{equation}
This diagram describes a correspondence between holomorphic
projective lines in $\CP^{2|4}$ and points in the Euclidean
superspace $\FR^{3|8}$ obtained by a dimensional reduction of the
superspace $\FR^{4|8}$ along the $x^4$-axis. The correspondence
space $\CF^{5|8}$ admits a so-called {\em Cauchy-Riemann} (CR)
{\em structure}, which can be considered as a generalization of a
complex structure (see e.g.\ \cite{LeBrun:1984} for the purely
bosonic case). After enlarging the integrable distribution
defining this CR structure by one real direction to a distribution
$\CT$, one is led to the notion of $\CT$-flat vector bundles over
$\CF^{5|8}$. These bundles take over the r{\^o}le of holomorphic
vector bundles, and they can be defined by a $\CT$-flat connection
one-form $\CA_\CT$ \cite{Rawnsley}. The condition of
$\CT$-flatness of $\CA_\CT$ can be derived as the equations of
motion of a theory we shall call {\em partially holomorphic
Chern-Simons} (phCS) {\em theory}. This theory can be obtained by
a dimensional reduction of hCS theory on the supertwistor space
$\CP^{3|4}$. We prove that there are one-to-one correspondences
between equivalence classes of holomorphic vector
bundles\footnote{obeying certain triviality
conditions}\addtocounter{footnote}{-1} over $\CP^{2|4}$,
equivalence classes of $\CT$-flat vector bundles\footnotemark{}
over $\CF^{5|8}$ and gauge equivalence classes of solutions to
supersymmetric Bogomolny equations on $\FR^3$. In other words, the
moduli spaces of all three theories are bijective. Thus, we show
that phCS theory is the connecting link between hBF theory on
$\CP^{2|4}$ and the supersymmetric Bogomolny model on $\FR^3$:
\begin{equation}\label{eq:1.1}
\begin{aligned}
\begin{picture}(180,70)(0,-5)
\put(0.0,0.0){\makebox(0,0)[c]{hBF theory on $\CP^{2|4}$}}
\put(160.0,14.0){\makebox(0,0)[c]{supersymmetric}}
\put(160.0,0.0){\makebox(0,0)[c]{ Bogomolny model on $\FR^3$}}
\put(80.0,50.0){\makebox(0,0)[c]{phCS theory on $\CF^{5|8}$}}
\put(40.0,40.0){\vector(-1,-1){30}}
\put(10.0,10.0){\vector(1,1){30}}
\put(100.0,40.0){\vector(1,-1){18}}
\put(118.0,22.0){\vector(-1,1){18}} \put(56,0){\vector(1,0){37}}
\put(86,0){\vector(-1,0){30}}
\end{picture}
\end{aligned}
\end{equation}
By a deformation of the complex structure on $\CP^{2|4}$, which in
turn induces a deformation of the CR structure on $\CF^{5|8}$, we
obtain a correspondence of the type \eqref{eq:1.1} with additional
mass terms for fermions and scalars in the supersymmetric
Bogomolny equations.

The twistorial description of the supersymmetric Bogomolny
equations on $\FR^3$ has the nice feature that it yields novel
methods for constructing explicit solutions. For simplicity, we
restrict our discussion to solutions where only fields with
helicity $\pm 1$ and a Higgs field are nontrivial. The
corresponding Abelian configurations give rise to the Dirac
monopole-antimonopole systems. For the non-Abelian case, we
present two ways of constructing solutions: First, by using a
dressed version of the Penrose-Ward transform and second, by
considering a nilpotent deformation of the holomorphic vector
bundle corresponding to an arbitrary seed solution of the ordinary
Bogomolny equations.

The organization of the paper is as follows. In section 2, we
review the geometry of the (super)manifolds and the field theories
involved in the $\CN=4$ supertwistor correspondence. In
particular, the equivalence of hCS theory on the supertwistor
space $\CP^{3|4}\cong \FR^{4|8}\times S^2$ and $\CN=4$ SDYM theory
on $\FR^4$ is recalled. Translations along the $x^4$-axis in
$\FR^4$ induce actions of a real and a complex one-parameter group
on the space $\CP^{3|4}$, which are described in section 3. Taking
the quotient of $\CP^{3|4}$ with respect to these groups yields
the orbit spaces $\CF^{5|8}$ and $\CP^{2|4}$. In section 4, the
partially holomorphic Chern-Simons theory (which is naturally
defined on $\CF^{5|8}$) is introduced and its equivalence to a
supersymmetric Bogomolny model on $\FR^3$ is proven. In section 5,
we extend this equivalence to a holomorphic BF theory on the
mini-supertwistor space $\CP^{2|4}$, thus completing the picture
\eqref{eq:1.1}. The deformations of the complex structure on the
mini-supertwistor space $\CP^{2|4}$ and of the CR structure on the
space $\CF^{5|8}$ which yield additional mass terms in the
Bogomolny equations together with a detailed analysis of the
geometric background are presented in section 6. Section 7 is
concerned with the construction of explicit solutions to the
supersymmetric Bogomolny equations: we describe two
solution-generating algorithms and give some examples. Eventually,
we briefly comment on a twistor correspondence for the full
$\CN=8$ SYM theory in section 8. While appendices A and B cover
some technical details, appendix C provides some remarks on the
supertwistor correspondence for the case of signature
$($$+$$+$$-$$)$.

\section{Geometry of the $\CN=4$ supertwistor space}

\subsection{Euclidean twistors in real and complex setting}

\paragraph{General case.} Let us consider a smooth oriented real
four-manifold $X$ with a metric $g$ of signature (++++) and the
principal bundle $P(X,\sSO(4))$ of orthonormal frames over $X$.
The twistor space $\CZ$ of $X$ can be defined\footnote{Further
(equivalent) definitions of the twistor space $\CZ$ can be found
in appendix C.} as the associated bundle \cite{Atiyah:wi}
\begin{equation}\label{eq:2.1}
\CZ\ :=\ P(X,\sSO(4))\times_{\sSO(4)}(\sSO(4)/\sU(2))
\end{equation}
with the canonical projection
\begin{equation}\label{eq:2.2}
\pi\,:\,\CZ\ \rightarrow\ X~.
\end{equation}
The fibres of this bundle are two-spheres
$S^2_x\cong\sSO(4)/\sU(2)$ which parametrize almost complex
structures on the tangent spaces $T_xX$. As a real manifold, $\CZ$
has dimension six.

Note that while a manifold $X$ admits in general no almost complex
structure, its twistor space $\CZ$ can always be equipped with an
almost complex structure $\CJ$ \cite{Atiyah:wi}. Furthermore,
$\CJ$ is integrable if and only if the Weyl tensor of $X$ is
self-dual \cite{Penrose:in,Atiyah:wi}. Then $\CZ$ is a complex
three-manifold with an antiholomorphic involution $\tau$ (a real
structure) which maps $\CJ$ to $-\CJ$ and the fibres of the bundle
\eqref{eq:2.2} over $x\in X$ are $\tau$-invariant projective lines
$\CPP^1_x$, each of which has normal bundle $\CO(1)\oplus\CO(1)$
in the complex manifold $\CZ$.

\paragraph{The projective space $\CPP^3$.} It follows from
\eqref{eq:2.1} that the twistor space of the four-sphere $S^4$
endowed with the canonical conformally flat metric is the complex
projective space $\CPP^3$ \cite{Atiyah:wi},
\begin{equation}
\CPP^3\ \cong\ P(S^4,\sSO(4))\times_{\sSO(4)}(\sSO(4)/\sU(2))~.
\end{equation}
In the following, we describe this space by the complex
homogeneous coordinates $(\omega^\alpha,\lambda_\ald)$ subject to
the equivalence relation
$(\omega^\alpha,\lambda_\ald)\sim(t\omega^\alpha,t\lambda_\ald)$
for any $t\in\FC^*$, where the spinor indices
$\alpha,\beta,\ldots$ and $\ald,\bed,\ldots$ run over $1,2$ and
$\dot{1},\dot{2}$, respectively. The real structure $\tau$ on
$\CPP^3$ is induced by the anti-linear transformations
\begin{equation}\label{eq:2.4}
\left(\begin{array}{c} \omega^1\\\omega^2
\end{array}\right)\ \mapsto\
\left(\begin{array}{c}-\bar{\omega}^2\\\bar{\omega}^1
\end{array}\right)\eand
\left(\begin{array}{c} \lambda_{\ed}\\\lambda_{\zd}
\end{array}\right)\ \mapsto\ \left(\begin{array}{c}-\bl_{\zd}\\
\bl_{\ed}
\end{array}\right)~.
\end{equation}
While there are no fixed points of $\tau$ in $\CPP^3$, there are
$\tau$-invariant rational curves $\CPP^1\embd \CPP^3$.

\paragraph{The twistor space $\CP^3$ of $\FR^4$.} By definition
\eqref{eq:2.1}, the twistor space of the Euclidean space $\FR^4$
is
\begin{equation}\label{eq:2.5}
\CZ\ =\ P(\FR^4,\sSO(4))\times_{\sSO(4)}(\sSO(4)/\sU(2))\ \cong \
\FR^4\times S^2~.
\end{equation}
Having in mind that $\FR^4\cong S^4\backslash\{\infty\}$, one can
identify the twistor space $\CZ$ of $\FR^4$ with the complex
three-manifold $\CP^3:=\CPP^3\backslash\CPP^1_\infty$. Here, the
point $\infty\in S^4$ corresponds to the projective line
$\CPP^1_\infty\subset\CPP^3$. Note that we can choose to
parametrize this sphere $\CPP^1_\infty$ by the homogeneous
coordinates $(\lambda_\ald)=(0,0)^T$ and $(\omega^\alpha)\neq
(0,0)^T$. Thus, one can obtain $\CP^3$ by taking the subset
$(\lambda_\ald)\neq (0,0)^T$ on $\CPP^3$ and a real structure
$\tau$ on $\CP^3$ is induced from the one on $\CPP^3$. This space
together with $\tau$ is diffeomorphic to the space \eqref{eq:2.5},
\begin{equation}
\CP^3\ =\ \CPP^3\backslash \CPP^1_\infty\ \cong\ \FR^4\times S^2
\end{equation}
and therefore $\CP^3$ is the twistor space of $\FR^4$ with the
canonical projection
\begin{equation}
\pi\ :\ \CP^3\ \rightarrow\ \FR^4~.
\end{equation}

We can cover $\CP^3$ by two patches $\CU_+$ and $\CU_-$ for which
$\lambda_{\ed}\neq 0$ and $\lambda_{\zd}\neq 0$, respectively, and
introduce the coordinates
\begin{equation}\label{eq:2.8}
\begin{aligned}
z_+^\alpha &\ =\ \frac{\omega^\alpha}{\lambda_{\dot{1}}}~,~~~
&z_+^3\ =\ \frac{\lambda_{\dot{2}}}{\lambda_{\dot{1}}}\ =:\
\lambda_+~~~~~ &\mbox{on}~~~~~\CU_+~,\\ z_-^\alpha&\ =\
\frac{\omega^\alpha}{\lambda_{\dot{2}}}~,~~~ &z_-^3\ =\
\frac{\lambda_{\dot{1}}}{\lambda_{\dot{2}}}\ =:\ \lambda_-~~~~~
&\mbox{on}~~~~~\CU_-~,
\end{aligned}
\end{equation}
which are related on $\CU_+\cap\CU_-$ by the equations
\begin{equation}
z^\alpha_+\ =\ \frac{1}{z_-^3}\, z^\alpha_-\eand z_+^3\ =\
\frac{1}{z^3_-}~.
\end{equation}
From this, it follows that $\CP^3$ coincides with the total space
of the rank 2 holomorphic vector bundle $\CO(1)\oplus\CO(1)$ over
the Riemann sphere $\CPP^1$, i.e.\
\begin{equation}\label{eq:2.10}
\CP^3\ \rightarrow\ \CPP^1~~~\mbox{with}~~~\CP^3\ =\
\CO(1)\oplus\CO(1)~.
\end{equation}
The base manifold $\CPP^1$ of this fibre bundle is covered by two
patches $U_\pm=\CU_\pm\cap \CPP^1$ with affine coordinates
$\lambda_\pm$.

The real structure $\tau$ on $\CP^3$ induced by the
transformations \eqref{eq:2.4} acts on the coordinates
\eqref{eq:2.8} as follows:
\begin{equation}\label{eq:2.11}
\tau(z_\pm^1,z_\pm^2,z^3_\pm)\ =\
\left(\pm\frac{\bz_\pm^2}{\bz^3_\pm},
\mp\frac{\bz_\pm^1}{\bz^3_\pm},-\frac{1}{\bz^3_\pm}\right)~.
\end{equation}
It is not difficult to see that (analogously to the case of
$\CPP^3$) $\tau$ has no fixed points in $\CP^3$ but leaves
invariant projective lines $\CPP^1$ joining the points $p$ and
$\tau(p)$ for any $p\in\CP^3$. For two other possible real
structures on $\CP^3$, see e.g.\ \cite{Popov:2004rb}.

\paragraph{Incidence relations.} Global holomorphic sections of
the bundle \eqref{eq:2.10} are locally polynomials of degree one
in $\lambda_\pm$. Introducing the spinorial notation
\begin{equation}\label{eq:2.12}
(\lambda_\ald^+)\ :=\  \left(\begin{array}{c} 1\\\lambda_+
\end{array}\right)\eand (\lambda_\ald^-)\ :=\ \left(\begin{array}{c}
\lambda_- \\ 1
\end{array}\right)~,
\end{equation}
one can parametrize these sections by the moduli
$x=(x^{\alpha\ald})\in\FC^4$ as
\begin{equation}\label{eq:2.13}
z^\alpha_\pm\ =\ x^{\alpha\ald}\lambda_\ald^\pm
\end{equation}
over the patches $U_\pm$. These sections describe a holomorphic
embedding of rational curves\footnote{By the Kodaira theorem
\cite{Kodaira}, the complex dimensions of the moduli space
parametrizing a family of rational curves embedded holomorphically
into $\CP^3$ is $\dim_\FC H^0(\CPP^1,\CO(1)\oplus\CO(1))=4$ and
there are no obstructions to these deformations since
$H^1(\CPP^1,\CO(1)\oplus\CO(1))=0$.} $\CPP^1_x\embd \CP^3$ for
fixed $x\in\FC^4$. On the other hand, for each point
$p=(z^\alpha_\pm,\lambda_\ald^\pm)\in\CP^3$, the incidence
relations \eqref{eq:2.13} define a null (anti-self-dual) two-plane
($\beta$-plane) in $\FC^4$. The correspondences
\begin{equation}
\begin{aligned}
\{\,\mbox{projective lines $\CPP^1_x$ in $\CP^3$}\}&\
\longleftrightarrow \
\{\,\mbox{points $x$ in $\FC^4$}\}~, \\
\{\,\mbox{points $p$ in $\CP^3$}\}&\ \longleftrightarrow\
\{\,\mbox{$\beta$-planes $\FC^2_p$ in $\FC^4$}\}
\end{aligned}
\end{equation}
between subspaces in $\CP^3$ and $\FC^4$ can be described by a
double fibration (see e.g.\ \cite{Popov:2004rb} and references
therein). Those curves $\CPP^1_x\embd\CP^3$ which are invariant
under the involution $\tau$ defined in \eqref{eq:2.11} are
parametrized by moduli $x\in\FC^4$ satisfying the equations
\begin{equation}\label{eq:2.15}
\tau\left(\begin{array}{cc} x^{1\ed} & x^{1\zd} \\ x^{2\ed} &
x^{2\zd}
\end{array}\right)\ =\ \left(\begin{array}{cc} \bar{x}^{2\zd} &
-\bar{x}^{2\ed}\\
-\bar{x}^{1\zd} & \bar{x}^{1\ed}
\end{array}\right)\ =\ \left(\begin{array}{cc} x^{1\ed} & x^{1\zd} \\
x^{2\ed} & x^{2\zd}
\end{array}\right)~,
\end{equation}
and therefore we can introduce real coordinates $(x^\mu)\in\FR^4$
with $\mu=1,\ldots,4$ by\footnote{Note that our choice of
relations between $x^{\alpha\ald}$ and $x^\mu$ differs from that
of \cite{Popov:2004rb}.}
\begin{equation}\label{eq:2.16}
x^{2\zd}\ =\ \bar{x}^{1\ed}\ =:\ -\di(x^1-\di x^2)\eand x^{2\ed}\
=\ -\bar{x}^{1\zd}\ =:\ -\di(x^3-\di x^4)~.
\end{equation}
These are coordinates on the base of the fibration
$\CP^3\rightarrow \FR^4$, which parametrize $\tau$-real
holomorphic curves $\CPP^1_x\embd \CP^3$, and come naturally with
the Euclidean metric
\begin{equation}
\dd s^2\ =\ \det(\dd x^{\alpha\ald})\ =\ \delta_{\mu\nu}\dd
x^\mu\dd x^\nu~.
\end{equation}
Other real structures on $\CP^3$ give rise to a metric with
signature $($$+$$+$$-$$-$$)$ on $\FR^4$.

\subsection{The twistor space of $\FR^4$ as a direct product of
complex manifolds}

\paragraph{Coordinate transformations.} Recall that the twistor
space of $\FR^4$ can be considered both as the smooth manifold
$\FR^4\times S^2$ and as the complex manifold $\CP^3$ because
there is a diffeomorphism between them. Switching to the
coordinates $x^{\alpha\ed}$ from \eqref{eq:2.16} and to
$\lambda_\pm=\mathrm{Re}\,\lambda_\pm+\di\,
\mathrm{Im}\,\lambda_\pm$, we identify $\FR^4$ with $\FC^2$ and
$S^2$ with $\CPP^1$, respectively. Thus, we have further
diffeomorphisms
\begin{equation}
\FR^4\times S^2\ \cong\ \FC^2\times\CPP^1\ \cong\ \CP^3~.
\end{equation}
The latter is defined by \eqref{eq:2.13} and \eqref{eq:2.15}, and
its inverse reads explicitly as
\begin{equation}\label{eq:2.19}
\begin{aligned}
x^{1\ed}\ =\ \frac{z^1_++z_+^3\bz^2_+}{1+z_+^3\bz_+^3}\ =\
\frac{\bz_-^3z_-^1+\bz_-^2}{1+z_-^3\bz_-^3}~,~~~ &x^{1\zd}\ =\
-\frac{\bz^2_+-\bz_+^3z^1_+}{1+z_+^3\bz_+^3}\ =\
-\frac{z_-^3\bz^2_--z_-^1}{1+z_-^3\bz_-^3}~,\\\lambda_\pm\ =\
&z_\pm^3~.
\end{aligned}
\end{equation}
Note that while the complex manifolds $\CP^3$ and
$\FC^2\times\CPP^1$ are diffeomorphic, they are not biholomorphic
as their complex structures obviously differ.

\paragraph{Vector fields.} On the complex manifold $\CP^3$, we
have the natural basis $\{\der{\bz^\alpha_\pm},\der{\bz^3_\pm}\}$
for the space of antiholomorphic vector fields. On the
intersection $\CU_+\cap\CU_-$, we find
\begin{equation}
\der{\bz^\alpha_+}\ =\ \bz_-^3\der{\bz^\alpha_-}\eand
\der{\bz_+^3}\ =\
-(\bz_-^3)^2\der{\bz_-^3}-\bz_-^3\bz_-^\alpha\der{\bz_-^\alpha}~.
\end{equation}
Using formul\ae{} \eqref{eq:2.19}, we can express these vector
fields in terms of the coordinates $(x^{\alpha\ed},\lambda_\pm)$
and their complex conjugates according to
\begin{equation}\label{eq:2.21}
\begin{aligned}
\der{\bz_\pm^1}&\ =\ -\gamma_\pm\lambda_\pm^\ald\der{x^{2\ald}} \
=:\ -\gamma_\pm\bar{V}_2^\pm~,~~~& \der{\bz_\pm^2}&\ =\
\gamma_\pm\lambda_\pm^\ald\der{x^{1\ald}} \ =:\
\gamma_\pm\bar{V}_1^\pm~,\\\der{\bz_+^3}&\ =\ \der{\bl_+}-\gamma_+
x^{\alpha\dot{1}}\bar{V}_\alpha^+~,~~~& \der{\bz_-^3}&\ =\
\der{\bl_-}-\gamma_- x^{\alpha\dot{2}}\bar{V}_\alpha^-~,
\end{aligned}
\end{equation}
where we have used
\begin{equation}\label{eq:2.22}
\lambda_\pm^\ald\ =\
\eps^{\ald\bed}\lambda_\bed^\pm~~~\mbox{with}~~~ \eps^{\ed\zd}\ =\
-\eps^{\zd\ed}\ =\ 1\eand\gamma_\pm\ :=\
\frac{1}{1+\lambda_\pm\bl_\pm}\ =\
\frac{1}{\hl_\pm^\ald\lambda^\pm_\ald}
\end{equation}
together with the convention $\eps_{\ed\zd}=-\eps_{\zd\ed}=-1$,
which implies $\eps_{\ald\bed}\eps^{\bed\gad}=\delta_\ald^\gad$.
Thus, the vector fields
\begin{equation}\label{eq:2.23}
\bar{V}_\alpha^\pm\ =\ \lambda_\pm^\ald\der{x^{\alpha\ald}}\eand
\bar{V}_3^\pm\ :=\ \der{\bl_\pm}
\end{equation}
form a basis of vector fields of type $(0,1)$ on
$\CU_\pm\subset\CP^3$ in the coordinates
$(x^{\alpha\ald},\lambda_\pm,\bl_\pm)$.

\paragraph{Forms.} It is easy to check that the basis of
(0,1)-forms on $\CU_\pm$, which are dual to the vector fields
\eqref{eq:2.23}, is given by
\begin{equation}
\bar{E}_\pm^\alpha\ =\ -\gamma_\pm\hat{\lambda}_\ald^\pm\dd
x^{\alpha\ald}\eand\bar{E}^3_\pm\ =\ \dd \bl_\pm~,
\end{equation}
where
\begin{equation}\label{eq:2.25}
\begin{aligned}
(\hl_\ald^+)&\ :=\ \left(\begin{array}{cc} 0 & -1 \\ 1 & 0
\end{array}\right)\left(\begin{array}{c} 1\\ \bl_+
\end{array}\right)\ =\ \left(\begin{array}{c}-\bl_+\\1
\end{array}\right)~,\\
(\hl_\ald^-)&\ :=\ \left(\begin{array}{cc} 0 & -1 \\ 1 & 0
\end{array}\right)\left(\begin{array}{c} \bl_-\\ 1
\end{array}\right)\ =\ \left(\begin{array}{c}-1\\\bl_-
\end{array}\right)~.
\end{aligned}
\end{equation}
One can easily verify that
\begin{equation}
\dparb|_{\CU_\pm}\ =\ \dd \bz_\pm^a\der{\bz^a_\pm}\ =\
\bar{E}^a_\pm\bar{V}_a^\pm~~~\mbox{for}~~~a\ =\ 1,2,3~,
\end{equation}
and $\bar{E}^a_+\bar{V}_a^+=\bar{E}^a_-\bar{V}_a^-$ on
$\CU_+\cap\CU_-$. More details on twistor theory can be found in
the books \cite{Penrose:ca}-\cite{Mason:rf}.

\subsection{Supertwistor spaces as complex supermanifolds}

\paragraph{The supermanifolds $\CPP^{3|4}$ and $\CP^{3|4}$.} An
extension of the twistor space $\CPP^3$ to a Calabi-Yau
supermanifold is the space $\CPP^{3|4}$ which is described by
homogeneous coordinates $(\omega^\alpha,$ $\lambda_\ald,\eta_i)$
$\in\FC^{4|4}\backslash\{0\}$ subject to the identification
$(\omega^\alpha,\lambda_\ald,\eta_i)\sim(t\omega^\alpha,
t\lambda_\ald,t\eta_i)$ for any $t\in\FC^*$ \cite{Witten:2003nn}.
Here, $(\omega^\alpha,\lambda_\ald)$ are the homogeneous
coordinates on the body $\CPP^3$ and $\eta_i$ with $i=1,\ldots,4$
are complex Gra{\ss}mann variables.

Similarly to the bosonic case, we introduce the space
$\CP^{3|4}:=\CPP^{3|4}\backslash \CPP_\infty^{1|4}$ by demanding
that the $\lambda_\ald$ are not simultaneously zero. This space is
an open subset of $\CPP^{3|4}$ covered by two patches
$\hat{\CU}_+$ and $\hat{\CU}_-$ with bosonic coordinates
\eqref{eq:2.8} and fermionic coordinates
\begin{equation}\label{eq:2.27}
\eta_i^+\ =\
\frac{\eta_i}{\lambda_{\dot{1}}}~~~~\mbox{on}~~~~\hat{\CU}_+
~~~~\mbox{and}~~~~\eta_i^-\ =\
\frac{\eta_i}{\lambda_{\dot{2}}}~~~~ \mbox{on}~~~~\hat{\CU}_-~.
\end{equation}
The latter are related by $\eta_i^+=(z_-^3)^{-1}\eta^-_i$ on the
intersection $\hat{\CU}_+\cap\hat{\CU}_-$. From this, it becomes
clear that $\CP^{3|4}$ is the following holomorphic vector
bundle\footnote{The operator $\Pi$ inverts the parity of the fibre
coordinates of a vector bundle.} over $\CPP^1$:
\begin{equation}\label{eq:2.28}
\CP^{3|4}\ \rightarrow\ \CPP^1~~~\mbox{with}~~~\CP^{3|4}\ =\
\CO(1)\otimes \FC^2\oplus\Pi\CO(1)\otimes\FC^4~.
\end{equation}
The fibres over $\lambda\in\CPP^{1|0}\equiv\CPP^1$ are the
superspaces $\FC_\lambda^{2|4}$. In the following, we will refer
to $\CP^{3|4}$ as the ($\CN=4$) {\em supertwistor space}.

\paragraph{Incidence relations.} Holomorphic sections of the
bundle \eqref{eq:2.28} are polynomials of degree one in
$\lambda_\pm$. They are defined by the equations
\begin{align}\label{eq:2.29}
z_\pm^\alpha\ =\ x^{\alpha\ald}_R\,\lambda_\ald^\pm~\eand~
\eta_i^\pm\ =\
\eta_i^\ald\lambda_\ald^\pm~~~~\mbox{on}~~~~\hat{\CU}_\pm
\end{align}
and parametrized by the moduli
$(x_R^{\alpha\ald},\eta_i^\ald)\in\FC^{4|8}$. The latter space is
called the $\CN=4$ (complex) anti-chiral superspace. In the
following, we omit the subscript $R$ for brevity.

Equations \eqref{eq:2.29} define a curve $\CPP^1_{x,\eta}\embd
\CP^{3|4}$ for fixed
$(x,\eta)=(x^{\alpha\ald},\eta_i^\ald)\in\FC^{4|8}$ and a null
$\beta$-superplane of complex dimension $(2|4)$ in $\FC^{4|8}$ for
fixed $p=(z_\pm^\alpha,\lambda_\pm,\eta_i^\pm)\in\CP^{3|4}$. Thus,
the incidence relations \eqref{eq:2.29} yield the correspondences
\begin{equation}
\begin{aligned}
\{\,\mbox{projective lines $\CPP^1_{x,\eta}$ in $\CP^{3|4}$}\}&\
\longleftrightarrow\
\{\,\mbox{points $(x,\eta)$ in $\FC^{4|8}$}\}~,\\
\{\,\mbox{points $p$ in $\CP^{3|4}$}\}&\ \longleftrightarrow\
\{\,\mbox{$\beta$-superplanes $\FC_p^{2|4}$ in $\FC^{4|8}$}\}~,
\end{aligned}
\end{equation}
which can be described by a double fibration (see e.g.\
\cite{Popov:2004rb}).

\subsection{Real structure on the $\CN=4$ supertwistor space}

\paragraph{Reality conditions.} The antiholomorphic involution
$\tau:\CP^3\rightarrow \CP^3$ given by \eqref{eq:2.11} can be
extended to an antiholomorphic involution
$\tau:\CP^{3|4}\rightarrow \CP^{3|4}$ by defining
\begin{equation}\label{eq:2.31}
\tau(\eta_i^\pm)\ =\
\pm\frac{1}{\bz^3_\pm}T_i{}^j\bar{\eta}^\pm_j~~~\mbox{with}~~~
(T_i{}^j)\ =\ \left(\begin{array}{cccc}
0 & -1 & 0 & 0 \\ 1 & 0 & 0 & 0 \\
0 & 0 & 0 & -1\\
0 & 0 & 1 & 0
\end{array}\right)~
\end{equation}
for the fermionic coordinates $\eta_i^\pm$ on $\CP^{3|4}$. On the
moduli space $\FC^{4|8}$, this corresponds to the involution
\eqref{eq:2.15} for the bosonic coordinates $x^{\alpha\ald}$ and
to
\begin{equation}\label{eq:2.32}
\tau(\eta_i^\ald)\ =\
\eps^{\ald\bed}T_i{}^j\bar{\eta}^\bed_j~~\Leftrightarrow~~
\tau\left(\begin{array}{cccc} \eta_1^\ed & \eta_2^\ed &\eta_3^\ed
&\eta_4^\ed \\\eta_1^\zd &\eta_2^\zd &\eta_3^\zd &\eta_4^\zd
\end{array}\right)\ =\ \left(\begin{array}{cccc}
-\bar{\eta}^{\dot{2}}_2 &\bar{\eta}^{\dot{2}}_1
&-\bar{\eta}^{\dot{2}}_4 &\bar{\eta}^{\dot{2}}_3 \\
\bar{\eta}^{\dot{1}}_2 &-\bar{\eta}^{\dot{1}}_1
&\bar{\eta}^{\dot{1}}_4 &-\bar{\eta}^{\dot{1}}_3 \\
\end{array}\right)
\end{equation}
for the fermionic coordinates $\eta_i^\ald$. From \eqref{eq:2.32},
one can directly read off the reality conditions
\begin{equation}\label{eq:2.33}
\tau(\eta_i^\ald)\ =\ \eta_i^\ald~~~\Leftrightarrow~~~ \eta_i^\zd\
=\ -T_i{}^j \etab_j^\ed~,
\end{equation}
which we impose on the Gra{\ss}mann variables $\eta_i^\ald$. These
reality conditions together with \eqref{eq:2.16} define the
$\CN=4$ (anti-chiral) superspace $\FR^{4|8}\cong\FC^{2|4}$. Thus,
$\CP^{3|4}$ is the supertwistor space for the Euclidean superspace
$\FR^{4|8}$.

\paragraph{Coordinate transformations.} Note that formul\ae{}
\eqref{eq:2.29} with $x^{\alpha\ald}$ and $\eta_i^\ald$ obeying
the reality conditions \eqref{eq:2.16} and \eqref{eq:2.33} define
the diffeomorphisms
\begin{equation}
\FR^{4|8}\times S^2\ \cong\ \FC^{2|4}\times\CPP^1\ \cong\
\CP^{3|4}~.
\end{equation}
The map from $\CP^{3|4}$ to the space $\FC^{2|4}\times\CPP^1$ with
complex coordinates $(x^{\alpha\ed},\eta^\ed_i,\lambda_\pm)$ is
given by \eqref{eq:2.19} and
\begin{equation}\label{eq:2.35}
\begin{aligned}
\eta_1^\ed&\ =\
\frac{\eta_1^+-z_+^3\bar{\eta}_2^+}{1+z_+^3\bz_+^3}\ =\
\frac{\bz_-^3\eta_1^--\bar{\eta}_2^-}{1+z_-^3\bz_-^3}~,&
\eta_2^\ed&\ =\
\frac{\eta_2^++z_+^3\bar{\eta}_1^+}{1+z_+^3\bz_+^3}\ =\
\frac{\bz_-^3\eta_2^-+\bar{\eta}_1^-}{1+z_-^3\bz_-^3}~,\\
\eta_3^\ed&\ =\
\frac{\eta_3^+-z_+^3\bar{\eta}_4^+}{1+z_+^3\bz_+^3}\ =\
\frac{\bz_-^3\eta_3^--\bar{\eta}_4^-}{1+z_-^3\bz_-^3}~,&
\eta_4^\ed&\ =\
\frac{\eta_4^++z_+^3\bar{\eta}_3^+}{1+z_+^3\bz_+^3}\ =\
\frac{\bz_-^3\eta_4^-+\bar{\eta}_3^-}{1+z_-^3\bz_-^3}~,
\end{aligned}
\end{equation}
together with \eqref{eq:2.33}. The formul\ae{} \eqref{eq:2.19} and
\eqref{eq:2.35} define also a (smooth) projection
\begin{equation}
\CP^{3|4}\ \rightarrow\ \FR^{4|8}~.
\end{equation}

\paragraph{Odd vector fields and forms.} The odd
antiholomorphic vector fields $\der{\etab^\pm_i}$ on $\CP^{3|4}$
can be expressed due to \eqref{eq:2.33} and \eqref{eq:2.35} in
terms of coordinates on $\FC^{2|4}\times \CPP^1$ as follows:
\begin{equation}\label{eq:2.37}
\begin{aligned}
\der{\etab_1^\pm}&\ =\
\gamma_\pm\lambda^\ald_\pm\der{\eta_2^\ald}\ =:\
\gamma_\pm\bV_\pm^2~,& \der{\etab_2^\pm}&\ =\
-\gamma_\pm\lambda^\ald_\pm\der{\eta_1^\ald}\ =:\
-\gamma_\pm\bV_\pm^1~,\\ \der{\etab_3^\pm}&\ =\
\gamma_\pm\lambda^\ald_\pm\der{\eta_4^\ald}\ =:\
\gamma_\pm\bV_\pm^4~,& \der{\etab_4^\pm}&\ =\
-\gamma_\pm\lambda^\ald_\pm\der{\eta_3^\ald}\ =:\
-\gamma_\pm\bV_\pm^3~,
\end{aligned}
\end{equation}
or
\begin{equation}
\der{\etab_i^\pm}\ =\ \gamma_\pm T_j{}^i\bV_\pm^j
\end{equation}
for short. Therefore, the odd vector fields
\begin{equation}\label{eq:2.38}
\bV_\pm^i\ =\ \lambda_\pm^\ald\der{\eta_i^\ald}
\end{equation}
complement the vector fields \eqref{eq:2.23} to a basis of vector
fields of type (0,1) on $\hat{\CU}_\pm\subset\CP^{3|4}$ in the
coordinates $(x^{\alpha\ald},\lambda_\pm,\bl_\pm,\eta^\ald_i)$.
The basis of odd (0,1)-forms dual to the vector fields
\eqref{eq:2.38} is given by
\begin{equation}\label{eq:2.39}
\bar{E}_i^\pm\ =\ -\gamma_\pm\hat{\lambda}^\pm_\ald\dd
\eta_i^\ald~.
\end{equation}
Note that on the supermanifold $\CP^{3|4}$, the transformations
for $\der{\bz_\pm^3}$ in \eqref{eq:2.21} are changed to
\begin{equation}
\der{\bz_+^3}\ =\
\der{\bl_+}-\gamma_+x^{\alpha\ed}\bV_\alpha^+-\gamma_+\eta_i^\ed
\bV^i_+\eand
\der{\bz_-^3}\ =\
\der{\bl_-}-\gamma_-x^{\alpha\zd}\bV_\alpha^-+\gamma_-\eta_i^\zd
\bV^i_-~,
\end{equation}
and we obtain
\begin{equation}
\dparb|_{\hat{\CU}_\pm}\ =\ \dd \bz_\pm^a\der{\bz^a_\pm}+\dd
\etab_i^\pm\der{\etab_i^\pm}\ =\
\bar{E}^a_\pm\bV^\pm_a+\bar{E}^\pm_i\bV^i_\pm~.
\end{equation}
For a discussion of other real structures on $\CP^{3|4}$ related
with signature $($$+$$+$$-$$-$$)$, see e.g.\ \cite{Popov:2004rb}.

\paragraph{Holomorphic integral form.} Let us furthermore
introduce the (nowhere vanishing) holomorphic volume element
\begin{equation}\label{eq:holvolform}
\hat{\Omega}|_{\hat{\CU}_\pm}\ :=\ \pm\dd z^1_\pm\wedge \dd
z^2_\pm\wedge \dd z^3_\pm\dd\eta_1^\pm\cdots\dd\eta_4^\pm
\end{equation}
on $\CP^{3|4}$. This holomorphic volume element exists since the
Berezinian of $T^{1,0}\CP^{3|4}$ is a trivial bundle, and this
implies that $\CP^{3|4}$ is a Calabi-Yau supermanifold. Note,
however, that $\hat{\Omega}$ is not a differential form in the
Gra{\ss}mann coordinates, since Gra{\ss}mann differential forms (as the
ones used e.g.\ in \eqref{eq:2.39}) are dual to Gra{\ss}mann vector
fields and thus transform contragrediently to them. Berezin
integration is, however, equivalent to differentiation, and thus a
volume element has to transform as a product of Gra{\ss}mann vector
fields, i.e.\ with the inverse of the Jacobian. Such forms are
called {\em integral forms} and for short, we will call
$\hat{\Omega}$ a {\em holomorphic volume form}, similarly to the
usual nomenclature for Calabi-Yau manifolds.

\subsection{Holomorphic Chern-Simons theory on $\CP^{3|4}$}

Recall that the open topological B-model on a complex
three-dimensional Calabi-Yau manifold with a stack of $n$
D5-branes is equivalent to holomorphic Chern-Simons theory and
describes holomorphic structures on a rank $n$ vector bundle over
the same space. In the following, we will study this setting on
the supertwistor space $\CP^{3|4}$.

\paragraph{Equations of motion.} Consider a trivial rank $n$
complex vector bundle $\CE$ over $\CP^{3|4}$ and a connection
one-form $\CA$ on $\CE$. We can use the holomorphic volume form
\eqref{eq:holvolform} to write down an action for holomorphic
Chern-Simons theory on $\CP^{3|4}$,
\begin{equation}\label{eq:2.41}
S_{\mathrm{hCS}}\ =\ \int_{\CCZ}\hat{\Omega}\wedge
\tr\left(\CA^{0,1}\wedge\bar{\dpar}\CA^{0,1}
+\tfrac{2}{3}\CA^{0,1}\wedge \CA^{0,1}\wedge \CA^{0,1}\right)~,
\end{equation}
where $\CCZ$ is the subspace of $\CP^{3|4}$ for which
$\etab_i^\pm=0$, $\dparb$ is the antiholomorphic part of the
exterior derivative on $\CP^{3|4}$ and $\CA^{0,1}$ is the $(0,1)$
component of $\CA$ which we assume to satisfy\footnote{Here,
``$\lrcorner$'' denotes the interior product of vector fields with
differential forms.}
$\bV^i_\pm(\bar{V}_a^\pm\lrcorner\CA^{0,1})=0$ and
$\bV_\pm^i\lrcorner \CA^{0,1}=0$. The equations of motion of this
theory are readily derived to be
\begin{equation}\label{eq:2.42}
\dparb\CA^{0,1}+\CA^{0,1}\wedge\CA^{0,1}\ =\ 0~.
\end{equation}

\paragraph{Equivalence to $\CN=4$ SDYM theory.} By linearizing
\eqref{eq:2.42} around the trivial solution $\CA^{0,1}=0$, Witten
has shown \cite{Witten:2003nn} that the equations \eqref{eq:2.42}
are equivalent to the field equations of $\CN=4$ self-dual
Yang-Mills (SDYM) theory\footnote{This theory was introduced in
\cite{Siegel:1992za}.} on $\FR^4$. On the full nonlinear level,
this equivalence was demonstrated in \cite{Popov:2004rb}. The
$\CN=4$ SDYM equations on the Euclidean space $\FR^4$ take the
form
\begin{equation}\label{eq:2.43}
\begin{aligned}
f_{\ald\bed}&\ =\ 0~,
\\\eps^{\alpha\beta}D_{\alpha\ald}\chi^i_\beta&\ =\ 0~,
\\\square\phi^{ij}&\ =\ -\eps^{\alpha\beta}
\{\chi^i_\alpha,\chi^j_\beta\}~,
\\\eps^{\ald\bed}D_{\alpha\ald}\tilde{\chi}_{i\bed}&
\ =\ 2[\phi_{ij},\chi^j_\alpha]~,\\
\eps^{\ald\bed}D_{\alpha\ald}G_{\bed\gad}&\ =\
\{\chi^i_\alpha,\tilde{\chi}_{i\gad}\}
-\tfrac{1}{2}[\phi_{ij},D_{\alpha\gad}\phi^{ij}]~,
\end{aligned}
\end{equation}
where $f_{\ald\bed}:=-\tfrac{1}{2}\eps^{\alpha\beta}
F_{\alpha\ald\beta\bed}$ denotes the anti-self-dual part of the
curvature $F_{\alpha\ald\beta\bed}=$\linebreak
$[D_{\alpha\ald},D_{\beta\bed}]$. The fields in the $\CN=4$
supermultiplet $(f_{\alpha\beta},\chi_\alpha^{
i},\phi^{ij},\tilde{\chi}_{i\ald},G_{\ald\bed})$ carry the
helicities $(+1,+\frac{1}{2},0,$ $-\frac{1}{2},-1)$, and
$A_{\alpha\ald}$ in
$D_{\alpha\ald}=\dpar_{\alpha\ald}+A_{\alpha\ald}$ are the
components of a (self-dual) gauge potential. Furthermore, we have
introduced the abbreviation
$\phi_{ij}:=\frac{1}{2!}\eps_{ijkl}\phi^{kl}$ as well as the
operator $\square:=\frac{1}{2}\eps^{\alpha\beta}
\eps^{\ald\bed}D_{\alpha\ald}D_{\beta\bed}$. Note that all the
fields live in the adjoint representation of the gauge group. For
proving this equivalence via a twistor correspondence, one
considers only those gauge potentials $\CA^{0,1}$ for which the
component $\dpar_{\bl_\pm}\lrcorner\CA^{0,1}$ can be gauged away
\cite{Popov:2004rb}. This means, that one works with a
subset\footnote{This subset contains in particular the vacuum
solution $\CA^{0,1}=0$, and therefore hCS theory is perturbatively
equivalent to $\CN=4$ SDYM theory.} in the set of all solutions of
hCS theory on $\CP^{3|4}$. In the following, we always imply this
restriction when speaking about a twistor correspondence.

\paragraph{\v{C}ech description.} Note that solutions $\CA^{0,1}$
to \eqref{eq:2.42} define holomorphic structures
$\dparb_\CA=\dparb+\CA^{0,1}$ on $\CE$ and thus $(\CE,\dparb_\CA)$
is a holomorphic vector bundle (with trivial transition functions)
in the Dolbeault description. To switch to the \v{C}ech
description, we consider the restrictions of $\CA^{0,1}$ to the
patches $\hat{\CU}_\pm$. Since $\CA^{0,1}$ is flat due to
\eqref{eq:2.42}, we can write it locally as a pure gauge
configuration,
\begin{equation}
\CA^{0,1}|_{\hat{\CU}_+}\ =\ \psi_+\dparb\psi_+^{-1}\eand
\CA^{0,1}|_{\hat{\CU}_-}\ =\ \psi_-\dparb\psi_-^{-1}~,
\end{equation}
where the $\psi_\pm$ are $\sGL(n,\FC)$-valued functions on
$\hat{\CU}_\pm$ such that $\bV^i_\pm\psi_\pm=0$. The nontriviality
of the flat (0,1)-connection arises from the gluing condition on
$\hat{\CU}_+\cap\hat{\CU}_-$ which reads as
\begin{equation}
\psi_+\dparb\psi_+^{-1}\ =\ \psi_-\dparb\psi_-^{-1}~,
\end{equation}
since $\CE$ is a trivial bundle. Upon using the identity
$\dparb\psi_\pm^{-1}=-\psi_\pm^{-1}(\dparb\psi_\pm)\psi_\pm^{-1}$,
one obtains
\begin{equation}
\dparb(\psi_+^{-1}\psi_-)\ =\ 0
\end{equation}
and thus, we can define a holomorphic vector bundle
$\tilde{\CE}\rightarrow \CP^{3|4}$ with the holomorphic structure
$\dparb$ and a transition function
\begin{equation}
\tilde{f}_{+-}\ :=\  \psi_+^{-1}\psi_-~.
\end{equation}
One should stress that the bundles $\CE$ and $\tilde{\CE}$ are
diffeomorphic but not biholomorphic. In this description, the
above assumed existence of a gauge in which $\CA_{\bl_\pm}=0$ is
equivalent to the holomorphic triviality of the bundle
$\tilde{\CE}$ when restricted to any projective line
$\CPP^1_{x,\eta}\embd\CP^{3|4}$.

Note that the condition
\begin{equation}\label{realitycond}
\begin{aligned}
\psi_+(x,\eta_i^\ald,\lambda_+)&\ =\
\left(\psi_-^{-1}(\tau(x,\eta_i^\ald,\lambda_-))\right)^\dagger~,\\
\Leftrightarrow~~~ \tilde{f}_{+-}(x,\eta_i^\ald,\lambda_+)&\ =\
\left(\tilde{f}_{+-}\left(\tau(x,\eta_i^\ald,\lambda_+)
\right)\right)^\dagger~
\end{aligned}
\end{equation}
corresponds in the Dolbeault description to the fact that all the
fields in the supermultiplet $(f_{\alpha\beta},\chi_\alpha^{
i},\phi^{ij},\tilde{\chi}_{i\ald},G_{\ald\bed})$ take values in
the Lie algebra $\ahu(n)$ (or $\asu(n)$ if
$\det\tilde{f}_{+-}=1$).

Summarizing, there is a one-to-one correspondence between gauge
equivalence classes of solutions to the $\CN=4$ SDYM equations on
$(\FR^4,\delta_{\mu\nu})$ and equivalence classes of holomorphic
vector bundles $\tilde{\CE}$ over the supertwistor space
$\CP^{3|4}$, which become holomorphically trivial when restricted
to any $\tau$-invariant projective line
$\CPP^1_{x,\eta}\embd\CP^{3|4}$. Furthermore, because of the
equivalence of the data $(\tilde{\CE},\tilde{f}_{+-},\dparb)$ and
$(\CE,f_{+-}\!=\!\unit_n,\dparb_{\CA})$, the moduli space of hCS
theory on $(\CP^{3|4},\tau)$ is bijective to the moduli space of
$\CN=4$ SDYM theory on $(\FR^4,\delta_{\mu\nu})$.

\section{Cauchy-Riemann supermanifolds and mini-supertwistors}

The supersymmetric Bogomolny monopole equations are obtained from
the four-dimensional supersymmetric self-dual Yang-Mills equations
by the dimensional reduction $\FR^{4}\rightarrow \FR^{3}$. In this
section, we study in detail the meaning of this reduction on the
level of the supertwistor space. We find that the supertwistor
space $\CP^{3|4}$, when interpreted as the real manifold
$\FR^{4|8}\times S^2$, reduces to the space $\FR^{3|8}\times S^2$.
As a complex manifold, however, $\CP^{3|4}$ reduces to the rank
$1|4$ holomorphic vector bundle
$\CP^{2|4}:=\CO(2)\oplus\Pi\CO(1)\otimes\FC^4$ over $\CPP^1$. Due
to this difference, the twistor correspondence gets more involved
(one needs a double fibration). We also show that on the space
$\FR^{3|8}\times S^2$, one can introduce a so-called
Cauchy-Riemann structure, which generalizes the notion of a
complex structure. This allows us to use tools familiar from
complex geometry.

In later sections, we will then discuss the supertwistor
description of the supersymmetric Bogomolny equations and, in
particular, the reductions of hCS theory \eqref{eq:2.41} and
\eqref{eq:2.42} to theories on $\CP^{2|4}$ and $\FR^{3|8}\times
S^2$, which lay the ground for future calculations using twistor
string techniques.

\subsection{The dimensional reduction $\FR^{4|8}\times
S^2\rightarrow \FR^{3|8}\times S^2$}

\paragraph{Supersymmetric Bogomolny equations.} It is well known
that the Bogomolny equations on $\FR^3$ describing BPS monopoles
\cite{Bogomolny:1975de,Prasad:1975kr} can be obtained from the
SDYM equations on $\FR^4$ by demanding the components $A_\mu$ with
$\mu=1,\ldots,4$ of a gauge potential to be independent of $x^4$
and by putting $\Phi:=A_4$
\cite{Manton:1977ht,Hitchin:1982gh,Atiyah:1988jp}. Here, $\Phi$ is
a Lie-algebra valued scalar field in three dimensions (the Higgs
field) which enters into the Bogomolny equations. Obviously, one
can similarly reduce the $\CN=4$ SDYM equations \eqref{eq:2.43} on
$\FR^4$ by imposing the $\der{x^4}$-invariance condition on all
the fields
$(f_{\alpha\beta},\chi^i_\alpha,\phi^{ij},\tilde{\chi}_{i\ald},
G_{\ald\bed})$ in the supermultiplet and obtain supersymmetric
Bogomolny equations on $\FR^3$. Recall that both $\CN=4$ SDYM
theory and $\CN=4$ SYM theory have an $\sSU(4)\cong \sSpin(6)$
R-symmetry group. In the case of the full $\CN=4$ super Yang-Mills
theory, the R-symmetry group and supersymmetry get enlarged to
$\sSpin(7)$ and $\CN=8$ supersymmetry by a reduction from four to
three dimensions, cf.\ \cite{Chiou:2005jn}. However, the situation
in the dimensionally reduced $\CN=4$ SDYM theory is more involved
since there is no parity symmetry interchanging left-handed and
right-handed fields, and only the $\sSU(4)$ subgroup of
$\sSpin(7)$ is manifest as an R-symmetry of the Bogomolny model.

\paragraph{The reduction $\FR^{4|8}\rightarrow \FR^{3|8}$.} Recall
that on $\FR^4\cong\FC^2$, we may use the complex coordinates
$x^{\alpha\ald}$ satisfying the reality conditions \eqref{eq:2.15}
or the real coordinates $x^\mu$ defined in \eqref{eq:2.16}.
Translations generated by the vector field $\CCT_4:=\der{x^4}$ are
isometries of $\FR^{4|8}$ and by taking the quotient with respect
to the action of the Abelian group $\CCG:=\{\exp(a
\CCT_4):x^4\mapsto x^4+a,a\in \FR\}$ generated by $\CCT_4$, we
obtain the superspace $\FR^{3|8}\cong\FR^{4|8}/\CCG$. Recall that
the eight odd complex coordinates $\eta_i^\ald$ satisfy the
reality conditions \eqref{eq:2.33}. The vector field $\CCT_4$ is
trivially lifted to $\FR^{4|8}\times S^2$ (see e.g.\
\cite{Wolf:2004hp}) and therefore the supertwistor space,
considered as the smooth supermanifold $\FR^{4|8}\times S^2$, is
reduced to $\FR^{3|8}\times S^2\cong \FR^{4|8}\times S^2/\CCG$. In
other words, smooth $\CCT_4$-invariant functions on
$\CP^{3|4}\cong\FR^{4|8}\times S^2$ can be considered as ``free''
smooth functions on the supermanifold $\FR^{3|8}\times S^2$.

\paragraph{Bosonic coordinates on $\FR^{3|8}$.} Recall that the
rotation group $\sSO(4)$ of $(\FR^4,\delta_{\mu\nu})$ is locally
isomorphic to $\sSU(2)_L\times \sSU(2)_R\cong\sSpin(4)$. Upon
dimensional reduction to three dimensions, the rotation group
$\sSO(3)$ of $(\FR^3,\delta_{rs})$ with $r,s=1,2,3$ is locally
$\sSU(2)\cong \sSpin(3)$, which is the diagonal group
$\diag(\sSU(2)_L\times \sSU(2)_R)$. Therefore, the distinction
between undotted, i.e.\ $\sSU(2)_L$, and dotted, i.e.\
$\sSU(2)_R$, indices disappears. This implies, that one can
relabel the bosonic coordinates $x^{\alpha\bed}$ from
\eqref{eq:2.15}, \eqref{eq:2.16} by $x^{\ald\bed}$ and split them
as
\begin{equation}
x^{\ald\bed}\ =\ x^{(\ald\bed)}+x^{[\ald\bed]}\ :=\
\tfrac{1}{2}(x^{\ald\bed}+x^{\bed\ald})+
\tfrac{1}{2}(x^{\ald\bed}-x^{\bed\ald})~,
\end{equation}
into symmetric
\begin{equation}\label{eq:3.2}
y^{\ald\bed}\ :=\  -\di x^{(\ald\bed)}~~\mbox{with}~~y^{\ed\ed}\
=\ -\bar{y}^{\zd\zd}\ =\ (x^1+\di x^2)\ =:\ y~,~~~y^{\ed\zd}\ =\
\bar{y}^{\ed\zd}\ =\ -x^3
\end{equation}
and antisymmetric
\begin{equation}
x^{[\ald\bed]}\ =\  \eps^{\ald\bed}x^4
\end{equation}
parts. More abstractly, this splitting corresponds to the
decomposition $\rfour=\rthree\oplus\rone$ of the irreducible real
vector representation $\rfour$ of the group
$\sSpin(4)\cong\sSU(2)_L\times \sSU(2)_R$ into two irreducible
real representations $\rthree$ and $\rone$ of the group
$\sSpin(3)\cong\sSU(2)=\diag(\sSU(2)_L\times\sSU(2)_R)$. For
future use, we also introduce the operators
\begin{subequations}\label{eq:3.4}
\begin{equation}
\dpar_{(\ald\bed)}\ :=\
\frac{\di}{2}\left(\der{x^{\ald\bed}}+\der{x^{\bed\ald}}\right)~,
\end{equation}
which read explicitly as
\begin{equation}
\dpar_{(\ed\ed)}\ =\ \der{y^{\ed\ed}}~,~~~\dpar_{(\ed\zd)}\ =\
\frac{1}{2}\der{y^{\ed\zd}}\eand \dpar_{(\zd\zd)}\ =\
\der{y^{\zd\zd}}~.
\end{equation}
\end{subequations}
Altogether, we thus have
\begin{equation}\label{eq:3.5}
\der{x^{\ald\bed}}\ =\
-\di\dpar_{(\ald\bed)}-\tfrac{1}{2}\eps_{\ald\bed}\der{x^4}~.
\end{equation}

\subsection{The holomorphic reduction $\CP^{3|4}\rightarrow
\CP^{2|4}$}

\paragraph{The complex Abelian group action.} The vector field
$\CCT_4=\der{x^4}$ yields a free twistor space action of the
Abelian group $\CCG\cong \FR$ which is the real part of the
holomorphic action of the complex group $\CCG_\FC\cong \FC$. In
other words, we have
\begin{equation}\label{eq:T4}
\begin{aligned}
\CCT_4&\ =\ \der{x^4}\ =\
\derr{z_+^a}{x^4}\der{z_+^a}+\derr{\bz_+^a}{x^4}\der{\bz_+^a}\\& \
=\
\left(-\der{z^2_+}+z_+^3\der{z^1_+}\right)+\left(-\der{\bz^2_+}+
\bz_+^3\der{\bz^1_+}\right)
\ =:\ \CCT_+'+\bar{\CCT}_+'
\end{aligned}
\end{equation}
in the coordinates $(z_+^a,\eta_i^+)$ on $\hat{\CU}_+$, where
\begin{equation}\label{eq:T4compl}
\CCT_+'\ :=\ \CCT'|_{\hat{\CU}_+}\ =\
-\der{z^2_+}+z_+^3\der{z_+^1}
\end{equation}
is a holomorphic vector field on $\hat{\CU}_+$. Similarly, we
obtain
\begin{equation}
\CCT_4\ =\ \CCT'_-+\bar{\CCT}'_-~~~\mbox{with}~~~\CCT'_-\ :=\
\CCT'|_{\hat{\CU}_-}\ =\ -z^3_-\der{z_-^2}+\der{z^1_-}
\end{equation}
on $\hat{\CU}_-$ and $\CCT_+'=\CCT_-'$ on
$\hat{\CU}_+\cap\hat{\CU}_-$. For holomorphic functions $f$ on
$\CP^{3|4}$ we have
\begin{equation}
\CCT_4f(z^a_\pm,\eta_i^\pm)\ =\ \CCT' f(z_\pm^a,\eta_i^\pm)
\end{equation}
and therefore $\CCT'$-invariant holomorphic functions on
$\CP^{3|4}$ can be considered as ``free'' holomorphic functions on
a reduced space $\CP^{2|4}\cong \CP^{3|4}/\CCG_\FC$ obtained as
the quotient space of $\CP^{3|4}$ by the action of the complex
Abelian group $\CCG_\FC$ generated by $\CCT'$.

\paragraph{Reduction diagram.} In the purely bosonic case,
the space $\CP^2\cong \CP^3/\CCG_\FC$ was called mini-twistor
space \cite{Hitchin:1982gh} and we shall refer to $\CP^{2|4}$ as
the {\em mini-supertwistor space}. To sum up, the reduction of the
supertwistor correspondence induced by the $\CCT_4$-action is
described by the following diagram:
\begin{equation}\label{mrsuperdblfibration2}
\begin{aligned}
\begin{picture}(150,100)
\put(0.0,0.0){\makebox(0,0)[c]{$\CP^{2|4}$}}
\put(140.0,0.0){\makebox(0,0)[c]{$\FR^{3|8}$}}
\put(0,90.0){\makebox(0,0)[c]{$\CP^{3|4}$}}
\put(30.0,88.0){\makebox(0,0)[c]{$\cong$}}
\put(73.0,90.0){\makebox(0,0)[c]{$\FR^{4|8}\times S^2$}}
\put(102.0,90.0){\vector(1,0){20}}
\put(140,90.0){\makebox(0,0)[c]{$\FR^{4|8}$}}
\put(73.0,50.0){\makebox(0,0)[c]{$\FR^{3|8}\times S^2$}}
\put(26.0,27.0){\makebox(0,0)[c]{$\pi_2$}}
\put(110.0,27.0){\makebox(0,0)[c]{$\pi_1$}}
\put(59.0,37.0){\vector(-4,-3){40}}
\put(82.0,37.0){\vector(4,-3){40}}
\put(0.0,80.0){\vector(0,-1){70}}
\put(72.0,80.0){\vector(0,-1){20}}
\put(140.0,80.0){\vector(0,-1){70}}
\end{picture}
\end{aligned}
\end{equation}
Here, $\downarrow$ symbolizes projections generated by the action
of the groups $\CCG$ or $\CCG_\FC$ and $\pi_1$ is the canonical
projection. The projection $\pi_2$ will be described momentarily.

\subsection{Geometry of the mini-supertwistor space}

\paragraph{Local coordinates.} It is not difficult to see that the
functions
\begin{equation}\label{eq:3.10}
\begin{aligned}
&w_+^1\ :=\ -\di(z_+^1+z_+^3z_+^2)~,~~~w_+^2\ :=\
z_+^3\eand\eta_i^+~~~\mbox{on}~~~\hat{\CU}_+~,\\&w_-^1\ :=\
-\di(z_-^2+z_-^3z_-^1)~,~~~w_-^2\ :=\
z_-^3\eand\eta_i^-~~~\mbox{on}~~~\hat{\CU}_-
\end{aligned}
\end{equation}
are constant along the $\CCG_\FC$-orbits in $\CP^{3|4}$ and thus
descend to the patches $\hat{\CV}_\pm:=\hat{\CU}_\pm\cap\CP^{2|4}$
covering the (orbit) space $\CP^{2|4}\cong \CP^{3|4}/\CCG_\FC$. On
the overlap $\hat{\CV}_+\cap\hat{\CV}_-$, we have
\begin{equation}
w_+^1\ =\ \frac{1}{(w_-^2)^2}w_-^1~,~~~w_+^2\ =\
\frac{1}{w_-^2}\eand \eta_i^+\ =\ \frac{1}{w_-^2}\eta_i^-
\end{equation}
which coincides with the transformation of canonical coordinates
on the total space
\begin{equation}
\CO(2)\oplus\Pi\CO(1)\otimes\FC^4\ =\ \CP^{2|4}
\end{equation}
of the holomorphic vector bundle
\begin{equation}
\CP^{2|4}\ \rightarrow\ \CPP^1~.
\end{equation}
This space is a Calabi-Yau supermanifold \cite{Chiou:2005jn} with
a holomorphic volume form
\begin{equation}\label{eq:3.14}
\Omega|_{\hat{\CV}_\pm}\ :=\ \pm\dd w_\pm^1\wedge \dd w_\pm^2
\dd\eta_1^\pm\cdots\dd\eta_4^\pm~.
\end{equation}
The body of this Calabi-Yau supermanifold is the mini-twistor
space \cite{Hitchin:1982gh}
\begin{equation}
\CP^2\ \cong\ \CO(2)\ \cong\ T^{1,0}\CPP^1~,
\end{equation}
where $T^{1,0}\CPP^1$ denotes the holomorphic tangent bundle of
$\CPP^1$. Note that the space $\CP^{2|4}$ can be considered as an
open subset of the weighted projective space
$W\CPP^{2|4}(2,1,1|1,1,1,1)$.

\paragraph{Incidence relations.} The real structure $\tau$ on
$\CP^{3|4}$ induces a real structure on $\CP^{2|4}$ acting on
local coordinates by the formula
\begin{equation}\label{eq:3.16}
\tau(w_\pm^1,w_\pm^2,\eta_i^\pm)\ =\
\left(-\frac{\bar{w}_\pm^1}{(\bar{w}_\pm^2)^2},
-\frac{1}{\bar{w}_\pm^2},\pm\frac{1}{\bar{w}^2_\pm}
T_i{}^j\etab_j^\pm\right)~,
\end{equation}
where the matrix $T=(T_i{}^j)$ has already been defined in
\eqref{eq:2.31}. From \eqref{eq:3.16}, one sees that $\tau$ has no
fixed points in $\CP^{2|4}$ but leaves invariant projective lines
$\CPP^1_{x,\eta}\embd\CP^{2|4}$ defined by the equations
\begin{equation}\label{eq:3.17}
\begin{aligned}
w_+^1&\ =\ y-2\lambda_+x^3-\lambda_+^2\bar{y}~,&\eta_i^+&\ =\
\eta_i^\ed+\lambda_+\eta_i^\zd&&\mbox{with}~~\lambda_+\ =\ w_+^2\
\in\
U_+~,\\
w_-^1&\ =\ \lambda_-^2y-2\lambda_-x^3-\bar{y}~,&\eta_i^-&\ =\
\lambda_-\eta_i^\ed+\eta_i^\zd&&\mbox{with}~~\lambda_-\ =\ w_-^2\
\in\ U_-~
\end{aligned}
\end{equation}
for fixed $(x,\eta)\in\FR^{3|8}$. Here, $y=x^1+\di x^2$,
$\bar{y}=x^1-\di x^2$ and $x^3$ are coordinates on $\FR^3$.

By using the coordinates \eqref{eq:3.2}, we can rewrite
\eqref{eq:3.17} as
\begin{equation}\label{eq:3.18}
w_\pm^1\ =\ \lambda_\ald^\pm\lambda_\bed^\pm
y^{\ald\bed}~,~~~w_\pm^2\ =\ \lambda_\pm\eand \eta_i^\pm\ =\
\eta_i^\ald\lambda_\ald^\pm~,
\end{equation}
where the explicit form of $\lambda_\ald^\pm$ has been given in
\eqref{eq:2.12}. In fact, the equations \eqref{eq:3.18} are the
incidence relations which lead to the double fibration
\begin{equation}\label{eq:3.19}
\begin{aligned}
\begin{picture}(50,40)
\put(0.0,0.0){\makebox(0,0)[c]{$\CP^{2|4}$}}
\put(64.0,0.0){\makebox(0,0)[c]{$\FR^{3|8}$}}
\put(34.0,33.0){\makebox(0,0)[c]{$\CF^{5|8}$}}
\put(7.0,18.0){\makebox(0,0)[c]{$\pi_2$}}
\put(55.0,18.0){\makebox(0,0)[c]{$\pi_1$}}
\put(25.0,25.0){\vector(-1,-1){18}}
\put(37.0,25.0){\vector(1,-1){18}}
\end{picture}
\end{aligned}
\end{equation}
where $\CF^{5|8}\cong \FR^{3|8}\times S^2$, $\pi_1$ is again the
canonical projection onto $\FR^{3|8}$ and the projection $\pi_2$
is defined by the formula
\begin{equation}\label{eq:3.20}
\pi_2(x^r,\lambda_\pm,\eta_i^\ald)\ =\
\pi_2(y^{\ald\bed},\lambda_\ald^\pm,\eta_i^\ald)\ =\
(w^1_\pm,w_\pm^2,\eta_i^\pm)~,
\end{equation}
where $r=1,2,3$, and $w_\pm^{1,2}$ and $\eta_i^\pm$ are given in
\eqref{eq:3.18}. The diagram \eqref{eq:3.19}, which is a part of
\eqref{mrsuperdblfibration2}, describes the one-to-one
correspondences
\begin{equation}
\begin{aligned}
\{\,\mbox{$\tau$-invariant projective lines $\CPP^1_{x,\eta}$ in
$\CP^{2|4}$}\}\ \longleftrightarrow\ \{\,\mbox{points $(x,\eta)$
in
$\FR^{3|8}$}\}~,\\
\{\,\mbox{points $p$ in $\CP^{2|4}$}\}\ \longleftrightarrow\
\{\,\mbox{oriented $(1|0)$-dimensional lines $\ell_p$ in
$\FR^{3|8}$}\}~.~~~
\end{aligned}
\end{equation}

\subsection{Cauchy-Riemann supertwistors}

\paragraph{Cauchy-Riemann structure.} Consider the double
fibration \eqref{eq:3.19}. For a moment, let us restrict ourselves
to the purely bosonic setup. The body of the space
$\FR^{3|8}\times S^2$ is the real five-dimensional manifold
$\FR^3\times S^2$. As an odd-dimensional space, this is obviously
not a complex manifold, but it can be understood as a so-called
{\em Cauchy-Riemann} (CR) {\em manifold}, i.e.\ as a {\em
partially complex manifold}. Recall that a {\em CR structure} on a
smooth manifold $X$ of real dimension $d$ is a complex subbundle
$\CCDb$ of rank $m$ of the complexified tangent bundle $T_\FC X$
such that $\CCD\cap\CCDb=\{0\}$ and $\CCDb$ is involutive
(integrable), i.e., the space of smooth sections of $\CCDb$ is
closed under the Lie bracket. Obviously, the
distribution\footnote{We use the same letter for the bundle $\CCD$
and a distribution generated by its sections.} $\CCD$ is
integrable if $\CCDb$ is integrable. The pair $(X,\CCDb)$ is
called a CR manifold of dimension $d=\dim_\FR X$, of rank
$m=\dim_\FC \CCDb$ and of codimension $d-2m$. In particular, a CR
structure on $X$ in the special case $d=2m$ is a complex structure
on $X$. Thus, the notion of CR manifolds generalizes that of
complex manifolds.

On the manifold $\FR^3\times S^2$, one can introduce several CR
structures. One of them, which we denote by $\CCDb_0$, is
generated by the vector fields $\{\dpar_\by,\dpar_{\bl_\pm}\}$ and
corresponds to the identification $\CF_0^5:=(\FR^3\times
S^2,\CCDb_0)\cong \FR\times\FC\times \CPP^1$. Another one, denoted
by $\CCDb$, is spanned by the basis sections
$\{\dpar_{\bw_\pm^1},\dpar_{\bw^2_\pm}\}$ of the bundle
$T_\FC(\FR^3\times S^2)$. Moreover, $\CCD\cap\CCDb=\{0\}$ and the
distribution $\CCDb$ is integrable since
$[\dpar_{\bw^1_\pm},\dpar_{\bw^2_\pm}]=0$. Therefore, the pair
$(\FR^3\times S^2,\CCDb)=:\CF^5$ is also a CR manifold. It is
obvious that there is a diffeomorphism between the manifolds
$\CF^5$ and $\CF_0^5$, but this is not a CR diffeomorphism since
it does not respect the chosen CR structures. Note that a CR
five-manifold generalizing the above manifold $\CF^5$ can be
constructed as a sphere bundle over an arbitrary three-manifold
with conformal metric \cite{LeBrun:1984}. Following
\cite{LeBrun:1984}, we shall call $\CF^5$ the {\em CR twistor
space}.

So far, we have restricted our attention to the purely bosonic
setup. However, the above definitions carry naturally over to the
case of supermanifolds (see e.g.\ \cite{Howe:1995md}). Namely, by
considering the integrable distribution $\hat{\CCDb}$ generated by
the vector fields $\dpar_{\bw^1_\pm}$, $\dpar_{\bw^2_\pm}$ and
$\dpar_{\etab_i^\pm}$, we obtain the CR supertwistor space
\begin{equation}
\CF^{5|8}\ :=\  (\FR^{3|8}\times S^2,\hat{\CCDb})
\end{equation}
with
\begin{equation}\label{eq:3.23}
\hat{\CCDb}\ =\
\spn\left\{\der{\bw^1_\pm},\der{\bw^2_\pm},\der{\etab_i^\pm}\right\}~.
\end{equation}
Similarly, we obtain the CR supermanifold
$\CF_0^{5|8}\!:=\!(\FR^{3|8}\times S^2,\hat{\CCDb}_0)$ for the
distribution
$\hat{\CCDb}_0=\spn\{\der{\by},\der{\bl^{\phantom{\dagger}}_\pm},
\der{\etab_i^\ed}\}$. In both cases, the CR structures have rank
$2|4$.

\paragraph{Real and complex coordinates on $\CF^{5|8}$.} Up to now,
we have used the coordinates
$(y,\by,x^3,\lambda_\pm,\bl_\pm,\eta_i^\ald)$ or
$(y^{\ald\bed},\lambda_\ald^\pm,\hl_\ald^\pm,\eta_i^\ald)$ on the
two patches $\tilde{\CV}_\pm$ covering the superspace
$\FR^{3|8}\times S^2$. More appropriate for the distribution
\eqref{eq:3.23} are, however, the coordinates \eqref{eq:3.17}
together with
\begin{equation}\label{eq:3.24}
\begin{aligned}
w_+^3&\ :=\ \tfrac{1}{1+\lambda_+\bl^{\phantom{\dagger}}_+}
\left[\bl_+y+(1-\lambda_+\bl_+)x^3+\lambda_+\by\right]
~~~\mbox{on}~~\tilde{\CV}_+~,\\
w_-^3&\ :=\ \tfrac{1}{1+\lambda_-\bl^{\phantom{\dagger}}_-}
\left[\lambda_-y+(\lambda_-\bl_--1)x^3+\bl_-\by\right]
~~~\mbox{on}~~\tilde{\CV}_-~.
\end{aligned}
\end{equation}
In terms of the coordinates \eqref{eq:3.2} and $\lambda_\ald^\pm$
from \eqref{eq:2.12}, we can rewrite \eqref{eq:3.17} and
\eqref{eq:3.24} concisely as
\begin{equation}\label{eq:3.25}
w_\pm^1\ =\ \lambda_\ald^\pm\lambda_\bed^\pm y^{\ald\bed}~,~~~
w_\pm^2\ =\ \lambda_\pm~,~~~ w_\pm^3\ =\
-\gamma_\pm\lambda_\ald^\pm\hat{\lambda}^\pm_\bed
y^{\ald\bed}\eand\eta_i^\pm\ =\ \eta_i^\ald\lambda_\ald^\pm~,
\end{equation}
where the factors $\gamma_\pm$ have been given in \eqref{eq:2.22}.
Note that the coordinates $w_\pm^3$ are real and all the other
coordinates in \eqref{eq:3.25} are complex. The relations between
the coordinates on $\tilde{\CV}_+\cap\tilde{\CV}_-$ follow
directly from their definitions \eqref{eq:3.25}.

\paragraph{Projection onto $\CP^{2|4}$.} The coordinates
$w_\pm^{1,2}$ and $\eta_i^\pm$ have already appeared in
\eqref{eq:3.18} since $\CP^{2|4}$ is a complex subsupermanifold of
$\CF^{5|8}$. Recall that the formul\ae{} \eqref{eq:3.20} together
with \eqref{eq:3.25} define a projection
\begin{equation}\label{eq:3.26}
\pi_2\ :\ \CF^{5|8}\ \rightarrow\ \CP^{2|4}
\end{equation}
onto the mini-supertwistor space $\CP^{2|4}$. The fibres over
points $p\in\CP^{2|4}$ of this projection are real one-dimensional
spaces $\od_p\cong \FR$ parametrized by the coordinates $w_\pm^3$.
Note that the pull-back to $\CF^{5|8}$ of the real structure
$\tau$ on $\CP^{2|4}$ given in \eqref{eq:3.16} reverses the
orientation of each line $\od_p$, since $\tau(w_\pm^3)=-w_\pm^3$.

In order to clarify the geometry of the fibration \eqref{eq:3.26},
we note that the body $\CF^5\cong\FR^3\times S^2$ of the
supermanifold $\CF^{5|8}$ can be considered as the sphere bundle
\begin{equation}
S(T\FR^3)\ =\ \{(x,u)\in T\FR^3\,|\,\delta_{rs}u^r u^s=1\}\ \cong\
\FR^3\times S^2
\end{equation}
whose fibres at points $x\in\FR^3$ are spheres of unit vectors in
$T_x\FR^3$ \cite{Hitchin:1982gh}. Since this bundle is trivial,
its projection onto $\FR^3$ is obviously $\pi_1(x,u)=x$. Moreover,
the complex two-dimensional mini-twistor space $\CP^2$ can be
described as the space of all oriented lines in $\FR^3$. That is,
any such line $\od$ is defined by a unit vector $u^r$ in the
direction of $\od$ and a shortest vector $v^r$ from the origin in
$\FR^3$ to $\od$, and one can show \cite{Hitchin:1982gh} that
\begin{equation}
\CP^2\ =\ \{(v,u)\in
T\FR^3\,|\,\delta_{rs}u^rv^s=0\,,~\delta_{rs}u^ru^s=1\}\ \cong\
T^{1,0}\CPP^1\ \cong\ \CO(2)~.
\end{equation}

The fibres of the projection $\pi_2:\FR^3\times S^2\rightarrow
\CP^2$ are the orbits of the action of the group $\CCG'\cong \FR$
on $\FR^3\times S^2$ given by the formula
$(v^r,u^s)\mapsto(v^r+tu^r,u^s)$ for $t\in\FR$ and
\begin{equation}
\CP^2\ \cong\ \FR^3\times S^2/\CCG'~.
\end{equation}

Recall that $\CF^5\cong \FR^3\times S^2$ is a (real) hypersurface
in the twistor space $\CP^3$. On the other hand, $\CP^2$ is a
complex two-dimensional submanifold of $\CF^5$ and therefore
\begin{equation}
\CP^2\ \subset\ \CF^5\ \subset\ \CP^3~.
\end{equation}
Similarly, we have in the supertwistor case
\begin{equation}
\CP^{2|4}\ \subset\ \CF^{5|8}\ \subset\ \CP^{3|4}~.
\end{equation}

\paragraph{Vector fields on $\CF^{5|8}$.} The formul\ae{}
\eqref{eq:3.17} and \eqref{eq:3.24} define a coordinate
transformation
$(y,\by,x^3,\lambda_\pm,\bl_\pm,\eta_i^\ald)\mapsto(w_\pm^a,\eta_i^\pm)$
on $\CF^{5|8}$. From corresponding inverse formul\ae{} defining
the transformation
$(w_\pm^a,\eta_i^\pm)\mapsto(y,\by,x^3,\lambda_\pm,\bl_\pm,\eta_i^\ald)$,
we obtain
\begin{subequations}
\begin{align}
\begin{aligned}
\der{w_+^1}&\ =\
\gamma_+^2\left(\der{y}-\bl_+\der{x^3}-\bl_+^2\der{\by}\right)\
=:\ \gamma_+^2
W_1^+~,\\
\der{w_+^2}&\ =\
W_2^++2\gamma_+^2(x^3+\lambda_+\by)W_1^+-\gamma_+^2(\by-2\bl_+
x^3-\bl_+^2 y) W_3^+-\gamma_+\etab_i^\ed V_+^i~,\\
\der{w_+^3}&\ =\
2\gamma_+\left(\lambda_+\der{y}+\bl_+\der{\by}+\frac{1}{2}
(1-\lambda_+\bl_+)\der{x^3}\right)\
=:\ W_3^+~,
\end{aligned}
\intertext{as well as}
\begin{aligned}
\der{w_-^1}&\ =\ \gamma_-^2\left(\bl_-^2\der{y}-\bl_-\der{x^3}
-\der{\by}\right)\ =:\ \gamma_-^2W_1^-~,\\
\der{w_-^2}&\ =\
W_2^-+2\gamma_-^2(x^3-\lambda_-y)W_1^-+\gamma_-^2(\bl_-^2\by
-2\bl_-x^3-y)W_3^-+\gamma_-\etab^\zd_i
V_-^i~,\\
\der{w_-^3}&\ =\
2\gamma_-\left(\bl_-\der{y}+\lambda_-\der{\by}+\frac{1}{2}
(\lambda_-\bl_--1)\der{x^3}\right)\
=:\ W_3^-~,
\end{aligned}
\end{align}
\end{subequations}
where $W_2^\pm:=\der{\lambda_\pm}$. Thus, when working in the
coordinates $(y,\by,x^3,\lambda_\pm,\bl_\pm,\eta_i^\ald)$ on
$\tilde{\CV}_\pm\subset \CF^{5|8}$, we will use the bosonic vector
fields $W_a^\pm$ with $a=1,2,3$ and the fermionic vector fields
$V_\pm^i$ together with their complex conjugates\footnote{Note
that the vector field $W_3^\pm$ is real.} $\bW_{1,2}^\pm$ and
$\bV_\pm^i$, respectively.

In the coordinates
$(y^{\ald\bed},\lambda_\ald^\pm,\hl_\ald^\pm,\eta_i^\ald)$, the
vector fields $W_a^\pm$, $V_\pm^i$ and their complex conjugates
read as
\begin{subequations}\label{eq:3.34all}
\begin{align}\label{eq:3.34}
\begin{aligned}
W_1^\pm\ =\
\hl^\ald_\pm\hl^\bed_\pm\dpar_{(\ald\bed)}~,~~~W_2^\pm\ =\
\dpar_{\lambda_\pm}~,~~~ W_3^\pm\ =\
2\gamma_\pm\hl^\ald_\pm\lambda^\bed_\pm\dpar_{(\ald\bed)}~,\\
V_\pm^i\ =\ -\hl_\pm^\ald T_j{}^i\der{\eta_j^\ald}~\hspace{4.0cm}~
\end{aligned}
 \intertext{as well
as} \label{eq:3.35}
\begin{aligned}
 \bW_1^\pm\ =\
-\lambda^\ald_\pm\lambda_\pm^\bed\dpar_{(\ald\bed)}~,~~~\bW_2^\pm\
=\ \dpar_{\bl_\pm}~,~~~ \bW_3^\pm\ =\ W_3^\pm\ =\
2\gamma_\pm\hl^\ald_\pm\lambda^\bed_\pm\dpar_{(\ald\bed)}~,\\
\bV_\pm^i\ =\ \lambda_\pm^\ald \der{\eta_i^\ald}~,\hspace{4.5cm}
\end{aligned}
\end{align}
\end{subequations}
where
\begin{equation}\label{eq:3.38}
(\hat{\lambda}^\ald_+)\ =\ \left(\begin{array}{c} 1\\ \bl_+
\end{array}\right)\eand
(\hat{\lambda}^\ald_-)\ =\ \left(\begin{array}{c} \bl_-\\
1\end{array}\right)~.
\end{equation}
Recall that the vector fields $\bW_1^\pm$, $\bW_2^\pm$ and
$\bV_\pm^i$ generate a distribution $\hat{\CCDb}$ (CR structure)
on $\FR^{3|8}\times S^2$, which is obviously integrable as these
vector fields commute with each other.

\paragraph{Forms on $\CF^{5|8}$.} It is not difficult to see that
forms dual to the vector fields \eqref{eq:3.34} and
\eqref{eq:3.35} are
\begin{subequations}\label{eq:3.37}
\begin{align}
\begin{aligned}
\Theta_\pm^1\ :=\  \gamma^2_\pm\lambda_\ald^\pm\lambda_\bed^\pm\dd
y^{\ald\bed}~,~~~ \Theta_\pm^2\ :=\  \dd \lambda_\pm~,~~~
\Theta_\pm^3\ :=\  -\gamma_\pm\lambda_\ald^\pm\hl_\bed^\pm\dd
y^{\ald\bed}~,\\ E_i^\pm\ :=\ \gamma_\pm\lambda_\ald^\pm
T_i{}^j\dd \eta_j^\ald\hspace{4cm}
\end{aligned}
\intertext{and}
\begin{aligned}
\Thetab_\pm^1\ =\ -\gamma^2_\pm\hl_\ald^\pm\hl_\bed^\pm\dd
y^{\ald\bed}~,~~~\Thetab_\pm^2\ =\ \dd \bl_\pm~,~~~\Thetab_\pm^3\
=\ \Theta^3_\pm~,\hspace{1cm}\\\bar{E}_i^\pm\ =\
-\gamma_\pm\hl^\pm_\ald\dd \eta^\ald_i~,\hspace{4cm}
\end{aligned}
\end{align}
\end{subequations}
where $T_i{}^j$ has been given in \eqref{eq:2.31}. The exterior
derivative on $\CF^{5|8}$ reads as
\begin{align}\nonumber
\dd|_{\tilde{\CV}_\pm}&\ =\ \dd w_\pm^1\der{w_\pm^1}+\dd
w_\pm^2\der{w_\pm^2}+\dd \bw_\pm^1\der{\bw_\pm^1}+\dd
\bw_\pm^2\der{\bw_\pm^2}+\dd
w_\pm^3\der{w_\pm^3}+\dd\eta_i^\pm\der{\eta_i^\pm}+\dd
\etab_i^\pm\der{\etab_i^\pm}\\
&\ =\ \Theta^1_\pm
W_1^\pm+\Theta_\pm^2W_2^\pm+\Thetab_\pm^1\bW_1^\pm+
\Thetab_\pm^2\bW_2^\pm+\Theta^3_\pm
W_3^\pm +E_i^\pm V^i_\pm+\bar{E}_i^\pm \bar{V}^i_\pm~.
\end{align}
Note again that $\Theta_\pm^3$ and $W_3^\pm$ are both
real.\footnote{To homogenize the notation later on, we shall also
use $\bW_3^\pm$ and $\dpar_{\bw^\pm_3}$ instead of $W_3^\pm$ and
$\dpar_{w^\pm_3}$, respectively.}

\section{Partially holomorphic Chern-Simons theory}

We have discussed how the mini-supertwistor and CR supertwistor
spaces arise via dimensional reductions from the supertwistor
space of four-dimensional spacetime. Subject of this section is
the discussion of a generalization of Chern-Simons theory and its
relatives to this setup. We call the theory we are about to
introduce {\em partially holomorphic Chern-Simons theory} or phCS
theory for short. Roughly speaking, this theory is a mixture of
Chern-Simons and holomorphic Chern-Simons theory on the CR
supertwistor space $\CF^{5|8}$ which has one real and two complex
bosonic dimensions. This theory is a reduction of hCS theory on
$\CP^{3|4}$. As we will show below, there is a one-to-one
correspondence between the moduli space of solutions to the
equations of motion of phCS theory on $\CF^{5|8}$ and the moduli
space of solutions to $\CN=4$ supersymmetric Bogomolny equations
on $\FR^3$, quite similar to the correspondence which exists
between hCS theory on the supertwistor space $\CP^{3|4}$ and
$\CN=4$ SDYM theory in four dimensions
\cite{Witten:2003nn,Popov:2004rb}.

\subsection{Partially flat connections}

In this subsection, we restrict ourselves to the purely bosonic
case since the extension of all definitions to supermanifolds is
straightforward.

\paragraph{Integrable distribution $\CT$.} Let $X$ be a smooth
manifold of real dimension $d$ and $T_\FC X$ the complexified
tangent bundle of $X$. A subbundle $\CT\subset T_\FC X$ is
integrable if $i$) $\CT\cap\bar{\CT}$ has constant rank $k$ and
$ii$) $\CT$ and $\CT\cap\bar{\CT}$ are closed under the Lie
bracket.\footnote{Again, we use the same letter for the bundle
$\CT$ and a distribution generated by its sections.} If $\CT$ is
such an integrable subbundle of $T_\FC X$ then $X$ can be covered
by open sets and on each open set $U$, there are coordinates
$u^1,\ldots,u^l,v^1,\ldots,v^k,x^1,\ldots,x^m,$ $y^1,\ldots,y^m$
such that $\CT$ is locally spanned by the vector fields
\begin{equation}
\der{v^1},\ldots,\der{v^k},\der{\bw^1},\ldots,\der{\bw^m}~,
\end{equation}
where $\bw^1=x^1-\di y^1,\ldots,\bw^m=x^m-\di y^m$
\cite{Nirenberg}. Note that a CR structure is the special case of
an integrable subbundle $\CT$ with $k=0$.

For any smooth function $f$ on $X$, let $\dd_\CT f$ denote the
restriction of $\dd f$ to $\CT$, i.e., $\dd_\CT$ is the
composition
\begin{equation}
C^\infty(X)\ \stackrel{\dd}{\longrightarrow}\ \Omega^1(X)\
\longrightarrow\ \Gamma(X,\CT^*)~,
\end{equation}
where $\Omega^1(X):=\Gamma(X,T^* X)$ and $\CT^*$ denotes the sheaf
of (smooth) one-forms dual to $\CT$ \cite{Rawnsley}. The operator
$\dd_\CT$ can be extended to act on relative $q$-forms from the
space $\Omega^q_\CT(X):=\Gamma(X,\Lambda^q\CT^*)$,
\begin{equation}
\dd_\CT\ :\ \Omega^q_\CT(X)\ \rightarrow\
\Omega_\CT^{q+1}(X)~,~~~\mbox{for}~~~q\ \geq\ 0~.
\end{equation}

\paragraph{Connection on $\CT$.} Let $\CE$ be a smooth complex
vector bundle over $X$. A covariant differential (or connection)
on $\CE$ along the distribution $\CT$ -- a $\CT$-connection
\cite{Rawnsley} -- is a $\FC$-linear mapping
\begin{equation}
D_\CT\ :\ \Gamma(X,\CE)\ \rightarrow\ \Gamma(X,\CT^*\otimes \CE)
\end{equation}
satisfying the Leibniz formula
\begin{equation}
D_\CT(f\sigma)\ =\ fD_\CT \sigma+\dd_\CT f\otimes \sigma~,
\end{equation}
where $\sigma\in\Gamma(X,\CE)$ is a local section of $\CE$ and $f$
is a local smooth function. This $\CT$-connection extends to a map
\begin{equation}
D_\CT\ :\ \Omega^q_\CT(X,\CE)\ \rightarrow\
\Omega_\CT^{q+1}(X,\CE)~,
\end{equation}
where $\Omega^q_\CT(X,\CE):=\Gamma(X,\Lambda^q\CT^*\otimes \CE)$.
Locally, $D_\CT$ has the form
\begin{equation}
D_\CT\ =\ \dd_\CT+\CA_\CT~,
\end{equation}
where the standard $\sEnd\CE$-valued $\CT$-connection one-form
$\CA_\CT$ has components only along the distribution $\CT$. As
usual, $D^2_\CT$ naturally induces
\begin{equation}
\CF_\CT\in\Gamma(X,\Lambda^2\CT^*\otimes \sEnd\CE)
\end{equation}
which is the curvature of $\CA_\CT$. We say that $D_\CT$ (or
$\CA_\CT$) is flat, if $\CF_\CT=0$. For a flat $D_\CT$, the pair
$(\CE,D_\CT)$ is called a $\CT$-flat vector bundle
\cite{Rawnsley}. In particular, if $\CT$ is a CR structure then
$(\CE,D_\CT)$ is a CR vector bundle. Moreover, if $\CT$ is the
integrable bundle $T^{0,1} X$ of vectors of type $(0,1)$ then the
$\CT$-flat complex vector bundle $(\CE,D_\CT)$ is a holomorphic
bundle.

\subsection{Field equations on the CR supertwistor space}

\paragraph{A distribution $\CT$ on $\CF^{5|8}$.} Consider the
CR supertwistor space $\CF^{5|8}$ and a distribution $\CT$
generated by the vector fields $\bW_1^\pm$, $\bW_2^\pm$,
$\bV^i_\pm$ from the CR structure $\hat{\CCDb}$ and $\bW_3^\pm$.
This distribution is integrable since all conditions described in
section 4.1 are satisfied, e.g.\ the only nonzero commutator is
\begin{equation}
[\bW_2^\pm,\bW_3^\pm]\ =\ \pm2 \gamma_\pm^2\bW_1^\pm
\end{equation}
and therefore $\CT$ is closed under the Lie bracket. Also,
\begin{equation}
\CCV\ :=\ \CT\cap\bar{\CT}
\end{equation}
is a real one-dimensional and hence integrable distribution. The
vector fields $\bW_3^\pm$ are a basis for $\CCV$ over the patches
$\tilde{\CV}_\pm\subset \CF^{5|8}$. Note that the
mini-supertwistor space $\CP^{2|4}$ is a subsupermanifold of
$\CF^{5|8}$ transversal to the leaves of $\CCV=\CT\cap\bar{\CT}$
and furthermore, $\CT|_{\CP^{2|4}}=\hat{\CCDb}$. Thus, we have an
integrable distribution $\CT=\hat{\CCDb}\oplus\CCV$ on the CR
supertwistor space $\CF^{5|8}$ and we will denote by $\CT_b$ its
bosonic part generated by the vector fields $\bW_1^\pm$,
$\bW_2^\pm$ and $\bW_3^\pm$,
\begin{equation}
\CT_b\ :=\ \mathrm{span}\{\bW_1^\pm,\bW_2^\pm,\bW_3^\pm\}~.
\end{equation}

\paragraph{Holomorphic integral form.} Let $\CE$ be a trivial
rank $n$ complex vector bundle over $\CF^{5|8}$ and $\CA_\CT$ a
$\CT$-connection one-form on $\CE$ with
$\CT=\hat{\CCDb}\oplus\CCV$. Consider now the subspace $\CCX$ of
$\CF^{5|8}$ which is parametrized by the same bosonic coordinates
but only the holomorphic fermionic coordinates of $\CF^{5|8}$,
i.e.\ on $\CCX$, all objects are holomorphic in $\eta_i^\pm$. As
it was already noted in \cite{Chiou:2005jn}, the $\CN=4$
mini-supertwistor space is a Calabi-Yau supermanifold. In
particular, this ensures the existence of a holomorphic volume
form on $\CP^{2|4}$. Moreover, $\CP^{2|4}\subset\CF^{5|8}$ and the
pull-back $\tilde{\Omega}$ of this form is globally defined
on\footnote{Recall that the integrable distribution $\hat{\CCD}$
and its dual are related to the holomorphic tangent and cotangent
bundles of $\CP^{2|4}$.} $\CF^{5|8}$. Locally, on the patches
$\tilde{\CV}_\pm\subset \CF^{5|8}$, one obtains
\begin{equation}\label{eq:4.12}
\tilde{\Omega}|_{\tilde{\CV}_\pm}\ =\ \pm\dd w_\pm^1\wedge \dd
w_\pm^2\dd \eta_1^\pm\cdots\dd\eta_4^\pm~.
\end{equation}
This well-defined integral form allows us to integrate on $\CCX$
by pairing it with elements from $\Omega^3_{\CT_b}(\CCX)$.

\paragraph{Action functional for phCS theory.} Let us assume
that $\CA_\CT$ contains no antiholomorphic fermionic components
and does not depend on $\etab_i^\pm$,
\begin{equation}\label{eq:4.13}
\bV_\pm^i\lrcorner \CA_\CT\ =\ 0\eand\bV_\pm^i(\CA^\pm_a)\ =\
0~~~\mbox{with}~~~ \CA^\pm_a\ :=\  \bW_a^\pm\lrcorner\CA_\CT~,
\end{equation}
i.e.\ $\CA_\CT\in\Omega^1_{\CT_b}(\CCX,\sEnd\CE)$. Now, we
introduce a CS-type action functional
\begin{equation}\label{eq:4.14}
S_{\mathrm{phCS}}\ =\
\int_\CCX\tilde{\Omega}\wedge\tr\left(\CA_\CT\wedge
\dd_\CT\CA_\CT+\tfrac{2}{3}
\CA_\CT\wedge\CA_\CT\wedge\CA_\CT\right)~,
\end{equation}
where
\begin{equation}
\dd_\CT|_{\tilde{\CV}_\pm}\ =\ \dd \bw_\pm^a\der{\bw_\pm^a}+\dd
\etab_i^\pm\der{\etab_i^\pm}
\end{equation}
is the $\CT$-part of the exterior derivative $\dd$ on $\CF^{5|8}$.

\paragraph{Field equations of phCS theory.} The action
\eqref{eq:4.14} leads to the CS-type field equations
\begin{equation}\label{eq:4.16}
\dd_\CT\CA_\CT+\CA_\CT\wedge\CA_\CT\ =\ 0~,
\end{equation}
which are the equations of motion of phCS theory. In the
nonholonomic basis $\{\bW_a^\pm, \bV_\pm^i\}$ of the distribution
$\CT$ over $\tilde{\CV}_\pm\subset\CF^{5|8}$, these equations read
as
\begin{subequations}\label{eq:4.17}
\begin{align}\label{eq:4.17a}
\bW_1^\pm\CA_2^\pm-\bW^\pm_2\CA_1^\pm+[\CA_1^\pm,\CA_2^\pm]&\ =\
0~,\\\label{eq:4.17b}
\bW_2^\pm\CA_3^\pm-\bW^\pm_3\CA_2^\pm+[\CA_2^\pm,\CA_3^\pm]\mp2
\gamma_\pm^2\CA_1^\pm&\ =\ 0~,\\\label{eq:4.17c}
\bW_1^\pm\CA_3^\pm-\bW^\pm_3\CA_1^\pm+[\CA_1^\pm,\CA_3^\pm]&\ =\
0~,
\end{align}
\end{subequations}
where the components $\CA_a^\pm$ have been defined in
\eqref{eq:4.13}.

\subsection{Equivalence to supersymmetric Bogomolny equations}

\paragraph{Dependence on $\lambda_\pm$, $\bl_\pm$.} Note that from
\eqref{eq:3.35}, it follows that
\begin{equation}
\bW_1^+\ =\ \lambda_+^2\bW_1^-~,~~~\bW_2^+\ =\
-\bl_+^{-2}\bW_2^-\eand \gamma_+^{-1}\bW_3^+\ =\
\lambda_+\bl_+\left(\gamma_-^{-1}\bW_3^-\right)
\end{equation}
and therefore $\CA_1^\pm$, $\CA_2^\pm$ and
$\gamma_\pm^{-1}\CA_3^\pm$ take values in the bundles $\CO(2)$,
$\COb(-2)$ and $\CO(1)\otimes\COb(1)$, respectively. Together with
the definitions \eqref{eq:4.13} of $\CA^\pm_a$ and
\eqref{eq:3.34all} of $\bW_a^\pm$ as well as the fact that the
$\eta_i^\pm$ are nilpotent and $\CO(1)$-valued, this determines
the dependence of $\CA_a^\pm$ on $\eta_i^\pm$, $\lambda_\pm$ and
$\bl_\pm$ to be
\begin{equation}
\CA_1^\pm\ =\ -\lambda_\pm^\ald \CB_\ald^\pm\eand \CA_3^\pm\ =\
2\gamma_\pm\hl^\ald_\pm\CB_\ald^\pm
\end{equation}
with the abbreviation
\begin{subequations}\label{eq:4.19}
\begin{align}
&\begin{aligned} \CB_\ald^\pm\ :=\  &\lambda^\bed_\pm
B_{\ald\bed}+\di\eta_i^\pm\chi^i_\ald+\tfrac{1}{2!}
\gamma_\pm\eta_i^\pm\eta_j^\pm\hl^\bed_\pm\phi^{ij}_{\ald\bed}
+\tfrac{1}{3!}\gamma_\pm^2\eta_i^\pm\eta_j^\pm\eta_k^\pm
\hl^\bed_\pm\hl^\gad_\pm\tilde{\chi}^{ijk}_{\ald\bed\gad}\ +
\\&+\tfrac{1}{4!}\gamma_\pm^3\eta_i^\pm\eta_j^\pm\eta_k^\pm
\eta_l^\pm\hl_\pm^\bed\hl_\pm^\gad\hl_\pm^\ded
G^{ijkl}_{\ald\bed\gad\ded}
\end{aligned}
\intertext{and}\label{eq:4.19c} &\CA_2^\pm\ =\
\pm\left(\tfrac{1}{2!}\gamma_\pm^2\eta_i^\pm\eta_j^\pm\phi^{ij}+
\tfrac{1}{3!}\gamma_\pm^3\eta_i^\pm\eta_j^\pm\eta_k^\pm
\hl^\ald_\pm\tilde{\chi}_\ald^{ijk}+\tfrac{1}{4!}\gamma_\pm^4
\eta_i^\pm\eta_j^\pm\eta_k^\pm\eta_l^\pm
\hl^\ald_\pm\hl^\bed_\pm G^{ijkl}_{\ald\bed}\right)~,
\end{align}
\end{subequations}
where $\lambda_\pm^\ald$, $\gamma_\pm$ and $\hl^\ald_\pm$ have
been given in \eqref{eq:2.22} and \eqref{eq:3.38}. The expansions
\eqref{eq:4.19} are defined up to gauge transformations generated
by group-valued functions which may depend on $\lambda_\pm$ and
$\bl_\pm$. In particular, it is assumed in this twistor
correspondence that for solutions to \eqref{eq:4.17}, there exists
a gauge in which terms of zeroth and first order in $\eta_i^\pm$
are absent in $\CA_2^\pm$. In the \v{C}ech approach, this
corresponds to the holomorphic triviality of the bundle
$\tilde{\CE}$ defined by such solutions when restricted to
projective lines. Put differently, we consider a subset in the set
of all solutions of phCS theory on $\CF^{5|8}$, and we will always
mean this subset when speaking of solutions to phCS theory. From
the properties of $\CA_a^\pm$ and $\eta_i^\pm$, it follows that
the fields with an odd number of spinor indices are fermionic
(odd) while those with an even number of spinor indices are
bosonic (even). Moreover, due to the symmetry of the
$\hl_\pm^\ald$ products and the antisymmetry of the $\eta_i^\pm$
products, all component fields are automatically symmetric in
(some of) their spinor indices and antisymmetric in their Latin
(R-symmetry) indices.

\paragraph{Supersymmetric Bogomolny equations.} Note that in
\eqref{eq:4.19}, all fields $B_{\ald\bed},\chi_\ald^i,\ldots$
depend only on the coordinates $(y^{\ald\bed})\in\FR^3$.
Substituting \eqref{eq:4.19} into \eqref{eq:4.17a} and
\eqref{eq:4.17b}, we obtain the equations
\begin{equation}\label{eq:4.constr}
\begin{aligned}
\phi^{ij}_{\ald\bed}\ =\
-\left(\dpar_{(\ald\bed)}\phi^{ij}+[B_{\ald\bed},\phi^{ij}]
\right)~,~~~
\tilde{\chi}^{ijk}_{\ald(\bed\gad)}\ =\ -\tfrac{1}{2}
\left(\dpar_{(\ald(\bed)}\tilde{\chi}^{ijk}_{\gad)}+
[B_{\ald(\bed},\tilde{\chi}^{ijk}_{\gad)}]\right)~,\\
G^{ijkl}_{\ald(\bed\gad\ded)}\ =\
-\tfrac{1}{3}\left(\dpar_{(\ald(\bed)}G^{ijkl}_{\gad\ded)}+
[B_{\ald(\bed},G^{ijkl}_{\gad\ded)}]\right)\hspace{3.5cm}
\end{aligned}
\end{equation}
showing that $\phi_{\ald\bed}^{ij}$,
$\tilde{\chi}_{\ald\bed\gad}^{ijk}$ and
$G^{ijkl}_{\ald\bed\gad\ded}$ are composite fields, which do not
describe independent degrees of freedom. Furthermore, the field
$B_{\ald\bed}$ can be decomposed into its symmetric part, denoted
by $A_{\ald\bed}=A_{(\ald\bed)}$, and its antisymmetric part,
proportional to $\Phi$, such that
\begin{equation}\label{eq:4.B}
B_{\ald\bed}\ =\ A_{\ald\bed}-\tfrac{\di}{2}\eps_{\ald\bed}\Phi~.
\end{equation}
Hence, we have recovered the covariant derivative $D_{\ald\bed}\
=\ \dpar_{(\ald\bed)}+A_{\ald\bed}$ and the (scalar) Higgs field
$\Phi$. Defining
\begin{equation}\label{eq:4.chi}
\tilde{\chi}_{i\ald}\ :=\
\tfrac{1}{3!}\eps_{ijkl}\tilde{\chi}_\ald^{jkl}\eand G_{\ald\bed}\
:=\  \tfrac{1}{4!}\eps_{ijkl}G^{ijkl}_{\ald\bed}~,
\end{equation}
we have thus obtained the supermultiplet in three dimensions
consisting of the fields
\begin{equation}
A_{\ald\bed},\chi^i_\ald,\Phi,\phi^{ij},\tilde{\chi}_{i\ald},
G_{\ald\bed}~.
\end{equation}
The equations \eqref{eq:4.17} together with the field expansions
\eqref{eq:4.19}, the constraints \eqref{eq:4.constr} and the
definitions \eqref{eq:4.B} and \eqref{eq:4.chi} yield the
following supersymmetric extension of the Bogomolny equations:
\begin{equation}\label{eq:4.24}
\begin{aligned}
f_{\ald\bed}&\ =\ -\tfrac{\di}{2}D_{\ald\bed}\Phi~,
\\\eps^{\bed\gad}D_{\ald\bed}\chi^i_\gad&\ =\
-\tfrac{\di}{2}[\Phi,\chi^i_\ald]~,
\\\triangle\phi^{ij}&\ =\ -\tfrac{1}{4}[\Phi,[\phi^{ij},\Phi]]+
\eps^{\ald\bed}\{\chi^i_\ald,\chi^j_\bed\}~,
\\\eps^{\bed\gad}D_{\ald\bed}\tilde{\chi}_{i\gad}&\ =\
-\tfrac{\di}{2}[\tilde{\chi}_{i\ald},\Phi]+
2\di[\phi_{ij},\chi^j_\ald]~,\\
\eps^{\bed\gad}D_{\ald\bed}G_{\gad\ded}&\ =\
-\tfrac{\di}{2}[G_{\ald\ded},\Phi]+
\di\{\chi^i_\ald,\tilde{\chi}_{i\ded}\}
-\tfrac{1}{2}[\phi_{ij},D_{\ald\ded}\phi^{ij}]+
\tfrac{\di}{4}\eps_{\ald\ded}[\phi_{ij},[\Phi,\phi^{ij}]]~,
\end{aligned}
\end{equation}
which can also be derived from the equations \eqref{eq:2.43} by
demanding that all the fields in \eqref{eq:2.43} are independent
of the coordinate $x^4$. In \eqref{eq:4.24}, we have used the fact
that we have a decomposition of the field strength in three
dimensions according to
\begin{equation}
F_{\ald\bed\gad\ded}\ =\ [D_{\ald\bed},D_{\gad\ded}]\ =:\
\eps_{\bed\ded}f_{\ald\gad}+\eps_{\ald\gad}f_{\bed\ded}
\end{equation}
with $f_{\ald\bed}=f_{\bed\ald}$. We have also introduced the
abbreviation
$\triangle:=\frac{1}{2}\eps^{\ald\bed}\eps^{\gad\ded}D_{\ald\gad}
D_{\bed\ded}$.

\paragraph{Action functional in component fields.} Note that
\eqref{eq:4.12} can be rewritten as
\begin{equation}
\tilde{\Omega}|_{\tilde{\CV}_\pm}\ =\ \pm\Theta_\pm^1\wedge
\Theta_\pm^2\dd\eta_1^\pm\cdots\dd\eta_4^\pm~,
\end{equation}
where the one-forms $\Theta_\pm^{1,2}$ have been given in
\eqref{eq:3.37}. Substituting this expression and the expansions
\eqref{eq:4.19} into the action \eqref{eq:4.14}, we arrive after a
straightforward calculation at the action
\begin{equation}\label{eq:4.27}
\begin{aligned}
S_{\mathrm{sB}}\ =\ \int\dd^3
x\tr\Big\{\Big.&\eps^{\ald\ded}\eps^{\bed\gad}G_{\gad\ded}
\left(f_{\ald\bed}+\tfrac{\di}{2}D_{\ald\bed}\Phi\right)+
\di\eps^{\ald\ded}\eps^{\bed\gad}\chi^i_\ald
D_{\ded\bed}\tilde{\chi}_{i\gad}+
\tfrac{1}{2}\phi_{ij}\triangle\phi^{ij}-\\&-\tfrac{1}{2}
\eps^{\ald\ded}\chi^i_\ald[\tilde{\chi}_{i\ded},\Phi]-
\eps^{\ald\gad}\phi_{ij}\{\chi^i_\ald,\chi^j_\gad\}+\tfrac{1}{8}
[\phi_{ij},\Phi][\phi^{ij},\Phi]\Big.\Big\}~,
\end{aligned}
\end{equation}
which yields the supersymmetric Bogomolny (sB) equations
\eqref{eq:4.24}. In this expression, we have again used the
shorthand notation $\phi_{ij}:=\frac{1}{2!}\eps_{ijkl}\phi^{kl}$.

\subsection{Partially holomorphic CS theory in the \v{C}ech
description}

\paragraph{Equivalent $\CT$-flat bundle.} Our starting point in
section 4.2 was to consider a trivial complex vector bundle $\CE$
over $\CF^{5|8}$ endowed with a $\CT$-connection. Such a
$\CT$-connection $D_\CT=\dd_\CT+\CA_\CT$ on $\CE$ is flat if
$\CA_\CT$ solves the equations \eqref{eq:4.16}, and then
$(\CE,f_{+-}\!\!=\!\unit_n,D_\CT)$ is a $\CT$-flat bundle in the
Dolbeault description. After one turns to the \v{C}ech description
of $\CT$-flat bundles, the connection one-form $\CA_\CT$
disappears and all the information is hidden in a transition
function. To achieve this, let us restrict a solution $\CA_\CT$ of
the equations \eqref{eq:4.16} to the patches $\tilde{\CV}_+$ and
$\tilde{\CV}_-$ covering $\CF^{5|8}$. Since $\CA_\CT$ is flat, it
is given as a pure gauge configuration on each patch and we have
\begin{equation}\label{eq:4.28}
\CA_\CT|_{\tilde{\CV}_\pm}\ =\ \psi_\pm\dd_\CT\psi_\pm^{-1}~,
\end{equation}
where the $\psi_\pm$ are smooth $\sGL(n,\FC)$-valued
superfunctions on $\tilde{\CV}_\pm$ such that
$\bV^i_\pm\psi_\pm=0$ (the existence of such a gauge was assumed
in the formulation of phCS theory). Due to the triviality of the
bundle $\CE$, we have
\begin{equation}\label{eq:4.29}
\psi_+\dd_\CT\psi_+^{-1}\ =\ \psi_-\dd_\CT\psi_-^{-1}
\end{equation}
on the intersection $\tilde{\CV}_+\cap\tilde{\CV}_-$. From
\eqref{eq:4.29}, one easily obtains
\begin{equation}\label{eq:4.30}
\dd_\CT(\psi_+^{-1}\psi_-)\ =\ 0~,
\end{equation}
and we can define a $\CT$-flat complex vector bundle
$\tilde{\CE}\rightarrow \CF^{5|8}$ with the canonical flat
$\CT$-connection $\dd_\CT$ and the transition function
\begin{equation}\label{eq:4.31}
\tilde{f}_{+-}\ :=\  \psi_+^{-1}\psi_-~.
\end{equation}
The bundles $\CE$ and $\tilde{\CE}$ are equivalent as smooth
bundles but not as $\CT$-flat bundles. However, we have an
equivalence of the following data:
\begin{equation}
(\CE,f_{+-}\!= \unit_n,\CA_\CT)\ \sim\
(\tilde{\CE},\tilde{f}_{+-},\tilde{\CA}_\CT= 0)~.
\end{equation}

\paragraph{Equivalent flat $\CT$-connection.} To improve our
understanding of the $\CT$-flatness of the bundle
$\tilde{\CE}\rightarrow \CF^{5|8}$ with the transition function
\eqref{eq:4.31}, we rewrite the conditions \eqref{eq:4.30} as
follows:
\begin{subequations}
\begin{align}\label{eq:4.33}
\bW_1^+\tilde{f}_{+-}\ =\ 0~,~~~\bW_2^+\tilde{f}_{+-}&\ =\ 0~,~~~
\bV_+^i\tilde{f}_{+-}\ =\ 0~,\\\label{eq:4.34}
\bW_3^+\tilde{f}_{+-}&\ =\ 0~.
\end{align}
\end{subequations}
Recall that $\CT=\hat{\CCDb}\oplus\CCV$ and the vector fields
appearing in \eqref{eq:4.33} generate the antiholomorphic
distribution $\hat{\CCDb}$, which is a CR structure. In other
words, the bundle $\tilde{\CE}$ is holomorphic along the
mini-supertwistor space $\CP^{2|4}\subset\CF^{5|8}$ and flat along
the fibres of the projection $\pi_2:\CF^{5|8}\rightarrow\CP^{2|4}$
as follows from \eqref{eq:4.34}. Let us now additionally assume
that $\tilde{\CE}$ is holomorphically trivial\footnote{This
assumption, which was already used in \eqref{eq:4.19c}, is crucial
in the twistor approach.} when restricted to any projective line
$\CPP^1_{x,\eta}\embd\CF^{5|8}$ given by \eqref{eq:3.18}. This
extra assumption guarantees the existence of a gauge in which the
component $\CA_2^\pm$ of $\CA_\CT$ vanishes. Hence, there exist
$\sGL(n,\FC)$-valued functions $\hat{\psi}_\pm$ such that
\begin{equation}
\tilde{f}_{+-}\ =\ \psi_+^{-1}\psi_-\ =\
\hat{\psi}_+^{-1}\hat{\psi}_-~~~\mbox{with}~~~~
\bW_2^\pm\hat{\psi}_\pm\ =\ 0
\end{equation}
and
\begin{equation}
g\ :=\  \psi_+\hat{\psi}^{-1}_+\ =\ \psi_-\hat{\psi}_-^{-1}
\end{equation}
is a matrix-valued function generating a gauge transformation
\begin{equation}
\psi_\pm\ \mapsto\ \hat{\psi}_\pm\ =\ g^{-1}\psi_\pm~,
\end{equation}
which acts on the gauge potential according to
\begin{equation}\label{eq:4.38}
\begin{aligned}
\CA_1^\pm&\ \mapsto\  \hat{\CA}_1^\pm\ =\ g^{-1}\CA_1^\pm
g+g^{-1}\bW_1^\pm
g\ =\ \hat{\psi}_\pm\bW_1^\pm\hat{\psi}_\pm^{-1}~,\\
\CA_2^\pm&\ \mapsto\  \hat{\CA}_2^\pm\ =\ g^{-1}\CA_2^\pm
g+g^{-1}\bW_2^\pm
g\ =\ \hat{\psi}_\pm\bW_2^\pm\hat{\psi}_\pm^{-1}\ =\ 0~,\\
\CA^\pm_3&\ \mapsto\  \hat{\CA}_3^\pm\ =\ g^{-1}\CA_3^\pm
g+g^{-1}\bW_3^\pm g\ =\ \hat{\psi}_\pm\bW_3^\pm
\hat{\psi}_\pm^{-1}~,\\
0\ =\ \CA_\pm^i\ :=\ \psi_\pm\bV^i_\pm\psi_\pm^{-1}&\ \mapsto\
\hat{\CA}_\pm^i\ =\ g^{-1}\bV^i_\pm g\ =\
\hat{\psi}_\pm\bV^i_\pm\hat{\psi}_\pm^{-1}~.
\end{aligned}
\end{equation}
In this new gauge, one generically has $\hat{\CA}^i_\pm\neq 0$.

\paragraph{Linear systems.} Note that \eqref{eq:4.28} can be
rewritten as the following linear system of differential
equations:
\begin{equation}\label{eq:4.39}
\begin{aligned}
(\bW_a^\pm+\CA_a^\pm)\psi_\pm&\ =\ 0~,\\ \bV_\pm^i\psi_\pm&\ =\
0~.
\end{aligned}
\end{equation}
The compatibility conditions of this linear system are the
equations \eqref{eq:4.17}. This means that for any solution
$\CA^\pm_a$ to \eqref{eq:4.17}, one can construct solutions
$\psi_\pm$ to \eqref{eq:4.39} and, conversely, for any given
$\psi_\pm$ obtained via a splitting \eqref{eq:4.31} of a
transition function $\tilde{f}_{+-}$, one can construct a solution
\eqref{eq:4.28} to \eqref{eq:4.17}.

Similarly, the equations \eqref{eq:4.38} can be rewritten as the
gauge equivalent linear system
\begin{subequations}\label{eq:4.40}
\begin{align}\label{eq:4.40a}
(\bW_1^\pm+\hat{\CA}_1^\pm)\hat{\psi}_\pm&\ =\
0~,\\\label{eq:4.40b}
 \bW_2^\pm\hat{\psi}_\pm&\ =\ 0~,\\\label{eq:4.40c}
 (\bW_3^\pm+\hat{\CA}_3^\pm)\hat{\psi}_\pm&\ =\ 0~,\\
 (\bV_\pm^i+\hat{\CA}_\pm^i)\hat{\psi}_\pm&\ =\ 0~.
\end{align}
\end{subequations}
Note that due to the holomorphicity of $\hat{\psi}_\pm$ in
$\lambda_\pm$ and the condition $\hat{\CA}_\CT^+=\hat{\CA}_\CT^-$
on $\tilde{\CV}_+\cap\tilde{\CV}_-$, the components
$\hat{\CA}_1^\pm$, $\gamma_\pm^{-1}\hat{\CA}_3^\pm$ and
$\hat{\CA}_\pm^i$ must take the form
\begin{equation}\label{eq:4.41}
\hat{\CA}^\pm_1\ =\ -\lambda_\pm^\ald\lambda_\pm^\bed
\CB_{\ald\bed}~,~~~ \gamma_\pm^{-1}\hat{\CA}_3^\pm\ =\
2\hl^\ald_\pm\lambda_\pm^\bed \CB_{\ald\bed}\eand \hat{\CA}^i_\pm\
=\ \lambda^\ald_\pm\CA^i_\ald~,
\end{equation}
with $\lambda$-independent superfields
$\CB_{\ald\bed}:=\CA_{\ald\bed}-\frac{\di}{2}\eps_{\ald\bed}\Phi$
and $\CA_\ald^i$. Defining the first-order differential operators
$\nabla_{\ald\bed}:=\dpar_{(\ald\bed)}+\CB_{\ald\bed}$ and
$D_\ald^i=\der{\eta_i^\ald}+\CA_\ald^i=:\dpar_\ald^i+\CA_\ald^i$,
we arrive at the following compatibility conditions of the linear
system \eqref{eq:4.40}:
\begin{equation}\label{eq:4.42}
\begin{aligned}
{}&[\nabla_{\ald\gad},\nabla_{\bed\ded}]+[\nabla_{\ald\ded},
\nabla_{\bed\gad}]\
=\ 0~,
~~~~[D^i_\ald,\nabla_{\bed\gad}]+[D_\gad^i,\nabla_{\bed\ald}]\ =\
0~,\\&\hspace{3cm} \{D_\ald^i,D_\bed^j\}+\{D_\bed^i,D_\ald^j\}\ =\
0~.
\end{aligned}
\end{equation}
These equations also follow from \eqref{eq:4.16} after
substituting the expansions \eqref{eq:4.41}.

\paragraph{Superfield equations.} The equations \eqref{eq:4.42} can
equivalently be rewritten as
\begin{equation}\label{constraint}
       [\nabla_{\ald\gad},\nabla_{\bed\ded}]\ =:\ \eps_{\gad\ded}
       \Sigma_{\ald\bed}~,
       ~~~
       {[D_\ald^i,\nabla_{\bed\gad}]}\ =:\ \di\eps_{\ald\gad}
       \Sigma^i_\bed
       \eand
       \{D^i_\ald,D^j_\bed\}\ =:\ \eps_{\ald\bed}\Sigma^{ij}~,
\end{equation}
where $\Sigma_{\ald\bed}=\Sigma_{\bed\ald}$ and
$\Sigma^{ij}=-\Sigma^{ji}$. Note that the first equation in
\eqref{eq:4.42} immediately shows that
$f_{\ald\bed}=-\tfrac{\di}{2}D_{\ald\bed}\Phi$ and thus the
contraction of the first equation of \eqref{constraint} with
$\eps^{\gad\ded}$ gives
$\Sigma_{\ald\bed}=f_{\ald\bed}-\tfrac{\di}{2}D_{\ald\bed}\Phi=
2f_{\ald\bed}$. The graded Bianchi identities for the differential
operators $\nabla_{\ald\bed}$ and $D^i_\ald$ yield in a
straightforward manner further field equations, which allow us to
identify the superfields $\Sigma^i_\ald$ and $\Sigma^{ij}$ with
the spinors $\chi^i_\ald$ and the scalars $\phi^{ij}$,
respectively. Moreover, $\tilde{\chi}_{i\ald}$ is given by
$\tilde{\chi}_{i\ald}:=\frac{1}{3}\eps_{ijkl}D^j_\ald\phi^{kl}$
and $G_{\ald\bed}$ is defined by
$G_{\ald\bed}:=-\frac{1}{4}D^i_{(\ald}\tilde{\chi}_{i\bed)}$.
Collecting the above information, one obtains the superfield
equations for $\CA_{\ald\bed}$, $\chi^i_\ald$, $\Phi$,
$\phi^{ij}$, $\tilde{\chi}_{i\ald}$ and $G_{\ald\bed}$ which take
the same form as \eqref{eq:4.24} but with all the fields now being
superfields.\footnote{In the following, the zeroth order component
in the $\eta$-expansion of a superfield (its body) is denoted by a
``$\circ$''.} Thus, the projection of the superfields onto the
zeroth order components of their $\eta$-expansions gives
\eqref{eq:4.24}.

\paragraph{Recursion relations.}
To extract the physical field content from the superfields, we
need their explicit expansions in powers of $\eta_i^\ald$. For
this, we follow the literature \cite{Harnad:1984vk} and impose the
transversal gauge condition
\begin{equation}\label{transversalgauge}
\eta^\ald_i\CA^i_\ald\ =\ 0~,
\end{equation}
which allows us to define the recursion operator
\begin{equation}
\CD\ :=\ \eta^\ald_iD^i_\ald\ =\
\eta^\ald_i\partial^i_\ald~~~\mbox{with}~~~\dpar^i_\ald\ :=\
\der{\eta^\ald_i}~.
\end{equation}
Note that this gauge removes the superfluous gauge degrees of
freedom associated with the fermionic coordinates. The constraint
equations \eqref{constraint} together with the graded Bianchi
identities yield the following recursion relations:
\begin{equation}\label{recursions}
\begin{aligned}
      (1+\CD)\CA^i_\ald &\ =\
      \eps_{\ald\bed}\eta^\bed_j \phi^{ij}~,\\
      \CD\CB_{\ald\bed} &\ =\
      -\di\eps_{\bed\gad}\eta^\gad_i\chi^i_\ald~,\\
      \CD \chi^i_\ald &\ =\
      \di\eta^\bed_j\nabla_{\ald\bed} \phi^{ij}~,\\
      \CD \phi_{ij} &\ =\
      -\eta^\ald_{[i}\tilde{\chi}_{j]\ald}~,\\
      \CD \tilde{\chi}_{i\ald} &\ =\
      -\eta^\bed_i G_{\ald\bed}+\eps_{\ald\bed}\eta^\bed_j
       [\phi^{jk},\phi_{ki}]~,\\
      \CD G_{\ald\bed} &\ =\
      \eta^\gad_i\eps_{\gad(\ald}[\tilde{\chi}_{j\bed)},
      \phi^{ij}]~.
\end{aligned}
\end{equation}
These equations define order by order all the superfield
expansions.

The explicit derivation of the expansions of the fields
$\CB_{\ald\bed}$ and $\CA_\ald^i$ is performed in appendix A.
Here, we just quote the result which we will need later on:
\begin{subequations}\label{superexp4}
\begin{align}\nonumber
\CB_{\ald\bed}\ =\ &
\ci{\CB}_{\ald\bed}-\di\eps_{\bed\gad_1}\eta_{j_1}^{\gad_1}
\ci{\chi}^{j_1}_\ald+
\tfrac{1}{2!}\eps_{\bed\gad_1}\eta_{j_1}^{\gad_1}
\eta_{j_2}^{\gad_2}
\nabla_{\ald\gad_2}\ci{\phi}^{j_1j_2}-\tfrac{1}{2\cdot3!}
\eps_{\bed\gad_1}\eta_{j_1}^{\gad_1}\eta_{j_2}^{\gad_2}
\eta_{j_3}^{\gad_3}\eps^{j_1j_2j_3k}\nabla_{\ald\gad_2}
\ci{\tilde{\chi}}_{k\gad_3}~-\\&
-\tfrac{1}{4!}\eps_{\bed\gad_1}\eta_{j_1}^{\gad_1}
\eta_{j_2}^{\gad_2}\eta_{j_3}^{\gad_3}\eta_{j_4}^{\gad_4}
\eps^{j_1j_2j_3j_4}\nabla_{\ald\gad_2}\ci{G}_{\gad_3\gad_4}+
\cdots
\end{align}
\begin{equation}
\CA^i_\ald\ =\ \tfrac{1}{2!}\eps_{\ald\gad_1}\eta_{j_1}^{\gad_1}
\ci{\phi}^{ij_1}-
\tfrac{1}{3!}\eps_{\ald\gad_1}\eta_{j_1}^{\gad_1}
\eta_{j_2}^{\gad_2}\eps^{ij_1j_2k}\ci{\tilde{\chi}}_{k\gad_2}
+\tfrac{3}{2\cdot4!}\eps_{\ald\gad_1}\eta_{j_1}^{\gad_1}
\eta_{j_2}^{\gad_2}\eta_{j_3}^{\gad_3}
\eps^{ij_1j_2j_3}\ci{G}_{\gad_2\gad_3}+\cdots
\end{equation}
\end{subequations}
The equations \eqref{constraint} are satisfied for these
expansions, if the supersymmetric Bogomolny equations
\eqref{eq:4.24} hold for the physical fields\footnote{The fields
$\ci{G}_{\ald\bed},\ci{\Phi},$ etc.\ are the same as the field
$G_{\ald\bed},\Phi,$ etc.\ in equations \eqref{eq:4.24}, but in
the present section, we need to clearly distinguish between
superfields and their bodies.} appearing in the above expansions
and vice versa.

\paragraph{Bijection between moduli spaces.} Summarizing the discussion of
this section, we have described a bijection between the moduli
space $\CM_{\mathrm{phCS}}$ of solutions\footnote{Recall that we
always consider only a subset of the full solution space as
discussed in section 4.3.} to the field equations \eqref{eq:4.17}
of phCS theory and the moduli space $\CM_{\mathrm{sB}}$ of
solutions to the supersymmetric Bogomolny equations
\eqref{eq:4.24},
\begin{equation}
\CM_{\mathrm{phCS}}\ \longleftrightarrow\ \CM_{\mathrm{sB}}~.
\end{equation}
The moduli spaces are obtained from the respective solution spaces
by taking the quotient with respect to the action of the
corresponding groups of gauge transformations. We also have shown
that there is a one-to-one correspondence between gauge
equivalence classes of solutions $\CA_\CT$ to the phCS field
equations \eqref{eq:4.16} and equivalence classes of topologically
trivial $\CT$-flat vector bundles $\tilde{\CE}$ over the CR
supertwistor space $\CF^{5|8}$. In other words, we have
demonstrated an equivalence of the Dolbeault and the \v{C}ech
descriptions of the moduli space of $\CT$-flat bundles.

\section{Holomorphic BF theory on the mini-supertwistor space}

In the preceding section, we have defined a theory on the CR
supertwistor space $\CF^{5|8}$ entering into the double fibration
\eqref{eq:3.19} which we called partially holomorphic Chern-Simons
theory. We have shown that this theory is equivalent to a
supersymmetric Bogomolny-type Yang-Mills-Higgs theory in three
Euclidean dimensions. The purpose of this section is to show, that
one can also introduce a theory (including an action functional)
on the mini-supertwistor space $\CP^{2|4}$, which is equivalent to
phCS theory on $\CF^{5|8}$. Thus, one can define at each level of
the double fibration \eqref{eq:3.19} a theory accompanied by a
proper action functional and, moreover, these three theories are
all equivalent.

\subsection{Field equations of hBF theory on $\CP^{2|4}$}

Consider the mini-supertwistor space $\CP^{2|4}$. Let $E$ be a
trivial rank $n$ complex vector bundle over $\CP^{2|4}$ with a
connection one-form $\CA$. Assume that its $(0,1)$ part
$\CA^{0,1}$ contains no antiholomorphic fermionic components and
does not depend on $\etab_i^\pm$, i.e.\
$\bV_\pm^i\lrcorner\CA^{0,1}=0$ and
$\bV_\pm^i(\dpar_{\bw_\pm^{1,2}}\lrcorner\CA^{0,1})=0$. Recall
that on $\CP^{2|4}$, we have a holomorphic volume form $\Omega$
which is locally given by \eqref{eq:3.14}. Hence, we can define a
holomorphic BF (hBF) type theory (cf.\
\cite{Popov:1999cq,Ivanova:2000xr,Baulieu:2004pv}) on $\CP^{2|4}$
with the action
\begin{equation}\label{eq:5.1}
S_{\mathrm{hBF}}\ =\ \int_{\CCY}\Omega\wedge\tr\{B(\dparb
\CA^{0,1}+\CA^{0,1}\wedge\CA^{0,1})\}\ =\ \int_{\CCY}\Omega\wedge
\tr\{B \CF^{0,2}\}~,
\end{equation}
where $B$ is a scalar field in the adjoint representation of the
group $\sGL(n,\FC)$, $\dparb$ is the antiholomorphic part of the
exterior derivative on $\CP^{2|4}$ and $\CF^{0,2}$ the $(0,2)$
part of the curvature two-form. The space $\CCY$ is the
subsupermanifold of $\CP^{2|4}$ constrained by $\etab_i^\pm=0$. In
fact, $\CCY$ is the worldvolume of a stack of $n$ not quite
space-filling D3-branes, as discussed in the introduction.

The equations of motion following from the action functional
\eqref{eq:5.1} are
\begin{subequations}\label{eq:5.2}
\begin{align}\label{eq:5.2a}
\dparb\CA^{0,1}+\CA^{0,1}\wedge\CA^{0,1}&\ =\ 0~,\\\label{eq:5.2b}
\dparb B+[\CA^{0,1},B]&\ =\ 0~.
\end{align}
\end{subequations}
These equations as well as the Lagrangian in \eqref{eq:5.1} can be
obtained from the equations \eqref{eq:4.16} and the Lagrangian in
\eqref{eq:4.14}, respectively, by imposing the condition
$\dpar_{\bw_\pm^3}\CA_{\bw_\pm^a}=0$ and identifying
\begin{equation}
\CA^{0,1}|_{\hat{\CV}_\pm}\ =\ \dd \bw_\pm^1\CA_{\bw_\pm^1}+\dd
\bw_\pm^2\CA_{\bw_\pm^2}\eand B^\pm\ :=\  B|_{\hat{\CV}_\pm}\ =\
\CA_{\bw_\pm^3}~.
\end{equation}
Note that $\CA_{\bw_\pm^3}$ behaves on $\CP^{2|4}$ as a scalar.
Thus, \eqref{eq:5.2} can be obtained from \eqref{eq:4.16} by
demanding invariance of all fields under the action of the group
$\CCG'$ from section 3.4 such that
$\CP^{2|4}\cong\CF^{5|8}/\CCG'$.

\subsection{\v{C}ech description}

When restricted to the patches $\hat{\CV}_\pm$, the equations
\eqref{eq:5.2} can be solved by
\begin{equation}\label{eq:5.4}
\CA^{0,1}|_{\hat{\CV}_\pm}\ =\
\tilde{\psi}_\pm\dparb\tilde{\psi}_\pm^{-1}\eand B^\pm\ =\
\tilde{\psi}_\pm B^\pm_0\tilde{\psi}_\pm^{-1}~,
\end{equation}
where $B_0^\pm$ is a holomorphic $\agl(n,\FC)$-valued function on
$\hat{\CV}_\pm$,
\begin{equation}
\dparb B_0^\pm\ =\ 0~.
\end{equation}
On the intersection $\hat{\CV}_+\cap\hat{\CV}_-$, we have the
gluing conditions
\begin{equation}\label{eq:5.6}
\tilde{\psi}_+\dparb\tilde{\psi}_+^{-1}\ =\
\tilde{\psi}_-\dparb\tilde{\psi}_-^{-1}\eand
\tilde{\psi}_+B_0^+\tilde{\psi}_+^{-1}\ =\
\tilde{\psi}_-B_0^-\tilde{\psi}_-^{-1}
\end{equation}
as $E$ is a trivial bundle. From \eqref{eq:5.6}, we learn that
\begin{equation}\label{eq:5.7}
\tilde{f}_{+-}\ :=\  \tilde{\psi}_+^{-1}\tilde{\psi}_-
\end{equation}
can be identified with the holomorphic transition function of a
bundle $\tilde{E}$ with the canonical holomorphic structure
$\dparb$, and
\begin{equation}
B_0^+\ =\ \tilde{f}_{+-}B_0^-\tilde{f}_{+-}^{-1}
\end{equation}
is a global holomorphic section of the bundle $\sEnd\tilde{E}$,
i.e.\ $B_0\in H^0(\CP^{2|4},\sEnd\tilde{E})$ and $B\in
H^0(\CP^{2|4},\sEnd E)$. Note that the pull-back
$\pi_2^*\tilde{E}$ of the bundle $\tilde{E}$ to the space
$\CF^{5|8}$ can be identified with the bundle $\tilde{\CE}$,
\begin{equation}
\tilde{\CE}\ =\ \pi_2^*\tilde{E}~,
\end{equation}
with the transition function
$\tilde{f}_{+-}=\psi_+^{-1}\psi_-=\tilde{\psi}_+^{-1}\tilde{\psi}_-$.
Recall that the transition functions of the bundle $\tilde{\CE}$
do not depend on $w_\pm^3$ and therefore they can always be
considered as the pull-backs of transition functions of a bundle
$\tilde{E}$ over $\CP^{2|4}$.

\subsection{Moduli space}

By construction, $B=\{B^\pm\}$ is a $\agl(n,\FC)$-valued function
generating trivial infinitesimal gauge transformations of
$\CA^{0,1}$ and therefore it does not contain any physical degrees
of freedom. Remember that solutions to the equations
\eqref{eq:5.2a} are defined up to gauge transformations
\begin{equation}\label{eq:5.10}
\CA^{0,1}\ \mapsto\ \tilde{\CA}^{0,1}\ =\ g\CA^{0,1}g^{-1}+g\dparb
g^{-1}
\end{equation}
generated by smooth $\sGL(n,\FC)$-valued functions $g$ on
$\CP^{2|4}$. The transformations \eqref{eq:5.10} do not change the
holomorphic structure $\dparb_\CA$ on the bundle $E$ and the two
$(0,1)$-connections in \eqref{eq:5.10} are considered as
equivalent. On infinitesimal level, the transformations
\eqref{eq:5.10} take the form
\begin{equation}
\delta\CA^{0,1}\ =\ \dparb B+[\CA^{0,1},B]
\end{equation}
with $B\in H^0(\CP^{2|4},\sEnd E)$ and such a field $B$ solving
\eqref{eq:5.2b} generates holomorphic transformations such that
$\delta \CA^{0,1}\ =\ 0$. Their finite version is
\begin{equation}
\tilde{\CA}^{0,1}\ =\ g\CA^{0,1}g^{-1}+g\dparb g^{-1}\ =\
\CA^{0,1}~,
\end{equation}
and for a gauge potential $\CA^{0,1}$ given by \eqref{eq:5.4},
such a $g$ takes the form
\begin{equation}
g_\pm\ =\
\tilde{\psi}_\pm\de^{B_0^\pm}\tilde{\psi}_\pm^{-1}~~~\mbox{with}~~~
g_+\ =\ g_-~~\mbox{on}~~\hat{\CV}_+\cap\hat{\CV}_-~.
\end{equation}

Thus, the hBF theory given by the action \eqref{eq:5.1} and the
field equations \eqref{eq:5.2} describes holomorphic structures on
the bundle $E\rightarrow \CP^{2|4}$ and its moduli space
$\CM_\mathrm{hBF}$ is bijective to the moduli space of holomorphic
bundles $\tilde{E}$ defined by the transition functions
\eqref{eq:5.7}. Furthermore, this moduli space is bijective to the
moduli space $\CM_{\mathrm{phCS}}$ of $\CT$-flat bundles
$\tilde{\CE}$ over the CR supertwistor space $\CF^{5|8}$.
Summarizing our above discussion, we have established the diagram
\begin{equation}
\begin{aligned}
\begin{picture}(180,70)(0,-5)
\put(0.0,0.0){\makebox(0,0)[c]{hBF theory on $\CP^{2|4}$}}
\put(160.0,14.0){\makebox(0,0)[c]{supersymmetric}}
\put(160.0,0.0){\makebox(0,0)[c]{ Bogomolny model on $\FR^3$}}
\put(80.0,50.0){\makebox(0,0)[c]{phCS theory on $\CF^{5|8}$}}
\put(40.0,40.0){\vector(-1,-1){30}}
\put(10.0,10.0){\vector(1,1){30}}
\put(100.0,40.0){\vector(1,-1){18}}
\put(118.0,22.0){\vector(-1,1){18}} \put(56,0){\vector(1,0){37}}
\put(86,0){\vector(-1,0){30}}
\end{picture}
\end{aligned}
\end{equation}
describing equivalent theories defined on different spaces. Here
it is again implied that the appropriate subsets of the solution
spaces to phCS and hBF theories are considered as discussed in
section 4.3 and above.

\section{Supersymmetric Bogomolny equations with massive fields}

In an interesting recent paper \cite{Chiou:2005jn}, a twistor
string theory corresponding to a certain massive super Yang-Mills
theory in three dimensions was developed. It was argued, that the
mass terms in this theory arise from coupling the R-symmetry
current to a constant background field when performing the
dimensional reduction. In this section, we want to study the
analogous construction for the supersymmetric Bogomolny model
which we discussed in the previous sections. We focus on the
geometric origin of the additional mass terms by discussing the
associated twistor description. More explicitly, we establish a
correspondence between holomorphic bundles over the deformed
mini-supertwistor space introduced in \cite{Chiou:2005jn} and
solutions to massive supersymmetric Bogomolny equations in three
dimensions.

\subsection{Mini-supertwistor and CR supertwistor spaces as vector
bundles}

\paragraph{$\CP^{2|4}$ as a supervector bundle.} We start from
the observation that the mini-supertwistor space $\CP^{2|4}$ can
be considered as the total space of a rank $0|4$ holomorphic
supervector bundle\footnote{A complex (real) supervector bundle of
rank $p|q$ is a vector bundle, whose typical fibre is the
superspace $\FC^{p|q}$ ($\FR^{p|q}$). We also refer to such a
supervector bundle simply as a vector bundle of rank $p|q$.} over
the mini-twistor space $\CP^2$, i.e.\ a holomorphic vector bundle
with Gra{\ss}mann odd fibres. More explicitly, we have
\begin{equation}
\CP^{2|4}\ =\ \CO(2)\oplus\Pi\CO(1)\otimes\FC^4
\end{equation}
together with a holomorphic projection
\begin{equation}\label{eq:6.2}
\CP^{2|4}\ \rightarrow\ \CP^2~.
\end{equation}
Recall that the mini-twistor space $\CP^2$ is covered by two
patches $\CV_\pm$ with coordinates $w_\pm^1$ and
$w_\pm^2=\lambda_\pm$. The additional fibre coordinates in the
supervector bundle $\CP^{2|4}$ over $\CP^2$ are the Gra{\ss}mann
variables $\eta_i^\pm$. For later convenience, we rearrange them
into the vector $\eta^\pm=(\eta_i^\pm)\in\FC^{0|4}$. On
$\CV_+\cap\CV_-$, we have the relation
\begin{equation}\label{eq:6.3}
\eta^+\ =\ \varphi_{+-}\eta^-
\end{equation}
with the transition function
\begin{equation}\label{eq:6.4}
\varphi_{+-}\ =\ w_+^2 (\delta_i{}^j)\ =\ w_+^2\unit_4~.
\end{equation}

\paragraph{$\CF^{5|8}$ as a supervector bundle.} The CR supertwistor
space $\CF^{5|8}$ is a CR supervector bundle\footnote{i.e., it has
a transition function annihilated by the vector fields
$\dpar_{\bw_\pm^1}$, $\dpar_{\bw_\pm^2}$ from the distribution
$\CCDb$ on $\CF^5$} over the CR twistor space
$\CF^5\cong\FR^3\times S^2$,
\begin{equation}\label{eq:6.5}
\CF^{5|8}\ \rightarrow\ \CF^5~,
\end{equation}
with complex coordinates $\eta_i^\pm$ on the fibres $\FC^{0|4}$
over the patches $\CV_\pm'$ covering $\CF^5$. Recall that we have
the double fibration
\begin{equation}\label{eq:6.5.2}
\begin{aligned}
\begin{picture}(50,40)
\put(0.0,0.0){\makebox(0,0)[c]{$\CP^{2}$}}
\put(64.0,0.0){\makebox(0,0)[c]{$\FR^{3}$}}
\put(34.0,33.0){\makebox(0,0)[c]{$\CF^{5}$}}
\put(7.0,18.0){\makebox(0,0)[c]{$\nu_2$}}
\put(55.0,18.0){\makebox(0,0)[c]{$\nu_1$}}
\put(25.0,25.0){\vector(-1,-1){18}}
\put(37.0,25.0){\vector(1,-1){18}}
\end{picture}
\end{aligned}
\end{equation}
in the purely bosonic case and the transition function of the
supervector bundle \eqref{eq:6.5} can be identified with
\begin{equation}
\nu_2^*\varphi_{+-}\ =\ \lambda_+(\delta_i{}^j)\ =\
\lambda_+\unit_4~,
\end{equation}
i.e., we have the same transformation \eqref{eq:6.3}
relating\footnote{Note that our notation often does not
distinguish between objects on $\CP^2$ and their pull-backs to
$\CF^5$.} $\eta^+$ to $\eta^-$ on $\CV_+'\cap \CV_-'$.

\paragraph{Combined double fibrations.} Finally, note that $\FC^{0|4}\cong\FR^{0|8}$
and the superspace $\FR^{3|8}\cong\FR^3\times\FR^{0|8}\cong
\FR^3\times \FC^{0|4}$ can be considered as a trivial supervector
bundle over $\FR^3$ with the canonical projection
\begin{equation}\label{eq:6.7}
\FR^{3|8}\ \rightarrow\ \FR^3~.
\end{equation}
Thus, we arrive at the diagram
\begin{equation}
\begin{aligned}
\begin{picture}(100,100)
\put(0.0,0.0){\makebox(0,0)[c]{$\CP^{2}$}}
\put(0.0,53.0){\makebox(0,0)[c]{$\CP^{2|4}$}}
\put(96.0,0.0){\makebox(0,0)[c]{$\FR^{3}$}}
\put(96.0,53.0){\makebox(0,0)[c]{$\FR^{3|8}$}}
\put(51.0,33.0){\makebox(0,0)[c]{$\CF^{5}$}}
\put(51.0,85.0){\makebox(0,0)[c]{$\CF^{5|8}$}}
\put(16.5,23.0){\makebox(0,0)[c]{$\nu_2$}}
\put(76.5,23.0){\makebox(0,0)[c]{$\nu_1$}}
\put(16.5,75.0){\makebox(0,0)[c]{$\pi_2$}}
\put(76.5,75.0){\makebox(0,0)[c]{$\pi_1$}}
\put(37.5,25.0){\vector(-3,-2){25}}
\put(55.5,25.0){\vector(3,-2){25}}
\put(37.5,78.0){\vector(-3,-2){25}}
\put(55.5,78.0){\vector(3,-2){25}}
\put(0.0,45.0){\vector(0,-1){37}}
\put(90.0,45.0){\vector(0,-1){37}}
\put(45.0,78.0){\vector(0,-1){37}}
\end{picture}
\end{aligned}
\end{equation}
combining the double fibrations \eqref{eq:3.19} and
\eqref{eq:6.5.2}.

\subsection{Deformed mini-supertwistor and CR supertwistor
spaces}

\paragraph{$\CP^{2|4}_M$ as a supervector bundle.}
Let us define a holomorphic supervector bundle
\begin{equation}\label{eq:6.9}
\CP^{2|4}_M\ \rightarrow\ \CP^2
\end{equation}
with complex coordinates $\tilde{\eta}^\pm\ =\
(\tilde{\eta}_i^\pm)\in\FC^{0|4}$ on the fibres over
$\CV_\pm\subset \CP^2$ which are related by the transition
function
\begin{equation}\label{eq:6.10}
\tilde{\varphi}_{+-}\ =\ w_+^2\de^{\frac{w^1_+}{w^2_+}M}
\end{equation}
on the intersection $\CV_+\cap\CV_-$, i.e.\
\begin{equation}\label{eq:6.11}
\tilde{\eta}^+\ =\ \tilde{\varphi}_{+-}\tilde{\eta}^-~.
\end{equation}
For reasons which will become more transparent in the later
discussion, we demand that $M$ is traceless and hermitean. The
matrix $M$ will eventually be the mass matrix of the fermions in
three dimensions.

This supermanifold $\CP^{2|4}_M$ was introduced in
\cite{Chiou:2005jn} as the target space of twistor string
theories\footnote{For this, $\CP^{2|4}_M$ has to be a Calabi-Yau
supermanifold, which is the reason underlying the above
restriction to $\tr M=0$, as we will discuss.} which correspond,
as proposed in this paper, to a supersymmetric Yang-Mills theory
in three dimensions with massive spinors and both massive and
massless scalar fields for hermitean matrices $M$. In the
following, we provide a twistorial derivation of analogous mass
terms in our supersymmetric Bogomolny model and explain their
geometric origin.

\paragraph{$\CF_M^{5|8}$ as a supervector bundle.}
Consider the rank $0|4$ holomorphic supervector bundle
\eqref{eq:6.9} and its pull-back
\begin{equation}\label{eq:6.12}
\CF^{5|8}_M\ :=\  \nu_2^*\CP^{2|4}\ \rightarrow\ \CF^5
\end{equation}
to the space $\CF^5$ from the double fibration \eqref{eq:6.5.2}.
Note that the supervector bundle $\CF_M^{5|8}\rightarrow \CF^5$ is
smoothly equivalent to the supervector bundle
$\CF^{5|8}\rightarrow \CF^5$ since in the coordinates
$(y^{\ald\bed},\lambda_\pm,\bl_\pm)=(y,\by,x^3,\lambda_\pm,\bl_\pm)$
on $\CF^5$, the pulled-back transition function
$\nu_2^*\tilde{\varphi}_{+-}$ can be split
\begin{equation}\label{eq:6.13}
\nu_2^*\tilde{\varphi}_{+-}\ =\
\lambda_+\de^{\frac{1}{\lambda_+}y^{\ald\bed}\lambda^+_\ald
\lambda^+_\bed M}\ =\ \varphi_+(\lambda_+\unit_4)\varphi_-^{-1}\
\sim\ \lambda_+\unit_4~.
\end{equation}
Here,
\begin{equation}
\varphi_+\ =\ \de^{-(x^3+\lambda_+\by)M}\ =\ \de^{\lambda^+_\ald
y^{\ald\zd}M}\eand \varphi_-\ =\ \de^{(x^3-\lambda_-y)M}\ =\
\de^{-\lambda^-_\ald y^{\ald\ed}M}
\end{equation}
are matrix-valued functions well-defined on the patches $\CV_+'$
and $\CV_-'$, respectively. Remember that $\etat_i^+$ and
$\etat_i^-$ are related by \eqref{eq:6.11} and their pull-backs to
$\CF^5$ (which we denote again by the same letter) are related by
the transition function \eqref{eq:6.13}. Therefore, we have
\begin{equation}\label{eq:6.15}
(\varphi_+^{-1}\tilde{\eta}^+)\ =\
\lambda_+(\varphi_-^{-1}\tilde{\eta}^-)
\end{equation}
From this, we conclude that
\begin{equation}\label{eq:6.16}
\etat^+\ =\ \varphi_+\eta^+\ =\
\de^{\lambda_\ald^+y^{\ald\zd}M}\eta^+\eand \etat^-\ =\
\varphi_-\eta^-\ =\ \de^{-\lambda_\ald^-y^{\ald\ed}M}\eta^-~,
\end{equation}
where $\eta^\pm=(\eta_i^\pm)$ are the fibre coordinates of the
bundle \eqref{eq:6.5} related by \eqref{eq:6.3} on the
intersection $\CV_+'\cap\CV_-'$.

\paragraph{The fibration $\CF_M^{5|8}\rightarrow \CF^5$ in the Dolbeault picture.}
It follows from \eqref{eq:6.16} that
\begin{equation}\label{eq:6.17}
\bW_1^\pm\etat^\pm_i\ =\ 0~,~~~\bW^\pm_2\etat_i^\pm\ =\ 0\eand
\bW_3^\pm\etat_i^\pm+M_i{}^j\etat^\pm_j\ =\ 0~.
\end{equation}
Recall that the vector fields $\bW_a^\pm$ generate an integrable
(bosonic) distribution $\CT_b=$\linebreak $\spn\{\bW_a^\pm\}$
together with the operator
\begin{equation}
\dd_{\CT_b}|_{\CV'_\pm}\ =\ \dd \bw_\pm^a\der{\bw_\pm^a}~,
\end{equation}
which annihilates the transition function \eqref{eq:6.13} of the
bundle \eqref{eq:6.12}. Due to formul\ae{} \eqref{eq:6.13} and
\eqref{eq:6.17}, the supervector bundle $\CF_M^{5|8}$ with
canonical $\CT_b$-flat connection $\dd_{\CT_b}$ is diffeomorphic
to the supervector bundle $\CF^{5|8}$ with the $\CT_b$-flat
connection $\dd_{\CT_b}+\hat{\CA}_{\CT_b}$ the components
$\hat{\CA}_a^\pm=\bW_a^\pm\lrcorner\hat{\CA}_{\CT_b}$ of which are
given by
\begin{equation}\label{eq:6.19}
\hat{\CA}^\pm_1\ =\ 0~,~~~ \hat{\CA}^\pm_2\ =\
0\eand\hat{\CA}^\pm_3\ =\ M~.
\end{equation}
In other words, we have an equivalence of the following data:
\begin{equation}\label{eq:6.20}
(\CF^{5|8}_M,\tilde{\varphi}_{+-},\dd_{\CT_b})\ \sim\
(\CF^{5|8},\varphi_{+-}\!=
\lambda_+\unit_4,\dd_{\CT_b}+\hat{\CA}_{\CT_b})~.
\end{equation}

By construction, the connection one-form $\hat{\CA}_{\CT_b}$,
given explicitly in \eqref{eq:6.19}, is a solution to the field
equations
\begin{equation}
\dd_{\CT_b}\hat{\CA}_{\CT_b}+\hat{\CA}_{\CT_b}
\wedge\hat{\CA}_{\CT_b}\ =\ 0
\end{equation}
of phCS theory on $\CF^5$, which are equivalent via the arguments
of section 4.3 to the Bogomolny equations on $\FR^3$. Due to this
correspondence, \eqref{eq:6.19} is equivalent to a solution of the
Bogomolny equations with vanishing Yang-Mills gauge potential
$a_{\ald\bed}$ and constant Higgs field
\begin{equation}\label{eq:6.22}
\phi\ =\ (\phi_i{}^j)\ =\ -\di(M_i{}^j)~,
\end{equation}
which takes values in the Lie algebra $\asu(4)$ of the R-symmetry
group $\sSU(4)$. Thus, the data \eqref{eq:6.20} are equivalent to
the trivial supervector bundle \eqref{eq:6.7} together with the
differential operators
\begin{equation}
\nabla_{\ald\bed}\ =\
\dpar_{(\ald\bed)}-\tfrac{1}{2}\eps_{\ald\bed}M
\end{equation}
encoding the information about the matrix $M$, i.e.\
\begin{equation}\label{eq:6.24}
(\CF_M^{5|8},\tilde{\varphi}_{+-},\dd_{\CT_b})\ \sim \
(\CF^{5|8},\varphi_{+-},\dd_{\CT_b}+\hat{\CA}_{\CT_b})\ \sim\
(\FR^{3|8},\nabla_{\ald\bed}=
\dpar_{(\ald\bed)}-\tfrac{1}{2}\eps_{\ald\bed}M)~.
\end{equation}

Note that the gauge potential $A_{\ald\bed}$ corresponding
to\footnote{Here, $A_r$ with $r=1,2,3$ are the components of the
ordinary gauge potential in three dimensions.} $A_r\in\ahu(n)$ in
a different basis and the Higgs fields $\Phi\in\ahu(n)$ considered
in section 4 can be combined with $a_{\ald\bed}$ and $\phi$ into
the fields
\begin{equation}
A_{\ald\bed}\otimes\unit_4+\unit_n\otimes a_{\ald\bed}\eand
\Phi\otimes\unit_4+\unit_n\otimes\phi
\end{equation}
acting on the tensor product $V_{\sU(n)}\otimes V_{\sSU(4)}$ of
the (adjoint) representation space $V_{\sU(n)}$ of the gauge group
and the representation space $V_{\sSU(4)}$ of the R-symmetry
group.

\paragraph{The fibration $\CP_M^{2|4}\rightarrow \CP^2$ in the Dolbeault picture.}
For completeness, we note that the deformed complex supervector
bundle $\CP^{2|4}_M\rightarrow \CP^2$ with the transition function
$\tilde{\varphi}_{+-}$ from \eqref{eq:6.10} and the holomorphic
structure
\begin{equation}
\dparb_b|_{\CV_\pm}\ =\
\dd\bw_\pm^1\der{\bw_\pm^1}+\dd\bw_\pm^2\der{\bw_\pm^2}
\end{equation}
is smoothly equivalent to the bundle $\CP^{2|4}\rightarrow \CP^2$
with the transition function $\varphi_{+-}$ from \eqref{eq:6.4}
and the holomorphic structure defined by the fields
$\hat{\CA}^{0,1}$ and $\hat{B}$ with the components\footnote{These
components can be derived from formula \eqref{eq:6.split} given
below.}
\begin{equation}
\hat{\CA}_{\bw^1_\pm}\ =\ 0~,~~~ \hat{\CA}_{\bw^2_\pm}\ =\
\mp\frac{w_\pm^1}{(1+w_\pm^2\bw_\pm^2)^2}M\eand \hat{B}_\pm\ =\
\hat{\CA}_{\bw^3_\pm}\ =\ M~.
\end{equation}
The fields $\hat{\CA}^{0,1}$ and $\hat{B}$ obviously satisfy the
field equations
\begin{equation}
\dparb_b\hat{\CA}^{0,1}+\hat{\CA}^{0,1}\wedge\hat{\CA}^{0,1}\ =\
0\eand \dparb_b\hat{B}+[\hat{\CA}^{0,1},\hat{B}]\ =\ 0
\end{equation}
of hBF theory on $\CP^2$. By repeating the discussion of section
5, one can show the equivalence of the data
\begin{equation}
(\CP^{2|4}_M,\tilde{\varphi}_{+-},\dparb_b)\ \sim\
(\CP^{2|4},\varphi_{+-}\!=\lambda_+\unit_4,\dparb_b+
\hat{\CA}^{0,1})\ \sim\
(\CF^{5|8}_M,\tilde{\varphi}_{+-},\dd_{\CT_b})~,
\end{equation}
which extends the equivalences described in \eqref{eq:6.24}.

\subsection{The deformed CR supertwistor space as a
supermanifold}

For developing a twistor correspondence involving the deformed CR
supertwistor space $\CF_M^{5|8}$, the description of $\CF_M^{5|8}$
as a rank $0|4$ complex supervector bundle with a constant gauge
potential \eqref{eq:6.19} which twists the direct product of even
and odd spaces is not sufficient. We rather have to interpret the
total space of $\CF_M^{5|8}$ as a supermanifold with deformed CR
structure and deformed distribution $\CT_M$.

\paragraph{Vector fields on $\CF_M^{5|8}$.}
Remember that a covariant derivative along a vector field on the
base space of a bundle can be lifted to a vector field on the
total space of the bundle. In our case of the bundle
\eqref{eq:6.12}, the lift of \eqref{eq:6.17} reads as
\begin{equation}
\bW_1^\pm\etat_i^\pm\ =\ 0~,~~~\bW_2^\pm\etat_i^\pm\ =\ 0\eand
\left(\bW_3^\pm+M_k{}^j\etat_j^\pm\der{\etat_k^\pm}\right)
\etat_i^\pm\ =\ 0~.
\end{equation}
To see the explicit form of the vector fields corresponding to the
integrable distribution
\begin{equation}\label{eq:6.30}
\CT_M\ =\
\spn\left\{\der{\bw_\pm^a},\der{\bar{\etat}_i^\pm}\right\}
\end{equation}
on $\CF^{5|8}_M$, it is convenient to switch to the coordinates
$(y^{\ald\bed},\lambda_\pm,\bl_\pm,\eta_i^\ald)$ by the
formul\ae{}
\begin{subequations}\label{eq:6.31}
\begin{align}
w_\pm^1\ =\ \lambda_\ald^\pm\lambda_\bed^\pm y^{\ald\bed}~,~~~
w_\pm^2\ =\ \lambda_\pm\eand
w_\pm^3\ =\ -\gamma_\pm\lambda_\ald^\pm\hl_\bed^\pm y^{\ald\bed}~,\\
\etat_i^+\ =\ \left(\de^{\lambda^+_\ald y^{\ald\zd}
M}\right)_i^{~j}\eta_j^\bed\lambda_\bed^+\eand \etat_i^-\ =\
\left(\de^{-\lambda^-_\ald y^{\ald\ed}
M}\right)_i^{~j}\eta_j^\bed\lambda_\bed^-~.
\end{align}
\end{subequations}
By a straightforward calculation, we obtain
\begin{equation}\label{eq:6.32}
\begin{aligned}
\dd_{\CT_M}|_{\tilde{\CV}_\pm}&\ =\ \dd
\bw_\pm^a\der{\bw_\pm^a}+\dd
\bar{\etat}_i^\pm\der{\bar{\etat}_i^\pm}\\
&\ =\ \Thetab^1_\pm \bar{\CW}_1^\pm+\Thetab^2_\pm
\bar{\CW}_2^\pm+(\Thetab^3_\pm
\mp\gamma_\pm^2\hl^\pm_\ald\hl^\pm_\bed
y^{\ald\bed}\Theta^2_\pm)\bar{\CW}_3^\pm+
\bar{\CE}_i^\pm\bV_\pm^i~,
\end{aligned}
\end{equation}
where
\begin{equation}\label{eq:6.33}
\begin{aligned}
\bar{\CW}_1^\pm&\ :=\ \bW_1^\pm\mp\lambda_\pm(T\bar{M}T)_i{}^j
\hl_\ald^\pm\eta_j^\ald\bV_\pm^i~,~~~~\bar{\CW}_2^\pm\ :=\
\bW_2^\pm~,\\
\bar{\CW}_3^\pm&\ :=\
\bW_3^\pm+\gamma_\pm(TM)_i{}^j\lambda_\ald^\pm\eta_j^\ald
V_\pm^i+\gamma_\pm(T\bar{M}T)_i{}^j\hl_\ald^\pm\eta_j^\ald
\bV_\pm^i~,\\
\bar{\CE}^+_i&\ :=\
\bar{E}_i^++\gamma_+\hl_\ald^+\hl^+_\bed\eta^\bed_j
(T\bar{M}T)_i{}^j\dd y^{\ed\ald}~,\\
\bar{\CE}^-_i&\ :=\
\bar{E}_i^-+\gamma_-\hl_\ald^-\hl^-_\bed\eta^\bed_j
(T\bar{M}T)_i{}^j\dd y^{\zd\ald}
\end{aligned}
\end{equation}
and $\bW_a^\pm,V_\pm^i,\bV_\pm^i$ and $\Thetab_\pm^a,\Theta_\pm^a$
were given in \eqref{eq:3.34all}--\eqref{eq:3.37} and
$(AB)_i{}^j:=A_i{}^kB_k{}^j$. In fact, the formul\ae{}
\eqref{eq:6.31} and their inverses define a diffeomorphism between
the supermanifolds $\CF_M^{5|8}=(\FR^{3|8}\times S^2,\CT_M)$ and
$\CF^{5|8}=(\FR^{3|8}\times S^2,\CT)$ which have different
integrable distributions $\CT_M$ and $\CT$ (and different CR
structures).

\paragraph{Vector fields on $\CP_M^{2|4}$.}
In the above discussion, we used a transformation from the
coordinates $\tilde{\eta}_i^\pm$ to the coordinates $\eta_i^\pm$
on $\CF^{5|8}_M$, which are (pulled-back) sections of $\Pi\CO(1)$.
The corresponding splitting of the transition function was given
in \eqref{eq:6.13}--\eqref{eq:6.15}. One can find a similar
splitting of the transition function \eqref{eq:6.10} also on the
complex supermanifold $\CP^{2|4}_M$ and obtain new coordinates
$\hat{\eta}_i^\pm$, which are sections of $\Pi \CO(1)$, as well.
Explicitly, we have
\begin{equation}\label{eq:6.split}
\de^{\frac{w_+^1}{w_+^2}M}\ =\
\de^{\left(1-\frac{1}{1+w_+^2\bw_+^2}\right)\frac{w_+^1}{w_+^2}
M}\de^{\frac{w_+^1}{w_+^2(1+w_+^2\bw_+^2)} M}\ =\
\de^{\frac{\bw_+^2 w_+^1}{1+w_+^2\bw_+^2} M}\de^{\frac{\bw_-^2
w_-^1}{1+w_-^2\bw_-^2}M}~,
\end{equation}
which yields the formul\ae{}
\begin{equation}
\tilde{\eta}^+\ =\ \de^{\frac{\bw_+^2 w_+^1}{1+w_+^2\bw_+^2}
M}\hat{\eta}^+\eand\tilde{\eta}^-\ =\ \de^{-\frac{\bw_-^2
w_-^1}{1+w_-^2\bw_-^2}M}\hat{\eta}^-~.
\end{equation}
From this and \eqref{eq:6.11}, it follows that
\begin{equation}
\hat{\eta}_i^+\ =\ w_+^2\hat{\eta}_i^-~
\end{equation}
and these coordinates have the desired property. Furthermore, in
the $(0,1)$ part of the differential
\begin{equation}
\begin{aligned}
&\dparb|_{\hat{\CV}_\pm}\ =\
\ddpart{\bw^1_\pm}+\ddpart{\bw^2_\pm}+
\ddpart{\bar{\tilde{\eta}}_i^\pm}\\
&\ =\ \dd \bar{\hat{w}}^1_\pm\dpar_{\bar{\hat{w}}^1_\pm}+\dd
\bar{\hat{w}}^2_\pm\left(\dpar_{\bar{\hat{w}}^2_\pm}\mp
\gamma^2_\pm\hat{w}^1_\pm
M_i{}^j\hat{\eta}_j^\pm\dpar_{\hat{\eta}^\pm_i}\right)+\left(\dd
\bar{\hat{\eta}}^\pm_i\pm\gamma^2_\pm\bar{\hat{w}}^1_\pm
\bar{M}_i{}^j\bar{\hat{\eta}}_j^\pm\dd
\hat{w}^2_\pm\right)\dpar_{\bar{\hat{\eta}}^\pm_i}~,
\end{aligned}
\end{equation}
where we introduced $\hat{w}^{1,2}_\pm=w_\pm^{1,2}$ for clarity,
we see explicitly the deformation of the complex structure from
$\CP^{2|4}$ to $\CP^{2|4}_M$. Note that the coordinates
$\hat{\eta}_i^\pm$ can be pulled-back to $\CF_M^{5|8}$ and are
there related to the coordinates $\eta^\pm$ by
\begin{equation}
\hat{\eta}^\pm\ =\ \de^{-w_\pm^3 M}\eta^\pm~.
\end{equation}

\subsection{Mass-deformed Bogomolny equations from phCS theory
on $\CF_M^{5|8}$}

\paragraph{Holomorphic integral forms.}
The deformed mini-supertwistor space $\CP^{2|4}_M$ fits into a
double fibration
\begin{equation}\label{eq:6.36}
\begin{aligned}
\begin{picture}(50,50)(0,-7)
\put(0.0,0.0){\makebox(0,0)[c]{$\CP_M^{2|4}$}}
\put(64.0,0.0){\makebox(0,0)[c]{$\FR^{3|8}$}}
\put(34.0,37.0){\makebox(0,0)[c]{$\CF_M^{5|8}$}}
\put(7.0,20.0){\makebox(0,0)[c]{$\pi_2$}}
\put(55.0,20.0){\makebox(0,0)[c]{$\pi_1$}}
\put(25.0,27.0){\vector(-1,-1){18}}
\put(37.0,27.0){\vector(1,-1){18}}
\end{picture}
\end{aligned}
\end{equation}
similarly to the undeformed case $M=0$. Recall that we had a
holomorphic integral form on $\CP^{2|4}$ locally defined by
\begin{equation}
\Omega|_{\hat{\CV}_\pm}\ =\ \pm\dd w_\pm^1\wedge \dd
w_\pm^2\dd\eta_1^\pm\cdots\dd\eta_4^\pm~.
\end{equation}
One can extend $\Omega$ to a nonvanishing holomorphic volume form
\begin{equation}\label{eq:6.37}
\Omega^M|_{\hat{\CV}_\pm}\ =\ \pm\dd w_\pm^1\wedge \dd
w_\pm^2\dd\etat_1^\pm\cdots\dd\etat_4^\pm~
\end{equation}
on $\CP^{2|4}_M$ if and only if $\tr M=0$ \cite{Chiou:2005jn}.
This is the reason, why we imposed this condition from the very
beginning. Similarly to the discussion of phCS theory on
$\CF^{5|8}$ in section 4, we consider a submanifold
$\CCX_M\subset\CF^{5|8}_M$ which is defined by the constraints
$\bar{\etat}_i^\pm=0$. Clearly, the latter equations are
equivalent to $\etab_i^\pm=0$ and therefore $\CCX_M$ is
diffeomorphic to $\CCX$. Note that the pull-back of the
holomorphic integral form \eqref{eq:6.37} to $\CF_M^{5|8}$
coincides with $\pi_2^*\Omega$,
\begin{equation}
\tilde{\Omega}^M|_{\tilde{\CV}_\pm}\ :=\
\pi_2^*\Omega^M|_{\hat{\CV}_\pm}\ =\
\pm\Theta^1_\pm\wedge\Theta^2_\pm\dd \etat_1^\pm\cdots\dd
\etat_4^\pm\ =\  \pm\Theta^1_\pm\wedge \Theta^2_\pm\dd
\eta_1^\pm\cdots\dd \eta_4^\pm
\end{equation}
which is due to \eqref{eq:6.16} and the tracelessness of $M$.

\paragraph{Action functional.}
From here on, we proceed as in section 4 and consider a trivial
rank $n$ complex vector bundle over the CR supertwistor space
$\CF_M^{5|8}$ with a connection $\CA_{\CT_M}$ along the integrable
distribution $\CT_M$ defined in \eqref{eq:6.30} and
\eqref{eq:6.32}. By assuming that
$\bV_\pm^i\lrcorner\CA_{\CT_M}=0$ and
$\bV_\pm^i(\bar{\CW}_a^\pm\lrcorner \CA_{\CT_M})=0$, we may define
the action functional
\begin{equation}\label{eq:6.39}
S^M_{\mathrm{phCS}}\ =\
\int_{\CCX_M}\tilde{\Omega}^M\wedge\tr\left(\CA_{\CT_M}\wedge
\dd_{\CT_M}\CA_{\CT_M}+\tfrac{2}{3}\CA_{\CT_M}\wedge\CA_{\CT_M}
\wedge\CA_{\CT_M}\right)
\end{equation}
of deformed phCS theory. The equations of motion keep the form
\eqref{eq:4.16} up to relabelling $\CT$ by $\CT_M$ and in
components $\CAt_a^\pm:=\bar{\CW}_a^\pm\lrcorner\CA_{\CT_M}$, we
have
\begin{equation}\label{eq:6.40}
\begin{aligned}
  \bar{W}_1^\pm\CAt^\pm_2-\bar{W}^\pm_2\CAt^\pm_1+
  [\CAt^\pm_1,\CAt^\pm_2]&\ =\ 0~,\\
  \bar{W}_2^\pm\CAt_3^\pm-\bar{W}^\pm_3\CAt^\pm_2+
  [\CAt^\pm_2,\CAt_3^\pm]\mp2\gamma_\pm^2\CAt^\pm_1
  &\ =\ M_j{}^i\eta_i^\pm\der{\eta_j^\pm}\CAt^\pm_2~,\\
  \bar{W}_1^\pm\CAt_3^\pm-\bar{W}_3^\pm\CAt^\pm_1+
  [\CAt^\pm_1,\CAt^\pm_3] &\
  =\
  M_j{}^i
      \eta_i^\pm\der{\eta_j^\pm}\CAt^\pm_1~,
\end{aligned}
\end{equation}
where the vector fields \eqref{eq:6.33} have already been
substituted. The dependence of the components $\CAt_a^\pm$ on
$\lambda_\pm,\bl_\pm$ and $\eta_i^\pm$ is of the same form as the
one for $\CA_a^\pm$ given in \eqref{eq:4.19} but with coefficient
functions obeying $M$-deformed equations. In the following, we
will not put tildes over the coefficient functions for simplicity.

\paragraph{Mass-deformed super Bogomolny equations.}
Substituting the expansions of the form \eqref{eq:4.19} for
$\CAt_a^\pm$ and our vector fields $\bar{W}_a^\pm$ into
\eqref{eq:6.40}, we obtain mass-deformed supersymmetric Bogomolny
equations
\begin{equation}\label{eq:6.41}
\begin{aligned}
  f_{\ald\bed}\ &=\ -\tfrac{\di}{2}D_{\ald\bed}\Phi~,\\
    \eps^{\bed\gad}D_{\ald\bed}\chi^i_\gad
             -\tfrac{1}{2}M_j{}^i\chi^j_\ald\ &=\
             -\tfrac{\di}{2}[\Phi,\chi^i_\ald]~,\\
\triangle \phi^{ij}+M_k{}^{[i}M_l{}^{[j]}\phi^{k]l}\ &=\
-\tfrac{1}{4}[\Phi,[\phi^{ij},\Phi]]
 -\di M_k{}^{[i}[\Phi,\phi^{j]k}]+\eps^{\ald\bed}
 \{\chi^i_\ald,\chi^j_\bed\}~,\\
\eps^{\bed\gad}D_{\ald\bed}\tilde{\chi}_{i\gad}-\tfrac{1}{2}
           M_i{}^j\tilde{\chi}_{j\ald}\ &=\ -\tfrac{\di}{2}
           [\tilde{\chi}_{i\ald},\Phi]+2\di[\phi_{ij},\chi^j_\ald]~,\\
  \eps^{\bed\gad}D_{\ald\bed}G_{\gad\ded}\ &=\
  -\tfrac{\di}{2}[G_{\ald\ded},\Phi]+
  \di\{\chi^i_\ald,\tilde{\chi}_{i\ded}\}
  -\tfrac{1}{2}[\phi_{ij},D_{\ald\ded}\phi^{ij}]\ +\ \\
 &\kern1cm +\tfrac{\di}{4}\eps_{\ald\ded}[\phi_{ij},
 [\Phi,\phi^{ij}]]+\tfrac{1}{2}
   \eps_{\ald\ded}M_m{}^k[\phi_{kl},\phi^{lm}]~,
\end{aligned}
\end{equation}
where we have again abbreviated
$\phi_{ij}:=\frac{1}{2!}\eps_{ijkl}\phi^{kl}$ and
$\nabla_{\ald\bed}=\dpar_{(\ald\bed)}+B_{\ald\bed}=
D_{\ald\bed}-\tfrac{\di}{2}\eps_{\ald\bed}\Phi$.

The equations \eqref{eq:6.40} show that, as in the undeformed case
\eqref{eq:4.17}, some of the fields appearing in the expansions of
$\CAt_a^\pm$ are not independent degrees of freedom but composite
fields:
\begin{equation}\label{eq:6.42}
\begin{aligned}
       \phi^{ij}_{\ald\bed}\ &=\ -(\partial_{(\ald\bed)}\phi^{ij}+
       [B_{\ald\bed},\phi^{ij}]
                              -\eps_{\ald\bed}M_k{}^{[i}\phi^{j]k})~,\\
       \tilde{\chi}^{ijk}_{\ald(\bed\gad)}\ &=\
       -\tfrac{1}{2}(\partial_{(\ald(\bed)}
              \tilde{\chi}^{ijk}_{\gad)}+
              [B_{\ald(\bed},\tilde{\chi}^{ijk}_{\gad)}]
               +\tfrac{3}{2}\eps_{\ald(\bed}
               M_l{}^{[i}\chi^{jk]l}_{\gad)})~,\\
       G^{ijkl}_{\ald(\bed\gad\ded)}\ &=\
       -\tfrac{1}{3}(\partial_{(\ald(\bed)}
          G^{ijkl}_{\gad\ded)}+[B_{\ald(\bed},G^{ijkl}_{\gad\ded)}]-
         2\eps_{\ald(\bed}M_m{}^{[i} G^{jkl]m}_{\gad\ded)})~.
\end{aligned}
\end{equation}

Finally, substituting our superfield expansions for $\CAt_a^\pm$
into the action \eqref{eq:6.39} and integrating over the odd
coordinates and over the Riemann sphere, we end up with
\begin{equation}\label{eq:6.43}
S^M_{\mathrm{sB}}\ =\ S_{\mathrm{sB}}-{\tfrac{1}{2}}\int\dd^3y\
{\rm tr}\left\{\di\eps^{\ald\bed}
      \tilde{\chi}_{i\ald}M_j{}^i\chi^j_\bed-\phi_{ij}
      M_k{}^iM_l{}^{[j}\phi^{k]l}+
       \di\Phi M_k{}^i[\phi_{ij},\phi^{jk}]\right\}~,
\end{equation}
where $S_{\mathrm{sB}}=S_{\mathrm{sB}}^{M=0}$ is the action
functional for the massless supersymmetric Bogomolny equations as
given in \eqref{eq:4.27}.

To sum up, we have described a one-to-one correspondence between
gauge equivalence classes of solutions to the supersymmetric
Bogomolny equations with massive fermions and scalar fields and
equivalence classes of $\CT_M$-flat bundles over the CR
supertwistor space $\CF_M^{5|8}$ which are holomorphically trivial
on each $\CPP^1_{x,\eta}\embd\CF^{5|8}_M$. We have also described
a one-to-one correspondence between the equivalence classes of
$\CT_M$-flat complex vector bundles over $\CF_M^{5|8}$ and of
holomorphic vector bundles over the deformed mini-supertwistor
space $\CP^{2|4}_M$. The assumption that these bundles become
holomorphically trivial on projective lines translates in the
Dolbeault description into a one-to-one correspondence between
gauge equivalence classes of solutions to the field equations of
$i$) hBF theory on the deformed mini-supertwistor space
$\CP_M^{2|4}$, $ii$) phCS theory on the CR supertwistor space
$\CF_M^{5|8}$ and $iii$) massive supersymmetric Bogomolny model on
$\FR^3$.

\section{Examples of solutions to the super Bogomolny equations}

In the preceding sections, we have presented in detail the
relations between the supersymmetric Bogomolny equations on the
Euclidean space $\FR^3$ and the field equations of phCS theory on
the CR supertwistor space $\CF^{5|8}$ as well as hBF theory on the
mini-supertwistor space $\CP^{2|4}$. We have shown that the moduli
spaces of solutions to the field equations of these three theories
are bijective. Furthermore, we introduced mass-deformed versions
of these field theories. In this section, we want to show how the
twistor correspondences described in the previous sections can be
used for constructing explicit solutions to the supersymmetric
Bogomolny equations. In fact, any solution to the standard
Bogomolny equations given as a pair $(A_{\ald\bed},\Phi)$ of a
gauge potential and a Higgs field can be extended to a solution
including the remaining fields of the supersymmetrically extended
Bogomolny equations in a nontrivial fashion. Here, we are not
considering this task in full generality but just want to give
some examples. For simplicity, we restrict ourselves to the case
when only the fields $A_{\ald\bed},\Phi$ and $G_{\ald\bed}$ are
non-zero. In this case, the supersymmetric Bogomolny equations
\eqref{eq:4.24} simplify to
\begin{subequations}\label{eq:7.1}
\begin{align}\label{eq:7.1a}
-\tfrac{1}{2}\eps^{\gad\ded}\left(\dpar_{(\ald\gad)}
A_{\bed\ded}-\dpar_{(\bed\ded)}A_{\ald\gad}+
[A_{\ald\gad},A_{\bed\ded}]\right)&\ =\ -\tfrac{\di}{2}
\left(\dpar_{(\ald\bed)}\Phi+[A_{\ald\bed},\Phi]\right)~,\\
\eps^{\gad\ded}\left(\dpar_{(\ald\gad)}G_{\bed\ded}+
[A_{\ald\gad},G_{\bed\ded}]\right)
&\ =\ -\tfrac{\di}{2}[G_{\ald\bed},\Phi]~.
\end{align}
\end{subequations}
First, we discuss Abelian solutions to these equations, which
correspond to Dirac mono\-pole-antimonopole systems. After this,
we present two algorithms which generate non-Abelian solutions.

\subsection{Abelian solutions}

\paragraph{Field equations.} In the Abelian case,
\eqref{eq:7.1} simplifies further to
\begin{equation}\label{eq:7.2}
\begin{aligned}
\eps^{\gad\ded}\left(\dpar_{(\ald\gad)}A_{\bed\ded}-
\dpar_{(\bed\ded)}A_{\ald\gad}\right)&\ =\ \di\dpar_{(\ald\bed)}\Phi~,\\
\eps^{\gad\ded}\dpar_{(\ald\gad)}G_{\bed\ded}&\ =\ 0~.
\end{aligned}
\end{equation}
It is convenient to rewrite these equations in terms of the real
coordinates $x^r$ on $\FR^3$ with $r=1,2,3$ as
\begin{subequations}\label{eq:7.4b}
\begin{align}\label{eq:7.4}
\tfrac{1}{2}\eps_{rst}(\dpar_sA_t-\dpar_tA_s)&\ =\
\dpar_r\Phi~,\\\label{eq:7.5} \dpar_r G_r&\ =\ 0~,\\\label{eq:7.6}
\eps_{rst}\dpar_sG_t&\ =\ 0~.
\end{align}
\end{subequations}
From \eqref{eq:7.5}, it follows that
\begin{equation}
G_r\ =\
\tfrac{1}{2}\eps_{rst}(\dpar_s\hat{A}_t-\dpar_t\hat{A}_s)~,
\end{equation}
and from \eqref{eq:7.6}, we obtain
\begin{equation}\label{eq:7.8}
G_r\ =\ -\dpar_r\hat{\Phi}~,
\end{equation}
where the sign in \eqref{eq:7.8} was chosen to match the fact that
in four dimensions, $G_r$ corresponds to an anti-self-dual
two-form with components $G_{\mu\nu}=\etab^r_{\mu\nu}G_r$ and
helicity $-1$, where $\etab^r_{\mu\nu}$ are the 't Hooft tensors.
Here, $\hat{A}_r$ and $\hat{\Phi}$ are a vector and a scalar,
respectively. Therefore, the equations \eqref{eq:7.4b} can be
rewritten as
\begin{subequations}
\begin{align}\label{eq:7.9}
\tfrac{1}{2}\eps_{rst}(\dpar_sA_t-\dpar_tA_s)&\ =\
\dpar_r\Phi~,\\\label{eq:7.10}
\tfrac{1}{2}\eps_{rst}(\dpar_s\hat{A}_t-\dpar_t\hat{A}_s)&\ =\
-\dpar_r\hat{\Phi}~.
\end{align}
\end{subequations}
It is well known that the equations \eqref{eq:7.9} describe Dirac
monopoles while \eqref{eq:7.10} describe Dirac antimonopoles (see
e.g.\ \cite{Atiyah:1988jp} and references therein). Thus, the
action \eqref{eq:4.27} with only the fields $f_{\ald\bed}$, $\Phi$
and $G_{\ald\bed}$ being non-zero can be considered as a proper
action for the description of monopole-antimonopole systems.

\paragraph{Abelian monopole-antimonopole configurations.}
Let us consider a configuration of $m_1$ Dirac monopoles and $m_2$
antimonopoles located at points $a_i=(a_i^1,a_i^2,a_i^3)$ with
$i=1,\ldots, m_1$ and $i=m_1+1,\ldots, m_1+m_2$, respectively.
Moreover, we assume for simplicity that $a_i^{1,2}\neq a_j^{1,2}$
for $i\neq j$. Such a configuration is then described by the
fields
\begin{equation}\label{eq:7.11}
\begin{aligned}
A^{N}&\ =\ \sum_{j=1}^{m_1}A^{N,j}~,&A^{S}&\ =\
\sum_{j=1}^{m_1}A^{S_,j}~,
&\Phi^N&\ =\ \Phi^S\ =\ \sum_{j=1}^{m_1}\frac{\di}{2r_j}~,\\
\hat{A}^{N}&\ =\
\sum_{j=m_1+1}^{m_1+m_2}\bar{A}^{N,j}~,&\hat{A}^{S}&\ =\
\sum_{j=m_1+1}^{m_1+m_2}\bar{A}^{S_,j}~,&\hat{\Phi}^N&\ =\
\hat{\Phi}^S\ =\ \sum_{j=m_1+1}^{m_1+m_2}\frac{\di}{2r_j}~,
\end{aligned}
\end{equation}
where $A^{N,j}= A^{N,j}_m\dd x^m$ and $A^{S,j}= A^{S,j}_m\dd x^m$
with
\begin{align}\label{eq:7.13}
\begin{aligned}
A_1^{N,j}&\ =\ \frac{\di x_j^2}{2r_j(r_j+x_j^3)}~,&A_2^{N,j}&\ =\
\frac{-\di
x_j^1}{2r_j(r_j+x_j^3)}~,&A_3^{N,j}&\ =\ 0~,\\
A_1^{S,j}&\ =\ -\frac{\di x_j^2}{2r_j(r_j-x_j^3)}~,&A_2^{S,j}&\ =\
\frac{\di x_j^1}{2r_j(r_j-x_j^3)}~,&A_3^{S,j}&\ =\ 0~,
\end{aligned}
\end{align}
\begin{equation}
x_j^s\ =\ x^s-a_j^s~,~~~r_j^2\ =\ \delta_{rs}x_j^rx_j^s~.
\end{equation}
Here, $N$ and $S$ denote the following two regions in $\FR^3:$
\begin{equation}
\begin{aligned}
\FR^3_{N,m_1+m_2}&\ :=\
\FR^3\backslash\bigcup_{i=1}^{m_1+m_2}
\left\{x^1=a_i^1,\,x^2=a_i^2,\,x^3\leq
a_i^3 \right\}~,\\
\FR^3_{S,m_1+m_2}&\ :=\
\FR^3\backslash\bigcup_{i=1}^{m_1+m_2}
\left\{x^1=a_i^1,\,x^2=a_i^2,\,x^3\geq
a_i^3 \right\}~
\end{aligned}
\end{equation}
and the bar stands for complex conjugation. Note that
\begin{align}
\FR^3_{N,m_1+m_2}\cup\FR^3_{S,m_1+m_2}\ =\
\FR^3\backslash\{a_1,\ldots,a_{m_1+m_2}\}
\end{align}
and the configuration \eqref{eq:7.11}, \eqref{eq:7.13} has
delta-function sources at the points $a_i$ with
$i=1,\ldots,m_1+m_2$.

\subsection{Non-Abelian solutions via a contour integral}

For the gauge group $\sSU(2)$, one can consider the Wu-Yang point
monopole \cite{wuyang} and its generalizations to configurations
describing $m_1$ monopoles and $m_2$ antimonopoles
\cite{Popov:2004rt}. This solution, which is singular at points
$a_i$, $i=1,\ldots,m_1+m_2$, is a solution to the equations
\eqref{eq:7.1} for $\asu(2)$-valued fields. However, it is just an
Abelian configuration in disguise, as it is equivalent to the
multi-monopole configuration \eqref{eq:7.11}, \eqref{eq:7.13}
\cite{Popov:2004rt}.

\paragraph{Solutions to linear equations.}
One can construct true non-Abelian solutions to \eqref{eq:7.1} as
follows. Let us consider a configuration $A_{\ald\bed}=0=\Phi$ and
$G_{\ald\bed}\neq 0$. Then from \eqref{eq:7.1}, one obtains the
equation
\begin{equation}\label{eq:7.9b}
\dpar_{(\ald\bed)}G^{\bed\gad}\ =\ 0~.
\end{equation}
A large class of solutions to this equation can be described in
the twistor approach \cite{Penrose:in} via the contour integral
\begin{equation}\label{eq:7.10b}
G_0^{\bed\gad}\ =\ \oint_{\gamma}\frac{\dd\lambda_+}{2\pi
\di}\lambda_+^\bed\lambda_+^\gad
\Upsilon(\lambda_\ald^+\lambda_\ded^+y^{\ald\ded},\lambda_+)~,
\end{equation}
where $\Upsilon(w_+^1,w_+^2)$ is a Lie-algebra valued meromorphic
function of $w_+^1=\lambda^+_\ald\lambda^+_\bed y^{\ald\bed}$ and
$w_+^2=\lambda_+$, holomorphic in the vicinity of the curve
$\gamma\cong S^1\subset\CPP^1$. From \eqref{eq:7.10b}, it follows
that nontrivial contributions to $G_0^{\bed\gad}$ are only given
by those $\Upsilon$ which are elements of the cohomology group
$H^1(\CP^2,\agl(n,\FC)\otimes\CO(-4))$. It is easy to see that
\eqref{eq:7.10b} satisfies \eqref{eq:7.9b} due to
\begin{equation}
\lambda_+^\bed\dpar_{(\ald\bed)}\Upsilon\ =\
\derr{\Upsilon}{w_+^1}\lambda_+^\bed\lambda_\bed^+\lambda_\ald^+\
=\ 0~,
\end{equation}
which appears after pulling the derivatives $\dpar_{(\ald\bed)}$
under the integral.

\paragraph{Dressed solutions.}
Consider now a fixed solution $(A_{\ald\bed},\Phi)$ of the
Bogomolny equations \eqref{eq:7.1a}, e.g.\ the $\sSU(2)$ BPS
monopole \cite{Prasad:1975kr,Bogomolny:1975de}. In the twistor
approach, we can find functions $\hat{\psi}_\pm$ solving the
linear system
\begin{equation}\label{eq:7.13a}
\lambda_\pm^\bed(\dpar_{(\ald\bed)}+A_{\ald\bed}-
\tfrac{\di}{2}\eps_{\ald\bed}\Phi)\hat{\psi}_\pm\
=\ 0\eand \dpar_{\bl_\pm}\hat{\psi}_\pm\ =\ 0~,
\end{equation}
which is equivalent to the linear system of phCS theory. These
$\hat{\psi}_\pm$ are known explicitly for many cases, e.g.\ for
our chosen example of the $\sSU(2)$ BPS monopole
\cite{Ward:1981jb}. Using $\hat{\psi}_\pm$, we can introduce
``dressed'' fields $G^{\bed\gad}$ by the formula
\begin{equation}\label{eq:7.13b}
G^{\bed\gad}\ =\ \oint_{\gamma}\frac{\dd \lambda_+}{2\pi\di}
\lambda_+^\bed\lambda_+^\gad\hat{\psi}_+(y,\lambda_+)
\Upsilon(y^{\ald\ded}\lambda_\ald^+\lambda_\ded^+,\lambda_+)
\hat{\psi}_-^{-1}(y,\lambda_+^{-1})~.
\end{equation}
One can check that with this choice,
\begin{equation}\label{eq:7.16}
\dpar_{(\ald\bed)}G^{\bed\gad}+[A_{\ald\bed}-
\tfrac{\di}{2}\eps_{\ald\bed}\Phi,G^{\bed\gad}]\
=\ 0~,
\end{equation}
and therefore the configuration $(A_{\ald\bed},\Phi,G_{\ald\bed})$
satisfies \eqref{eq:7.1}. The explicit form of a $G_{\ald\bed}$
for a given $A_{\ald\bed}$ and $\Phi$ is obtained from performing
the contour integral \eqref{eq:7.13b} along $\gamma$ after a
proper choice of the Lie-algebra valued function $\Upsilon$.
Recall that the configuration $(A_r,\Phi)$ will be real, i.e., the
fields will take values in the Lie algebra $\asu(n)$, if the
matrix-valued functions $\hat{\psi}_\pm$ in \eqref{eq:7.13a}
satisfy the reality condition \eqref{realitycond} and $\det
(\hat{\psi}_+^{-1}\hat{\psi}_-)=1$. Imposing a proper reality
condition on the function $\Upsilon$ will ensure the
antihermiticity of $G_r$.

\subsection{Solutions via nilpotent dressing transformations}

In this section, we will present a novel algorithm for
constructing solutions to the equations \eqref{eq:7.1} based on
the twistor description of hidden symmetry algebras in the
supersymmetric SDYM theory in four dimensions
\cite{Wolf:2004hp}.\footnote{For an earlier account of hidden
symmetry algebras in the context of SDYM theory, see e.g.\
\cite{Pohlmeyer:1979ya}--\cite{Popov:1995qb}.} Recall that we have
described a one-to-one correspondence between equivalence classes
$[\tilde{f}]$ of transition functions of $\CT$-flat vector bundles
$\tilde{\CE}$ over the CR supertwistor space $\CF^{5|8}$ obeying
certain triviality conditions and gauge equivalence classes
$[\hat{\CA}]$ of solutions to the supersymmetric Bogomolny
equations on $\FR^3$. We can, however, associate with any open
subset $\tilde{\CV}_+\cap\tilde{\CV}_-\subset\CF^{5|8}$ an
infinite number of such $[\tilde{f}]\in\CM_{\mathrm{phCS}}$, which
in turn yields an infinite number of
$[\hat{\CA}]\in\CM_{\mathrm{sB}}$. Therefore, one naturally meets
with a possibility of constructing new solutions from a given one
(dressing transformation). In the following, we will discuss an
example of such a construction but first we briefly recall the
necessary background (for details, see e.g.\
\cite{Popov:1995qb,Popov:1998fb,Wolf:2004hp}).

\paragraph{Linear system.} We consider the linear
system \eqref{eq:4.40}, which can be rewritten as\footnote{Note
that \eqref{eq:4.40a} and \eqref{eq:4.40c} are equivalent to
$\lambda_\pm^\ald(\bV_\ald^\pm+\hat{\CA}_\ald^\pm)\hat{\psi}_\pm=0$
and
$\hl_\pm^\ald(\bV_\ald^\pm+\hat{\CA}_\ald^\pm)\hat{\psi}_\pm=0$,
respectively, which together imply
$(\bV_\ald^\pm+\hat{\CA}_\ald^\pm)\hat{\psi}_\pm=0$.}
\begin{equation}\label{eq:7.15}
(\bV_\ald^\pm+\hat{\CA}_\ald^\pm)\hat{\psi}_\pm\ =\ 0~,~~~
\dpar_{\bl_\pm}\hat{\psi}_\pm\ =\ 0\eand
(\bV_\pm^i+\hat{\CA}_\pm^i)\hat{\psi}_\pm\ =\ 0~,
\end{equation}
where we have defined
\begin{equation}
\bV_\ald^\pm\ :=\  \lambda_\pm^\bed\dpar_{(\ald\bed)}\eand
\hat{\CA}_\ald^\pm\ :=\  \bV_\ald^\pm\lrcorner\hat{\CA}_{\CT}~.
\end{equation}
From arguments similar to those used subsequent to
\eqref{eq:4.40}, we have
$\hat{\CA}_\ald^\pm=\lambda_\pm^\bed\CB_{\ald\bed}$ and
$\hat{\CA}_\pm^i=\lambda_\pm^\ald\CA_\ald^i$ with
$\lambda$-independent superfields $\CB_{\ald\bed}$ and
$\CA^i_\ald$. The compatibility conditions for the linear system
\eqref{eq:7.15} are the equations \eqref{eq:4.42}. From this
linear system, one also derives that
$\tilde{f}_{+-}=\hat{\psi}_+^{-1}\hat{\psi}_-$ is $\CT$-flat,
i.e.\
\begin{equation}\label{eq:7.17}
\bV_\ald^+\tilde{f}_{+-}\ =\ 0~,~~~ \dpar_{\bl_+}\tilde{f}_{+-}\
=\ 0\eand \bV_+^i\tilde{f}_{+-}\ =\ 0~.
\end{equation}

\paragraph{Infinitesimal Riemann-Hilbert problem.}
The key idea is to study infinitesimal perturbations of the
transition function $\tilde{f}_{+-}$ of the $\CT$-flat vector
bundle preserving \eqref{eq:7.17} and the triviality properties
discussed above. More explicitly, given such a function
$\tilde{f}_{+-}=\hat{\psi}_+^{-1}\hat{\psi}_-$ (with
$\dpar_{\bl_\pm}\hat{\psi}_\pm=0$), we consider
\begin{equation}\label{eq:7.18}
\tilde{f}_{+-}+\delta \tilde{f}_{+-}\ =\
(\hat{\psi}_++\delta\hat{\psi}_+)^{-1}
(\hat{\psi}_-+\delta\hat{\psi}_-)~,
\end{equation}
where $\delta$ represents some generic infinitesimal perturbation.
Note that any infinitesimal $\CT$-flat perturbation (i.e.\
preserving \eqref{eq:7.17}) is allowed since for small
perturbations, the trivia\-lizability property of the bundle
$\tilde{\CE}$ on the holomorphic curves $\CPP^1_{x,\eta}\embd
\CF^{5|8}$ is preserved (by a variant of Kodaira's theorem). Upon
introducing the Lie-algebra valued function
\begin{equation}\label{M:5.2}
\phi_{+-}\ := \ \hat{\psi}_+(\delta
\tilde{f}_{+-})\hat{\psi}_-^{-1}
\end{equation}
and linearizing \eqref{eq:7.18}, we have to find a splitting
\begin{equation}\label{M:5.3}
\phi_{+-}\ =\ \phi_+-\phi_-~,
\end{equation}
where the Lie-algebra valued functions $\phi_\pm$ can be extended
to holomorphic functions in $\lambda_\pm$, which yields
\begin{equation}
\delta\hat{\psi}_\pm\ =\ -\phi_\pm\hat{\psi}_\pm~.
\end{equation}
To find these $\phi_\pm$ from $\phi_{+-}$ means to solve the
infinitesimal Riemann-Hilbert problem. Clearly, such solutions are
not unique, as we have the freedom
\begin{equation}\label{M:5.5}
\phi_{+-}\ =\ \phi_+-\phi_-\ =\ (\phi_++\omega)-(\phi_-+\omega)\
=:\
        \tphi_+-\tphi_-~,
\end{equation}
with $\tphi_\pm:=\phi_\pm+\omega$, where the function $\omega$
does not depend on $\lambda_\pm$. This freedom can be used to
preserve the transversal gauge condition \eqref{transversalgauge},
which is discussed in detail in appendix B.

\paragraph{Solutions to the linearized equations.}
Linearizing \eqref{eq:7.15}, we get
\begin{equation}\label{M:5.6}
\delta\hat{\CA}_\ald^+\ =\ \bar{\nabla}_\ald^+\phi_\pm \qquad{\rm
and}\qquad
       \delta\hat{\CA}^i_+\ =\ \bar{\nabla}^i_+\phi_\pm~,
\end{equation}
where we have introduced the operators
$\bar{\nabla}_\ald^+:=\bV^+_\ald+\hat{\CA}^+_\ald$ and
$\bar{\nabla}^i_+:=\bV^i_++\hat{\CA}^i_+$. From \eqref{eq:7.15},
\eqref{eq:7.17} and \eqref{eq:7.18}, it follows that
\begin{equation}
\bar{\nabla}^+_\ald\phi_{+-}\ =\ 0\ =\ \bar{\nabla}^i_+\phi_{+-}~,
\end{equation}
and we eventually arrive at the formul\ae{}
\begin{equation}\label{M:5.8}
\delta\CB_{\ald\bed}\ =\
\oint_{\gamma}\frac{\dd\lambda_+}{2\pi\di\lambda_+\lambda^\bed_+}\,
       \bar{\nabla}^+_\ald\phi_+\qquad{\rm and}\qquad
       \delta\CA_\ald^i\ =\ \oint_{\gamma}\frac{\dd\lambda_+}
       {2\pi\di\lambda_+\lambda^\ald_+}\,
       \bar{\nabla}^i_+\phi_+~,
\end{equation}
where the contour is $\gamma=\{\lambda_+\in
\CPP^1\,|\,|\lambda_+|=1\}$. Thus, the consideration of
infinitesimal perturbations of the transition function of some
$\CT$-flat vector bundle over the CR supertwistor space
$\CF^{5|8}$ obeying certain triviality conditions gives by virtue
of the integral formul\ae{} \eqref{M:5.8} infinitesimal
deformations of the components $\CB_{\ald\bed}$ and $\CA^i_\ald$,
which satisfy -- by construction -- the linearized supersymmetric
Bogomolny equations \eqref{eq:4.24}. Once again, we have a
one-to-one correspondence between equivalence classes of
solutions, with equivalence induced on the gauge theory side by
infinitesimal gauge transformations and on the twistor side by
transformations of the form
$\phi_\pm=\psi_\pm\chi_\pm\psi_\pm^{-1}$, where the $\chi_\pm$ are
functions globally defined on $\tilde{\CV}_\pm\subset\CF^{5|8}$
and annihilated by all vector fields from the distribution $\CT$.

\paragraph{Nilpotent deformation of $\tilde{f}_{+-}$.}
Let us now exemplify our discussion by describing how to construct
explicit solutions to \eqref{eq:7.1}. Consider a $\CT$-flat vector
bundle $\tilde{\CE}\to\CF^{5|8}$ of rank $n$ which is
holomorphically trivial when restricted to any projective line
$\CPP^1_{x,\eta}\hookrightarrow\CF^{5|8}$. Assume further that a
transition function $\tilde{f}_{+-}$ of $\tilde{\CE}$ is chosen
such that all the fields $\ci{\chi}^i_\ald$, $\ci{\phi}\
\!\!^{ij}$, $\ci{\tilde{\chi}}_{i\ald}$ and $\ci{G}_{\ald\bed}$
vanish identically, i.e., we start with the field equation
\begin{equation}
\ci{f}_{\ald\bed}\ =\ -\tfrac{\di}{2}\ci{D}_{\ald\bed}\ci{\Phi}~.
\end{equation}
Without loss of generality, we may assume that the transition
function of $\tilde{\CE}$ can be split as
$\tilde{f}_{+-}=\hat{\psi}_+^{-1}\hat{\psi}_-$, where the
$\hat{\psi}_\pm$ do not depend on the fermionic coordinates
$\eta^\pm_i$.

Consider now the perturbation
\begin{equation}\label{M:5.13}
   \delta \tilde{f}_{+-}\ :=\ -\tfrac{1}{4!}
                  \eps^{j_1\cdots j_4}\eta^+_{j_1}\cdots\eta^+_{j_4}
                      [K,\tilde{f}_{+-}]~,
\end{equation}
where\footnote{Considering vector bundles subject to the reality
conditions induced by \eqref{realitycond}, one restricts the
perturbations to those preserving these conditions. In our
subsequent example, they read explicitly
\begin{equation*}
   \delta \tilde{f}_{+-}\ =\ -\tfrac{1}{4!}
                  \eps^{j_1\cdots j_4}\left(\eta^+_{j_1}\cdots\eta^+_{j_4}
                      +\eta^-_{j_1}\cdots\eta^-_{j_4}
                      \right)[K,\tilde{f}_{+-}]~
\end{equation*}
with $K\in\asu(n)$. } $K\in\agl(n,\FC)$. Then a short calculation
reveals that any splitting \eqref{M:5.3} is of the form
\begin{equation}\label{M:5.14}
\phi_{+-}\ =\ \phi_+-\phi_-\ =\ -\tfrac{1}{4!}
       \eps^{j_1\cdots j_4}\eta^+_{j_1}\cdots\eta^+_{j_4}(\ci{\phi}_+-
       \ci{\phi}_-)~,
\end{equation}
with $\ci{\phi}_\pm:= -[K,\hat{\psi}_\pm]\hat{\psi}_\pm^{-1}$.
Introducing the shorthand notation $\eta^{\gad_1\cdots\gad_4} :=
-\frac{1}{4!}\eps^{j_1\cdots j_4}
        \eta^{\gad_1}_{j_1}\cdots\eta^{\gad_4}_{j_4}$,
we find
\begin{equation}\label{M:5.15}
\phi_{+-}\ =\
    \eta^{\dot{2}\dot{2}\dot{2}\dot{2}}\ci{\phi}\ \!\!^4_{+-}+
    4\eta^{\dot{2}\dot{2}\dot{2}\dot{1}}\ci{\phi}\ \!\!^3_{+-}+
    6\eta^{\dot{2}\dot{2}\dot{1}\dot{1}}\ci{\phi}\ \!\!^2_{+-}+
    4\eta^{\dot{2}\dot{1}\dot{1}\dot{1}}\ci{\phi}\ \!\!^1_{+-}+
    \eta^{\dot{1}\dot{1}\dot{1}\dot{1}}\ci{\phi}\ \!\!^0_{+-}~,
\end{equation}
where we have used the fact that $\eta^{\gad_1\cdots\gad_4}$ is
totally symmetric and defined
\begin{equation}
\ci{\phi}\ \!\!^m_{+-}\ :=\
\lambda_+^m\ci{\phi}_+-\lambda_+^m\ci{\phi}_-\ :=\
       \ci{\phi}\ \!\!^m_+-\ci{\phi}\ \!\!^m_-~.
\end{equation}
The functions $\ci{\phi}\ \!\!^m_\pm$ can be expanded as
($m\geq0$)
\begin{equation}
\ci{\phi}\ \!\!^m_\pm\ =\ \sum_{n=0}^\infty\lambda_+^{\pm
n}\ci{\phi}\
       \!\!^{m(n)}_\pm
\end{equation}
with
\begin{equation}\label{M:5.18}
\ci{\phi}\ \!\!^{m(n)}_+\ =\
        \begin{cases}
         \delta_{m,0}\ci{\phi}\ \!\!^{0(0)}_+ & n=0\\
        \ci{\phi}\ \!\!^{0(n-m)}_+-\ci{\phi}\ \!\!^{0(m-n)}_- & n>0
        \end{cases}\qquad{\rm and}\qquad
       \ci{\phi}\ \!\!^{m(n)}_-\ =\ \ci{\phi}\ \!\!^{0(m+n)}_-~.
\end{equation}
Combining the expansion
\begin{equation}
\phi_\pm\ =\ \sum_{n=0}^\infty\lambda_+^{\pm n}\phi_\pm^{(n)}
\end{equation}
with \eqref{M:5.14}--\eqref{M:5.18}, we therefore find
\begin{equation}\label{M:5.19}
\phi_-^{(n)}\ =\
    \eta^{\dot{2}\dot{2}\dot{2}\dot{2}}\ci{\phi}\ \!\!^{0(4+n)}_-+
    4\eta^{\dot{2}\dot{2}\dot{2}\dot{1}}\ci{\phi}\ \!\!^{0(3+n)}_-+
    6\eta^{\dot{2}\dot{2}\dot{1}\dot{1}}\ci{\phi}\ \!\!^{0(2+n)}_-+
    4\eta^{\dot{2}\dot{1}\dot{1}\dot{1}}\ci{\phi}\ \!\!^{0(1+n)}_-+
    \eta^{\dot{1}\dot{1}\dot{1}\dot{1}}\ci{\phi}\
    \!\!^{0(n)}_-~,
\end{equation}
and a similar expression for $\phi_+^{(n)}$. At this point, we
have to choose an $\omega$ which guarantees that the transversal
gauge condition is satisfied. Explicitly, a possible $\omega$ is
given by
\begin{equation}
\omega\ =\
        -\eta^{\dot{2}\dot{2}\dot{2}\dot{1}}\ci{\phi}\ \!\!^{0(3)}_--
        3\eta^{\dot{2}\dot{2}\dot{1}\dot{1}}\ci{\phi}\ \!\!^{0(2)}_--
        3\eta^{\dot{2}\dot{1}\dot{1}\dot{1}}\ci{\phi}\ \!\!^{0(1)}_--
         \eta^{\dot{1}\dot{1}\dot{1}\dot{1}}\ci{\phi}\ \!\!^{0(0)}_-~,
\end{equation}
which is derived in appendix B, where also a detailed discussion
of this point is found.

\paragraph{Towards explicit solutions.}
The infinitesimal perturbations $\delta\CB_{\ald\bed}$ and
$\delta\CA^i_\ald$ are obtained upon integration of the
formul\ae{} \eqref{M:5.8}:
\begin{subequations}\label{M:5.22}
\begin{equation}
\begin{aligned}
\delta\CB_{\ald\dot{1}}\ &=\ \ci{\nabla}_{\ald\dot{1}}
        (\phi^{(0)}_-+\omega)\ =\ \ci{\nabla}_{\ald\dot{1}}
        (\phi^{(0)}_++\omega)-\ci{\nabla}_{\ald\dot{2}}\phi^{(1)}_+~,\\
      \delta\CB_{\ald\dot{2}}\ &=\ -\ci{\nabla}_{\ald\dot{1}}
       \phi_-^{(1)}+\ci{\nabla}_{\ald\dot{2}}(\phi^{(0)}_-+\omega)
          \ =\ \ci{\nabla}_{\ald\dot{2}}(\phi^{(0)}_++\omega)
\end{aligned}
\end{equation}
and
\begin{equation}
\begin{aligned}
\delta\CA^i_{\dot{1}}\ &=\ \dpar_{\dot{1}}^i
    (\phi_-^{(0)}+\omega)\ =\ \dpar_{\dot{1}}^i(\phi_+^{(0)}+\omega)-
    \dpar_{\dot{2}}^i\phi_+^{(1)}~,\\
    \delta\CA^i_{\dot{2}}\ &=\
    -\partial_{\dot{1}}^i\phi_-^{(1)}+
    \partial_{\dot{2}}^i(\phi_-^{(0)}+\omega)
    \ =\ \partial_{\dot{2}}^i(\phi_+^{(0)}+\omega)~.
\end{aligned}
\end{equation}
\end{subequations}
Consider now the expansions
\begin{equation}\label{M:5.23}
\begin{aligned}
\delta\CB_{\ald\bed}\ &=\ \delta\ci{\CB}_{\ald\bed}+
    \sum_{k\geq1}\frac{1}{k!}\,
    \eta^{\gad_1}_{j_1}\cdots\eta^{\gad_k}_{j_k}\,
    \delta\ci{[\ald\bed]}\ \!\!_{\gad_1\cdots\gad_k}^{j_1\cdots j_k}~,\\
    \delta\CA^i_\ald\ &=\
    \sum_{k\geq1}\frac{k}{(k+1)!}\,
    \eta^{\gad_1}_{j_1}\cdots\eta^{\gad_k}_{j_k}\,
    \delta\ci{ [\,{}^{\,i}_\ald]}\
    \!\!_{\gad_1\cdots\gad_k}^{j_1\cdots j_k}~,
\end{aligned}
\end{equation}
where the brackets $[\ \ ]_{\gad_1\cdots\gad_k}^{j_1\cdots j_k}$
are composite expressions of some superfields, cf.\ also appendix
A. Since our particular deformation of the transition function
implies that $\phi_\pm+\omega=\CO(\eta^4)$, the resulting
deformations of $\CB_{\ald\bed}$ and $\CA^i_\ald$ are of the form
$\CB_{\ald\bed}=\CO(\eta^4)$ and $\CA^i_\ald=\CO(\eta^3)$,
respectively. In transversal gauge, the explicit superfield
expansions \eqref{superexp4} show that
$\delta\ci{\CB}_{\ald\bed}=\delta\ci{\chi}\ \!\!^i_\ald=
 \delta\ci{\phi}\ \!\!^{ij}=\delta\ci{\tilde{\chi}}_{i\ald}=0$.
Together with the recursion relations \eqref{recursions}, they
moreover imply that the variation of all higher order terms (than
those given in \eqref{superexp4}) in the $\eta$-expansions vanish.
Hence, from \eqref{M:5.23} we find
\begin{equation}
\begin{aligned}
\delta\CB_{\ald\bed}\ &=\ \tfrac{1}{2\cdot 4!}\eps^{j_1j_2j_3j_4}
\eta^{\gad_1}_{j_1}\eta^{\gad_2}_{j_2}\eta^{\gad_3}_{j_3}
\eta^{\gad_4}_{j_4}
\eps_{\bed\gad_1}\ci{\nabla}_{\ald\gad_2}
\delta\ci{G}_{\gad_3\gad_4}~,\\
\delta\CA^i_\ald\ &=\
\tfrac{3}{4!}\eps^{ij_1j_2j_3}\eta^{\gad_1}_{j_1}
\eta^{\gad_2}_{j_2}\eta^{\gad_3}_{j_3}\eps_{\ald\gad_1}
\delta\ci{G}_{\gad_2\gad_3}~.
\end{aligned}
\end{equation}
 Comparing these equations with \eqref{M:5.22}
and the $\eta$-expansions of $\phi_\pm^{(0)}$, $\phi_\pm^{(1)}$
and $\omega$ given earlier, we arrive at
\begin{equation}
\delta\ci{G}_{\dot{1}\dot{1}}\ =\ 2\ci{\phi}\
\!\!^{0(1)}_-~,\qquad
       \delta\ci{G}_{\dot{1}\dot{2}}\ =\ 2\ci{\phi}\
       \!\!^{0(2)}_-\eand
       \delta\ci{G}_{\dot{2}\dot{2}}\ =\ 2\ci{\phi}\
       \!\!^{0(3)}_-
\end{equation}
together with the field equations
\begin{equation}
\ci{f}_{\ald\bed}\ =\
-\tfrac{\di}{2}\ci{D}_{\ald\bed}\ci{\Phi}\qquad{\rm and}\qquad
       \eps^{\bed\gad}\ci{D}_{\ald\bed}\delta\ci{G}_{\gad\ded}\
       =\ -\tfrac{\di}{2}
       [\delta\ci{G}_{\ald\ded},\ci{\Phi}]~.
\end{equation}
Since the equations \eqref{eq:7.1} are linear in $G_{\ald\bed}$,
we hence have generated a solution
\begin{equation}
(A_{\ald\bed}:=\ci{A}_{\ald\bed},\Phi:=\ci{\Phi},
G_{\ald\bed}:=\delta\ci{G}_{\ald\bed})
\end{equation}
to \eqref{eq:7.1} starting from a solution to \eqref{eq:7.1a}.
Thus, knowing the explicit splitting
$\tilde{f}_{+-}=\hat{\psi}_+^{-1}\hat{\psi}_-$, we can define
functions $\ci{\phi}_\pm=-[K,\hat{\psi}_\pm]\hat{\psi}_\pm^{-1}$
which then in turn yield $G_{\ald\bed}$.

\section{Comments on $\CN=8$ SYM theory in three
dimensions}

The full $\CN=8$ SYM theory in three dimensions is slightly out of
the scope of this paper, but as an outlook, we would like to
sketch the construction of a supertwistor correspondence for this
theory and leave the details to future work.

\paragraph{Twistor description of $\CN=3$ SYM theory.}
Recall that there is a one-to-one correspondence between gauge
equivalence classes of solutions to the $\CN=3$ SYM equations in
four complex dimensions and equivalence classes of holomorphic
vector bundles $\CE$ over a quadric $\CL^{5|6}$ in (an open subset
of) the product space $\CPP^{3|3}\times \CPP^{3|3}_*$
\cite{Witten:1978xx} such that the bundles $\CE$ are
holomorphically trivial on each submanifold
$\CL^{2|0}_{x,\theta,\eta}\cong\CPP^1\times\CPP^1_*\embd
\CL^{5|6}$ with $(x,\eta,\theta)\in\FC^{4|12}$. The space
$\CL^{5|6}$ is also known as the super-ambitwistor
space\footnote{See also \cite{Isenberg:kk} for the bosonic case.
For a recent review and further references, see
\cite{Popov:2004rb}.} and enters into the double fibration
\begin{equation}\label{eq:8.1}
\begin{aligned}
\begin{picture}(80,40)
\put(0.0,0.0){\makebox(0,0)[c]{$\CL^{5|6}$}}
\put(64.0,0.0){\makebox(0,0)[c]{$\FC^{4|12}$}}
\put(32.0,33.0){\makebox(0,0)[c]{$\CF^{6|12}$}}
\put(25.0,25.0){\vector(-1,-1){18}}
\put(37.0,25.0){\vector(1,-1){18}}
\end{picture}
\end{aligned}
\end{equation}
where $\CF^{6|12}\cong \FC^{4|12}\times \CPP^1\times \CPP^1_*$ is
the correspondence space with coordinates
$(x^{\alpha\ald},\theta^{\alpha i},\eta_i^\ald)$ on $\FC^{4|12}$
and homogeneous coordinates $\lambda_\ald$ on $\CPP^1$ and
$\mu_\alpha$ on $\CPP^1_*$, respectively. The tangent spaces to
the $(1|6)$-dimensional leaves of the fibration
$\CF^{6|12}\rightarrow \CL^{5|6}$ are spanned by the holomorphic
vector fields
\begin{equation}\label{eq:8.2}
\mu^\alpha\lambda^\ald\der{x^{\alpha\ald}}~,~~~
\lambda^\ald\left(\der{\eta_i^\ald}+\theta^{\alpha
i}\der{x^{\alpha\ald}}\right)\eand\mu^\alpha\left(\der{\theta^{\alpha
i}}+\eta_i^\ald\der{x^{\alpha\ald}}\right)
\end{equation}
which enter into the linear system
\begin{equation}\label{eq:8.3}
\begin{aligned}
\mu^\alpha\lambda^\ald(\dpar_{\alpha\ald}+\CA_{\alpha\ald})\psi
\ =\ 0~,\hspace{4cm}\\
\lambda^\ald\left(\der{\eta_i^\ald}+\theta^{\alpha
i}\dpar_{\alpha\ald}+ \CA_\ald^i\right)\psi\ =\ 0~,~~~
\mu^\alpha\left(\der{\theta^{\alpha
i}}+\eta_i^\ald\dpar_{\alpha\ald}+\CA_{\alpha i}\right)\psi\ =\
0~,
\end{aligned}
\end{equation}
where $\psi$ is a locally defined $\sGL(n,\FC)$-valued function on
$\CF^{6|12}$. The compatibility conditions of \eqref{eq:8.3} are
equivalent to the $\CN=3$ SYM field equations. On the other hand,
these equations are also equivalent to the hCS field equations on
$\CL^{5|6}$ whose solutions describe holomorphic structures on the
complex vector bundle $\CE\rightarrow \CL^{5|6}$
\cite{Witten:2003nn,Popov:2004rb}.

\paragraph{Reduced twistor correspondence.}
Similarly to the supersymmetric Bogomolny model which was obtained
by a dimensional reduction of $\CN=4$ SDYM theory, one can
establish a twistor correspondence for the full $\CN=8$ SYM theory
in three (complex) dimensions by using a dimensional reduction of
$\CL^{5|6}$. Taking the quotient of the spaces in the diagram
\eqref{eq:8.1} with respect to the actions of the Abelian groups
generated by the vector field $\CCT_4$, we arrive at the diagram
\begin{equation}\label{ambdouble}
\begin{aligned}
\begin{picture}(100,95)(0,-5)
\put(0.0,0.0){\makebox(0,0)[c]{$\CL^{4|6}$}}
\put(0.0,52.0){\makebox(0,0)[c]{$\CL^{5|6}$}}
\put(96.0,0.0){\makebox(0,0)[c]{$\FC^{3|12}$}}
\put(96.0,52.0){\makebox(0,0)[c]{$\FC^{4|12}$}}
\put(51.0,33.0){\makebox(0,0)[c]{$\CF^{5|12}$}}
\put(51.0,85.0){\makebox(0,0)[c]{$\CF^{6|12}$}}
\put(37.5,25.0){\vector(-3,-2){25}}
\put(55.5,25.0){\vector(3,-2){25}}
\put(37.5,77.0){\vector(-3,-2){25}}
\put(55.5,77.0){\vector(3,-2){25}}
\put(0.0,45.0){\vector(0,-1){37}}
\put(90.0,45.0){\vector(0,-1){37}}
\put(45.0,78.0){\vector(0,-1){37}}
\end{picture}
\end{aligned}
\end{equation}
describing the reduction of the supertwistor correspondence
\eqref{eq:8.1}. Recall that in three dimensions, we have the
vector fields \eqref{eq:3.4} and the decomposition \eqref{eq:3.5}.
Substituting the latter into \eqref{eq:8.2} and assuming
independence of all functions of $x^4$, we obtain the vector
fields
\begin{equation}\label{eq:8.5}
\mu^\ald\lambda^\bed\dpar_{(\ald\bed)}~,~~~
\lambda^\bed\left(\der{\eta_i^\bed}+\theta^{\ald
i}\dpar_{(\ald\bed)}\right)\eand\mu^\ald\left(\der{\theta^{\ald
i}}+\eta_i^\bed\dpar_{(\ald\bed)}\right)~,
\end{equation}
which are tangent to the $(1|6)$-dimensional leaves of the
fibration $\CF^{5|12}\rightarrow \CL^{4|6}$, where
$\CF^{5|12}\cong \FC^{3|12}\times \CPP^1\times \CPP^1_*$.

\paragraph{Constraint equations.}
Similarly, the linear system \eqref{eq:8.3} is reduced to
\begin{equation}\label{eq:8.6}
\begin{aligned}
\mu^\ald\lambda^\bed(\dpar_{(\ald\bed)}+
\CB_{\ald\bed})\psi\ =\ 0~,\hspace{4.2cm}\\
\lambda^\bed\left(\der{\eta_i^\bed}+\theta^{\ald
i}\dpar_{(\ald\bed)}+ \CA_\bed^i\right)\psi\ =\ 0~,~~~
\mu^\ald\left(\der{\theta^{\ald
i}}+\eta_i^\bed\dpar_{(\ald\bed)}+\CA_{\ald i}\right)\psi\ =\ 0~.
\end{aligned}
\end{equation}
The corresponding compatibility conditions read as
\begin{equation}\label{eq:8.7}
\begin{aligned}
\{\nabla_{\ald
i},\nabla^j_{\bed}\}-2\delta_i^j\nabla_{\ald\bed}\ =\ 0~,\hspace{3.5cm}\\
\{\nabla_\ald^i,\nabla_\bed^j\}+ \{\nabla_\bed^i,\nabla_\ald^j\}\
=\ 0~,~~~ \{\nabla_{\ald i},\nabla_{\bed j}\}+\{\nabla_{\bed i},
\nabla_{\ald j}\}\ =\ 0~,\\
\end{aligned}
\end{equation}
where we introduced the differential operators
\begin{equation}
\begin{aligned}
\nabla_{\ald\bed}\ :=\
\dpar_{(\ald\bed)}+\CB_{\ald\bed}~,\hspace{3.8cm}\\
\nabla^i_\bed\ :=\ \der{\eta_i^\bed}+\theta^{\ald
i}\dpar_{(\ald\bed)}+ \CA_\bed^i~,~~~\nabla_{\ald i}\ :=\
\der{\theta^{\ald i}}+\eta_i^\bed\dpar_{(\ald\bed)}+\CA_{\ald i}~.
\end{aligned}
\end{equation}
The equations \eqref{eq:8.7} are the constraint equations of
$\CN=3$ SYM theory in four dimensions reduced to three dimensions.

\paragraph{Outlook.}
It remains to clarify the geometry of the mini-superambitwistor
space $\CL^{4|6}$ together with its real structure $\tau$. In case
that  $\CL^{4|6}$ is a Calabi-Yau supermanifold (for which there
are certain hints), one can define an open topological B-model on
this space, which is supposed to describe holomorphic structures
on complex vector bundles over $\CL^{4|6}$. These can then be
linked to solutions of $\CN=3$ SYM theory reduced to three
dimensions. The latter theory should (under an additional
assumption) be equivalent to $\CN=8$ SYM theory similarly to the
equivalence of $\CN=3$ and $\CN=4$ SYM theories in four
dimensions. Note that on $\CL^{4|6}$, one cannot impose the
Euclidean reality condition \eqref{eq:2.31} on the fermionic
coordinates. Only two other conditions leading to Kleinian and
Minkowski signature, respectively, are consistent
\cite{Popov:2004rb}.

\section*{Acknowledgements} We would like to thank O. Lechtenfeld, S.
Uhlmann and R. Wimmer for useful discussions. This work was
partially supported by the Deutsche Forschungsgemeinschaft (DFG).

\newpage

 \addtocontents{toc}{\hspace{-0.61cm}Appendices} \vspace{1.0cm}
\appendix
\noindent {\bf \Large Appendices}

\renewcommand{\thesection}{\Alph{section}.}
\setcounter{subsection}{0} \setcounter{equation}{0}
\renewcommand{\thesubsection}{\Alph{subsection}}
\renewcommand{\theequation}{\thesubsection.\arabic{equation}}

\subsection{Superfield expansions}

\paragraph{Field equations.}
In section 4.4, we obtained the constraint equations
\eqref{constraint} for the differential operators
$\nabla_{\ald\bed}:=\dpar_{(\ald\bed)}+\CB_{\ald\bed}$ and
$D_\ald^i=\der{\eta_i^\ald}+\CA_\ald^i$ which are equivalent to
the supersymmetric Bogomolny equations
\begin{equation}\label{A:1}
\begin{aligned}
  f_{\ald\bed}\ &=\ -\tfrac{\di}{2}D_{\ald\bed}\Phi~,\\
    \eps^{\bed\gad}D_{\ald\bed}\chi^i_\gad
              \ &=\ -\tfrac{\di}{2}[\Phi,\chi^i_\ald]~,\\
\triangle \phi^{ij}\ &=\ -\tfrac{1}{4}[\Phi,[\phi^{ij},\Phi]]
  +\eps^{\ald\bed}\{\chi^i_\ald,\chi^j_\bed\}~,\\
\eps^{\bed\gad}D_{\ald\bed}\tilde{\chi}_{i\gad}
   \ &=\ -\tfrac{\di}{2}[\tilde{\chi}_{i\ald},\Phi]+
   2\di[\phi_{ij},\chi^j_\ald]~,\\
  \eps^{\bed\gad}D_{\ald\bed}G_{\gad\ded}\ &=\
  -\tfrac{\di}{2}[G_{\ald\ded},\Phi]+
  \di\{\chi^i_\ald,\tilde{\chi}_{i\ded}\}
  -\tfrac{1}{2}[\phi_{ij},D_{\ald\ded}\phi^{ij}]
 +\tfrac{\di}{4}\eps_{\ald\ded}[\phi_{ij},[\Phi,\phi^{ij}]]~
\end{aligned}
\end{equation}
with all the fields being superfields and defined by
\begin{equation}
\begin{aligned}\label{A:2}
\CA_{\ald\bed}-\tfrac{\di}{2}\eps_{\ald\bed}\Phi&\ :=\
\CB_{\ald\bed}~,\\
\chi^i_\ald&\ :=\ \Sigma^i_\ald~,\\
\phi^{ij}&\ :=\ \Sigma^{ij}~,\\
\tilde{\chi}_{i\ald}&\ :=\
\tfrac{1}{3}\eps_{ijkl}D^j_\ald\phi^{kl}~,\\
G_{\ald\bed}&\ :=\ -\tfrac{1}{4}D^i_{(\ald}\tilde{\chi}_{i\bed)}~.
\end{aligned}
\end{equation}

\paragraph{Superfield expansions.}
To prove that the equations \eqref{A:1} are equivalent to the
supersymmetric Bogomolny equations \eqref{eq:4.24}, we need the
explicit superfield expansions of $\CB_{\ald\bed}$ and
$\CA^i_\ald$ from which all remaining expansions can be derived by
using the constraint equations \eqref{constraint}. Since the
proofs of the subsequent general assertions are rather
straightforward, we leave them to the interested reader.

Consider some generic superfield $\varphi$. Its explicit
$\eta$-expansion looks as
\begin{equation}
\varphi\ =\ \ci{ \varphi}+\sum_{k\geq1}\eta^{\gad_1}_{j_1}\cdots
\eta^{\gad_k}_{j_k}\,  \varphi_{\gad_1\cdots\gad_k}^{j_1\cdots
j_k}~.
\end{equation}
Furthermore, we have $\CD \varphi=\eta^{\gad_1}_{j_1}[\ \
]_{\gad_1}^{j_1}$, where the bracket $[\ \ ]_{\gad_1}^{j_1}$ is a
composite expression of some superfields.\footnote{For example,
$\CD\CB_{\ald\bed}=\eta^{\gad_1}_{j_1} [\ald\bed]_{\gad_1}^{j_1}$,
with $[\ald\bed]_{\gad_1}^{j_1}=-
\eps_{\bed\gad_1}\chi^{j_1}_\ald$.} Let
\begin{equation}
\CD[\ \ ]_{\gad_1\cdots\gad_k}^{j_1\cdots j_k}\ =\
\eta^{\gad_{k+1}}_{j_{k+1}}[\ \
]_{\gad_1\cdots\gad_{k+1}}^{j_1\cdots
       j_{k+1}}~.
\end{equation}
Then we have
\begin{equation}
 \varphi\ =\ \ci{ \varphi}+\sum_{k\geq1}\frac{1}{k!}\,
\eta^{\gad_1}_{j_1}\cdots\eta^{\gad_k}_{j_k}\, \ci{[\ \ ]}\
\!\!_{\gad_1\cdots\gad_k}^{j_1\cdots j_k}~,
\end{equation}
which follows by induction. In case the recursion relation of
$\varphi$ was given by $(1+\CD)\varphi=\eta^{\gad_1}_{j_1}[\ \
]_{\gad_1}^{j_1}$, as it happens to be for $\CA^i_\ald$, then
$\ci{\varphi}=0$ and the superfield expansion is of the form
\begin{equation}
 \varphi\ =\ \sum_{k\geq1}\frac{k}{(k+1)!}\,
\eta^{\gad_1}_{j_1}\cdots\eta^{\gad_k}_{j_k}\, \ci{[\ \ ]}\
\!\!_{\gad_1\cdots\gad_k}^{j_1\cdots j_k}~.
\end{equation}
From this, it is a bit tedious but straightforward to determine
the superfield expansions of $\CB_{\ald\bed}$ and $\CA^i_\ald$.
For
\begin{equation}\label{expAI}
\CB_{\ald\bed}\ =\ \ci{\CB}_{\ald\bed}+
\sum_{k\geq1}\frac{1}{k!}\,
\eta^{\gad_1}_{j_1}\cdots\eta^{\gad_k}_{j_k}\, \ci{[\ald\bed]}\
\!\!_{\gad_1\cdots\gad_k}^{j_1\cdots j_k}~,
\end{equation}
we obtain as the first four coefficients
\begin{equation}\label{expAIa}
\begin{aligned}
   {[\ald\bed]_{\gad_1}^{j_1}}\ &=\
   -\di\eps_{\bed\gad_1}\chi^{j_1}_\ald~,\\
   [\ald\bed]_{\gad_1\gad_2}^{j_1j_2}\ &=\
   \eps_{\bed\gad_1}\nabla_{\ald\gad_2}
     \phi^{j_1j_2}~,\\
   [\ald\bed]_{\gad_1\gad_2\gad_3}^{j_1j_2j_3}
   \ &=\ -\tfrac{1}{2}\eps_{\bed\gad_1}
     \eps^{j_1j_2j_3k}\nabla_{\ald\gad_2}
     \tilde{\chi}_{k\gad_3}-\di
     \eps_{\bed\gad_1}\eps_{\gad_2\gad_3}
     [\chi^{j_3}_\ald,\phi^{j_1j_2}]~,\\
   [\ald\bed]_{\gad_1\gad_2\gad_3\gad_4}^{j_1j_2j_3j_4}
   \ &=\ -\tfrac{1}{2}
     \eps_{\bed\gad_1}\eps^{j_1j_2j_3j_4}
     \nabla_{\ald\gad_2}G_{\gad_3\gad_4}+
    \tfrac{1}{2}\eps_{\bed\gad_1}\eps_{\gad_3\gad_4}
    \eps^{j_1j_2j_3k}
    \nabla_{\ald\gad_2}[\phi^{j_4l},\phi_{kl}]\ +\\
    &\kern1cm+\tfrac{\di}{2}\eps_{\bed\gad_1}
    \eps_{\gad_2\gad_4}
      \eps^{j_1j_2j_3k}\{\chi^{j_4}_\ald,\tilde{\chi}_{k\gad_3}\}+
      \eps_{\bed\gad_1}\eps_{\gad_2\gad_3}
      [\nabla_{\ald\gad_4}\phi^{j_3j_4},
      \phi^{j_1j_2}]\ -\\
    &\kern1.5cm-\ \tfrac{\di}{2}\eps_{\bed\gad_1}\eps_{\gad_2\gad_3}
     \eps^{j_1j_2j_3k}\{\chi^{j_3}_\ald,\tilde{\chi}_{k\gad_4}\}~,
\end{aligned}
\end{equation}
where we have heavily used the recursion relations
\eqref{recursions}. Quite similarly, we find for
\begin{equation}\label{expAII}
\begin{aligned}
\CA^i_\ald\ =\
             \sum_{k\geq1}\frac{k}{(k+1)!}\,
             \eta^{\gad_1}_{j_1}\cdots\eta^{\gad_k}_{j_k}\,
             \ci{[\,{}^{\,i}_\ald]}\
             \!\!_{\gad_1\cdots\gad_k}^{j_1\cdots j_k}
\end{aligned}
\end{equation}
the following coefficients:
\begin{equation}\label{expAIIa}
\begin{aligned}
{[\,{}^{\,i}_\ald]^{j_1}_{\gad_1}}\ &=\
\eps_{\ald\gad_1}\phi^{ij_1}~,\\
{[\,{}^{\,i}_\ald]^{j_1j_2}_{\gad_1\gad_2}}\ &=\
-\tfrac{1}{2}\eps_{\ald\gad_1}\eps^{ij_1j_2k}\tilde{\chi}_{k \gad_2}~,\\
{[\,{}^{\,i}_\ald]^{j_1j_2j_3}_{\gad_1\gad_2\gad_3}}
    \ &=\ \tfrac{1}{2}\eps_{\ald\gad_1}\eps^{ij_1j_2j_3}G_{\gad_2\gad_3}
      +\tfrac{1}{2}\eps_{\ald\gad_1}\eps_{\gad_2\gad_3}\eps^{ij_1j_2k}
       [\phi^{j_3l},\phi_{kl}]~,\\
  {[\,{}^{\,i}_\ald]^{j_1j_2j_3j_4}_{\gad_1\gad_2
   \gad_3\gad_4}}\ &=\ \tfrac{1}{2}\eps_{\ald\gad_1}\eps_{\gad_4(\gad_2}
   \eps^{ij_1j_2j_3}[\tilde{\chi}_{k\gad_3)},\phi^{j_4k}]-\tfrac{1}{4}
   \eps_{\ald\gad_1}\eps_{\gad_2\gad_3}\eps^{ij_1j_2k}
   \eps^{j_3lj_4m}[\tilde{\chi}_{m\,\gad_4},\phi_{kl}]\ +\\
   &\kern1cm+\ \tfrac{1}{2}\eps_{\ald\gad_1}\eps_{\gad_2\gad_3}
   \eps^{ij_1j_2k}\delta^{j_4}_{[k}[\tilde{\chi}_{l]\gad_4},\phi^{j_3l}]~.
\end{aligned}
\end{equation}

Putting everything together, the above expansions yield the
remaining expansions for the other superfields and moreover show
that the superfield equations \eqref{A:1} are true order by order
in the $\eta$-expansion. Thus, the supersymmetric Bogomolny
equations \eqref{eq:4.24} on $\FR^3$ are indeed equivalent to the
compatibility conditions \eqref{eq:4.42} of the linear system
\eqref{eq:4.40}.

\subsection{Transversal gauge}
\setcounter{equation}{0}

\paragraph{The general case.} In \eqref{M:5.5}, we
noticed a freedom in splitting the Lie-algebra valued function
$\phi_{+-}$ defined in \eqref{M:5.2} according to
\begin{equation}\label{B:1}
\phi_{+-}\ =\ \phi_+-\phi_-\ =\ (\phi_++\omega)-(\phi_-+\omega)\
=:\ \tphi_+-\tphi_-~.
\end{equation}
This freedom can be used to guarantee that the infinitesimal
deformations of the gauge potential obtained by
\begin{equation}\label{B:2}
\delta\hat{\CA}_\ald^+\ =\ \bar{\nabla}_\ald^+\phi_\pm \qquad{\rm
and}\qquad
       \delta\hat{\CA}^i_+\ =\ \bar{\nabla}^i_+\phi_\pm~,
\end{equation}
are in transversal gauge, i.e., the condition $\eta^\ald_i\delta
\CA^i_\ald=0$ is satisfied, cf.\ \eqref{transversalgauge}. Upon
expanding the functions $\phi_\pm$ in powers of $\lambda_\pm$ in
their respective domains, i.e.\
\begin{equation}\label{B:3}
\phi_\pm\ =\ \sum_{n=0}^\infty\lambda_+^{\pm n}\phi_\pm^{(n)}~,
\end{equation}
we deduce from \eqref{B:2}
\begin{equation}
\begin{aligned}\label{B:4}
\delta\CA^i_{\dot{1}}\ &=\ D_{\dot{1}}^i\phi_+^{(0)}-
          D_{\dot{2}}^i\phi_+^{(1)}\ =\ D_{\dot{1}}^i\phi_-^{(0)}~,\\
        \delta\CA^i_{\dot{2}}\ &=\ D_{\dot{2}}^i\phi_+^{(0)}\ =\
        -D_{\dot{1}}^i\phi_-^{(1)}+D_{\dot{2}}^i\phi_-^{(0)}~.
\end{aligned}
\end{equation}
The contraction of these equations with $\eta^\ald_i$ yields the
constraints
\begin{equation}\label{B:5}
    \CD\phi^{(0)}_-+\CD\omega\ =\ \eta^{\dot{2}}_i
    D^i_{\dot{1}}\phi_-^{(1)}
    \qquad{\rm and}\qquad
    \CD\phi^{(0)}_++\CD\omega\ =\ \eta^{\dot{1}}_i
    D^i_{\dot{2}}\phi_+^{(1)}~,
\end{equation}
where we have used the fact that
$\tphi^{(0)}_\pm=\phi_\pm^{(0)}+\omega$ and
$\tphi^{(1)}_\pm=\phi_\pm^{(1)}$, respectively. Thus, a splitting
\eqref{B:1} with an $\omega$ satisfying \eqref{B:5} yields a
deformation of the super gauge potential which respects the
transversal gauge condition.

\paragraph{The example.}
Let us briefly comment on the transversal gauge condition
discussed in the previous paragraph in the case of the example
presented in section 7.2. Equation \eqref{B:5} simplifies in our
case \eqref{M:5.14} to
\begin{equation}\label{M:5.20}
   \CD\phi^{(0)}_-+\CD\omega\ =\
   \eta^{\dot{2}}_i\partial^i_{\dot{1}}\phi_-^{(1)}
   \qquad{\rm and}\qquad
   \CD\phi^{(0)}_++\CD\omega\ =\
   \eta^{\dot{1}}_i\partial^i_{\dot{2}}\phi_+^{(1)}~.
\end{equation}
Since our particular deformation \eqref{M:5.13} is of fourth order
in the fermionic coordinates, we may assume that
$\omega=\eta^{\gad_1\cdots\gad_4}\omega_{\gad_1\cdots\gad_4}$.
Then, after some algebraic manipulations, the expansions of
$\phi_\pm^{(n)}$ (see e.g.\ \eqref{M:5.19}) together with
\eqref{M:5.20} and the ansatz for $\omega$ lead to
\begin{equation}\label{B.7}
\omega\ =\ -\eta^{\dot{2}\dot{2}\dot{2}\dot{1}}\ci{\phi}\
\!\!^{0(3)}_-- 3\eta^{\dot{2}\dot{2}\dot{1}\dot{1}}\ci{\phi}\
\!\!^{0(2)}_-- 3\eta^{\dot{2}\dot{1}\dot{1}\dot{1}}\ci{\phi}\
\!\!^{0(1)}_-- \eta^{\dot{1}\dot{1}\dot{1}\dot{1}}\ci{\phi}\
\!\!^{0(0)}_-~,
\end{equation}
i.e., this particular choice of $\omega$ ensures the preservation
of transversal gauge for the perturbation \eqref{M:5.13}.

\subsection{Signature $($$+$$+$$-$$)$}
\setcounter{equation}{0}

In section 2.1, we defined the twistor space of a real
four-dimensional manifold $X$ as the bundle \eqref{eq:2.1} of
almost complex structures compatible with a metric $g$, which
yielded \eqref{eq:2.10} as the twistor space of the Euclidean
space $(\FR^4,\delta_{\mu\nu})$. In fact, in Euclidean signature,
one can define the twistor space of $X$ in three equivalent ways,
the first being the above one.

\paragraph{Two further definitions of the twistor space.}
The second definition assumes that $X$ admits a spin structure.
Then one can introduce the vector bundle
\begin{equation}
\CS\ :=\ P(X,\sSpin(4))\times_{\sSpin(4)}\FC^4
\end{equation}
of Dirac spinors on $X$. Since $\sSpin(4)\cong
\sSU(2)_L\times\sSU(2)_R$, this bundle decomposes into a direct
sum $\CS=\CS_L\oplus\CS_R$ of bundles of left- and right-handed
Weyl spinors. Using the latter bundle, one can define the twistor
space of $X$ as the projectivization \cite{Atiyah:wi}
\begin{equation}\label{eq:B.1}
\CZ\ :=\ P(\CS_R)
\end{equation}
of the bundle $\CS_R\rightarrow X$ and $\CZ$ has again projective
lines $\CPP^1\cong S^2$ as fibres. Note that in this definition,
the spin structure on $X$ was only needed for introducing the
bundle $\CS_R\rightarrow X$ but not for its projectivization
$P(\CS_R)$.

The third definition is obtained by considering the vector bundle
$\Lambda^2 T^*X$ of two-forms on $X$. Using the Hodge star
operator, one can split $\Lambda^2 T^*X$ into the direct sum
$\Lambda^2 T^*X=\Lambda^2_+ T^*X\oplus\Lambda^2_- T^*X$ of the
subbundles of self-dual and anti-self-dual two-forms on $X$. Then
the twistor space $\CZ$ of $X$ can be defined as the unit sphere
bundle
\begin{equation}\label{eq:B.2}
\CZ\ :=\ S(\Lambda_-^2 T^*X)
\end{equation}
in the vector bundle $\Lambda_-^2 T^*X$.

\paragraph{Kleinian case.}
Although these definitions are all equivalent in the Euclidean
case, only the latter two are equivalent in the Kleinian case of
signature $($$+$$+$$-$$-$$)$ and they differ from definition
\eqref{eq:2.1}. In particular, for the space
$\FR^{2,2}:=(\FR^4,g_{\mu\nu})$ with
$(g_{\mu\nu})=\diag(+1,+1,-1,-1)$ we obtain both from
\eqref{eq:B.1} and \eqref{eq:B.2} the space
\begin{equation}\label{eq:B.3}
\CZ \cong\ \FR^{2,2}\times\CPP^1~.
\end{equation}
On the other hand, the definition \eqref{eq:2.1} from section 2
yields the space
\begin{equation}\label{eq:B.4}
\tilde{\CZ} \cong\ \FR^{2,2}\times H^2\ =\ (\FR^{2,2}\times
H^2_+)\cup(\FR^{2,2}\times H_-^2)\ =:\ \CZ_+\cup\CZ_-~,
\end{equation}
which is an open subset of $\CZ$. Here, $H^2=H_+^2\cup H_-^2$ is
the two-sheeted hyperboloid and $H^2_\pm=\{\lambda_\pm\in
U_\pm\,|\,|\lambda_\pm|<1\}\cong \sSU(1,1)/\sU(1)$ are open discs.
In fact,
\begin{equation}
\CZ\ =\ \CZ_+\cup \CZ_0\cup \CZ_-~,
\end{equation}
where $\CZ_0\cong \FR^{2,2}\times S^1$ is a boundary for both
$\CZ_+$ and $\CZ_-$.

The twistor space of $\FR^{2,2}$ is again the space
\eqref{eq:2.10}, which can be written as the union
\begin{equation}\label{eq:C7}
\CP^3\ =\ \CO(1)\oplus\CO(1)\ =\ \CP_+^3\cup\CP_0\cup\CP_-^3\ =\
\tilde{\CP}^3\cup\CP_0
\end{equation}
of three disjoint domains, $\CP^3_\pm=\CP^3|_{H^2_\pm}$ and
$\CP_0=\CP^3|_{S^1}$ since $\CPP^1=H_+^2\cup S^1\cup H_-^2$. There
is a natural map from $\CZ$ into $\CP^3$ which is a real-analytic
bijection between $\CZ_\pm$ and $\CP^3_\pm$,
\begin{equation}\label{eq:C8}
\tilde{\CP}^3\ \cong\ \tilde{\CZ}~,
\end{equation}
but this map becomes the real fibration $\CZ_0\rightarrow
\CT^3\subset \CP_0$ over a real three-dimensional submanifold
$\CT^3$ of $\CP_0$ (see e.g.\ \cite{Woodhouse,Lerner:1992ag}).

\paragraph{Reduction to mini-twistor spaces.}
This directly carries over to the mini-twistor space, which is
again obtained by taking the quotient of the twistor space with
respect to the Abelian group $\CCG_\FC\cong \FC$ defined in
section 3.2. From \eqref{eq:C7}, we thus have
$\CP^2:=\CP^3/\CCG_\FC \cong\CO(2)$ while \eqref{eq:C8} yields an
open subset $\tilde{\CP}^2=\tilde{\CP}^3/\CCG_\FC$ of $\CP^2$.
These considerations readily generalize to the supertwistor spaces
and we obtain
\begin{equation}
\CP^{2|4}\ =\ \CP^{3|4}/\CCG_\FC
\end{equation}
and an open subset
\begin{equation}
\tilde{\CP}^{2|4}\ =\ \tilde{\CP}^{3|4}/\CCG_\FC
\end{equation}
in $\CP^{2|4}$ together with the open subset of the CR
supertwistor space,
\begin{equation}
\tilde{\CF}^{5|8}\ \cong\ \FR^{2,2|8}\times H^2\ \subset\
\CF^{5|8}\ \ \cong\ \FR^{2,2|8}\times S^2~.
\end{equation}

\paragraph{Modifications in signature $($$+$$+$$-$$)$.}
In fact, all of these spaces appear in the twistor correspondence
between hBF theory, phCS theory and a supersymmetric Bogomolny
model on the space $\FR^{2,1}=(\FR^3,g)$ with the metric
$g=\diag(+1,+1,-1)$. Namely, one uses the spaces
$\tilde{\CP}^{2|4}$ and $\tilde{\CF}^{5|8}$ in the Dolbeault
description of these correspondences and the spaces $\CP^{2|4}$
and $\CF^{5|8}$ in the \v{C}ech description via transition
functions. There are only minor modifications to be made to all
the formul\ae{} of the Euclidean case to hold also here. First,
one replaces the reality condition \eqref{eq:2.16} and
\eqref{eq:2.33} by\footnote{See e.g.\ \cite{Popov:2004rb} for more
details.}
\begin{equation}
x^{2\zd}\ =\ \bar{x}^{1\ed}\ =\ -\di(x^1-\di x^2)~,~~~x^{2\ed}\ =\
\bar{x}^{1\zd}\ =\ -\di(x^4-\di x^3)\eand \eta_i^\zd\ =\
\etab_i^\ed~,
\end{equation}
which corresponds together with \eqref{eq:3.10} to the anti-linear
involution
\begin{equation}
\hat{\tau}(w_\pm^1,w_\pm^2,\eta_i^\pm)\ =\
\left(\frac{\bw_\pm^1}{(\bw_\pm^2)^2},\frac{1}{\bw_\pm^2},
\frac{1}{\bw_\pm^2}\etab_i^\pm\right)
\end{equation}
on the mini-supertwistor space. Second, the hBF theory is
considered on the supermanifold $\tilde{\CP}^{2|4}\subset
\CP^{2|4}$ and the phCS theory on the open subset
$\tilde{\CF}^{5|8}$ of the CR supertwistor space $\CF^{5|8}$.
Thus, one substitutes the space $\CPP^1$ by the two-sheeted
hyperboloid $H^2=\CPP^1\backslash S^1=H_+^2\cup H_-^2$ and uses
\begin{equation}
\begin{aligned}
(\hl_\ald^+)\ =\ \left(\begin{array}{c}\bl_+\\1
\end{array}\right)~,~~~
(\hl_\ald^-)\ =\ \left(\begin{array}{c}1\\\bl_-
\end{array}\right)~,~~~
\hl^\ald_\pm\ =\ \eps^{\ald\bed}\hl_\bed^\pm~,\\
\gamma_\pm\ =\ \pm\frac{1}{1-\lambda_\pm\bl_\pm}~,~~~\lambda_\pm\
\in\ H_\pm^2\hspace{2.6cm}
\end{aligned}
\end{equation}
instead of $\hl_\ald^\pm$, $\hl_\pm^\ald$ and $\gamma_\pm$ as
given in \eqref{eq:2.25}, \eqref{eq:3.38} and \eqref{eq:2.22}. All
other formul\ae{} including the equations of motion for phCS
theory on $\tilde{\CF}^{5|8}$ and the field expansions of
$\CA_\CT$ keep their form. The resulting Bogomolny-type field
equations on $\FR^{2,1}$ will only differ by some signs in front
of the interaction terms. All this also holds for the $M$-deformed
case, which eventually involves the spaces $\tilde{\CP}^{2|4}_M$
and $\tilde{\CF}^{5|8}_M$.


\begin{thebibliography}{99}
\setlength{\itemsep}{-2mm}

\bibitem{Witten:2003nn}
E.~Witten, {\em Perturbative gauge theory as a string theory in
twistor space,} Commun.\ Math.\ Phys.\  {\bf 252} (2004) 189
[hep-th/0312171].

\bibitem{Popov:2004rb}
A.~D.~Popov and C.~Saemann, {\em On supertwistors, the
Penrose-Ward transform and $\CN=4$ super Yang-Mills theory,}
hep-th/0405123.

\bibitem{Siegel:1992za}
W.~Siegel, {\em The N=2 (4) string theory is self-dual $\CN=4$
Yang-Mills theory,} Phys.\ Rev.\ D {\bf 46} (1992) R3235
[hep-th/9205075].

\bibitem{Nair:bq}
V.~P.~Nair, {\em A current algebra for some gauge theory
amplitudes,} Phys.\ Lett.\ B {\bf 214} (1988) 215.

\bibitem{webpage}
The web-page of the ``London Mathematical Society Workshop on
Twistor String Theory,'' Oxford 10-14 January 2005, {\tt
http://www.maths.ox.ac.uk/$\sim$lmason/Tws/}.

\bibitem{Cachazo:2005ga}
F.~Cachazo and P.~Svr\v{c}ek, {\em Lectures on twistor strings and
perturbative Yang-Mills theory,} hep-th/0504194.

\bibitem{supergravity}
S.~Giombi, R.~Ricci, D.~Robles-Llana and D.~Trancanelli, {\em A
note on twistor gravity amplitudes,} JHEP {\bf 0407} (2004) 059
[hep-th/0405086];
N.~Berkovits and E.~Witten, {\em Conformal supergravity in
twistor-string theory,} JHEP {\bf 0408} (2004) 009
[hep-th/0406051];
C.~H.~Ahn, {\em $\CN=1$ conformal supergravity and twistor-string
theory,} JHEP {\bf 0410} (2004) 064 [hep-th/0409195];
C.~H.~Ahn, {\em $\CN=2$ conformal supergravity from twistor-string
theory,} hep-th/0412202;
J.~Bedford, A.~Brandhuber, B.~Spence and G.~Travaglini, {\em A
recursion relation for gravity amplitudes,} hep-th/0502146;
N.~E.~J.~Bjerrum-Bohr, D.~C.~Dunbar and H.~Ita, {\em Six-point
one-loop $\CN=8$ supergravity NMHV amplitudes and their IR
behaviour,} hep-th/0503102;
Y.~Abe, {\em An interpretation of multigraviton amplitudes,}
hep-th/0504174.

\bibitem{mirror}
A.~Neitzke and C.~Vafa, {\em N = 2 strings and the twistorial
Calabi-Yau,} hep-th/0402128;
N.~Nekrasov, H.~Ooguri and C.~Vafa, {\em S-duality and topological
strings,} JHEP {\bf 0410} (2004) 009 [hep-th/0403167];
M.~Aganagic and C.~Vafa, {\em Mirror symmetry and supermanifolds,}
hep-th/0403192;
C.~H.~Ahn, {\em Mirror symmetry of Calabi-Yau supermanifolds,}
Mod.\ Phys.\ Lett.\ A {\bf 20} (2005) 407 [hep-th/0407009];
A.~Belhaj, L.~B.~Drissi, J.~Rasmussen, E.~H.~Saidi and A.~Sebbar,
{\em Toric Calabi-Yau supermanifolds and mirror symmetry,}
hep-th/0410291.

\bibitem{Popov:2004nk}
A.~D.~Popov and M.~Wolf, {\em Topological B-model on weighted
projective spaces and self-dual models in four dimensions,} JHEP
{\bf 0409} (2004) 007 [hep-th/0406224];
C.~Saemann, {\em The topological B-model on fattened complex
manifolds and subsectors of $\CN=4$ self-dual Yang-Mills theory,}
JHEP {\bf 0501} (2005) 042 [hep-th/0410292];
J.~Park and S.~J.~Rey, {\em Supertwistor orbifolds: Gauge theory
amplitudes and topological strings,} JHEP {\bf 0412} (2004) 017
[hep-th/0411123];
S.~Giombi, M.~Kulaxizi, R.~Ricci, D.~Robles-Llana, D.~Trancanelli
and K.~Zoubos, {\em Orbifolding the twistor string,}
hep-th/0411171.

\bibitem{Wolf:2004hp}
M.~Wolf, {\em On hidden symmetries of a super gauge theory and
twistor string theory,} JHEP {\bf 0502} (2005) 018
[hep-th/0412163].

\bibitem{Berkovits:2004hg}
N.~Berkovits, {\em An alternative string theory in twistor space
for $\CN=4$ super-Yang-Mills,} Phys.\ Rev.\ Lett.\  {\bf 93}
(2004) 011601 [hep-th/0402045];
N.~Berkovits and L.~Motl, {\em Cubic twistorial string field
theory,} JHEP {\bf 0404} (2004) 056 [hep-th/0403187];
W.~Siegel, {\em Untwisting the twistor superstring,}
hep-th/0404255;
O.~Lechtenfeld and A.~D.~Popov, {\em Supertwistors and cubic
string field theory for open N = 2 strings,} Phys.\ Lett.\ B {\bf
598} (2004) 113 [hep-th/0406179].


\bibitem{Abe:2004ep}
Y.~Abe, V.~P.~Nair and M.~I.~Park, {\em Multigluon amplitudes,
$\CN=4$ constraints and the WZW model,} Phys.\ Rev.\ D {\bf 71}
(2005) 025002 [hep-th/0408191];
M.~Kulaxizi and K.~Zoubos, {\em Marginal deformations of $\CN=4$
SYM from open/closed twistor strings,} hep-th/0410122.

\bibitem{Burinskii:2004tt}
A.~Sinkovics and E.~Verlinde, {\em A six dimensional view on
twistors,} Phys.\ Lett.\ B {\bf 608} (2005) 142 [hep-th/0410014];
A.~Burinskii, {\em Rotating black hole, twistor-string and
spinning particle,} hep-th/0412195;
L.~B.~Anderson and J.~T.~Wheeler, {\em Yang-Mills gravity in
biconformal space,} hep-th/0412293;
S.~Seki and K.~Sugiyama, {\em Gauged linear sigma model on
supermanifold,} hep-th/0503074.

\bibitem{Chiou:2005jn}
D.~W.~Chiou, O.~J.~Ganor, Y.~P.~Hong, B.~S.~Kim and I.~Mitra, {\em
Massless and massive three dimensional super Yang-Mills theory and
mini-twistor string theory,} hep-th/0502076.

\bibitem{Hitchin:1982gh}
N.~J.~Hitchin, {\em Monopoles and geodesics,} Commun.\ Math.\
Phys.\ {\bf 83} (1982) 579.

\bibitem{Murray:1985ji}
M.~K.~Murray, {\em Nonabelian magnetic monopoles,} Commun.\ Math.\
Phys.\  {\bf 96} (1984) 539.

\bibitem{Popov:1999cq}
A.~D.~Popov, {\em Holomorphic analogs of topological gauge
theories,} Phys.\ Lett.\ B {\bf 473} (2000) 65 [hep-th/9909135].

\bibitem{Ivanova:2000xr}
T.~A.~Ivanova and A.~D.~Popov, {\em Dressing symmetries of
holomorphic BF theories,} J.\ Math.\ Phys.\  {\bf 41} (2000) 2604
[hep-th/0002120];
{\em \v{C}ech, Dolbeault and de Rham cohomologies in Chern-Simons
and BF theories,} Proc.\ of the Intern. Colloquium on Group
Theoretical Methods in Physics, Eds. A.~N.~Sissakian and
G.~S.~Pogosian, vol.~1, p.138, Dubna, 2002 [hep-th/0101150].

\bibitem{Baulieu:2004pv}
L.~Baulieu and A.~Tanzini, {\em Topological symmetry of forms, N =
1 supersymmetry and S-duality on special manifolds,}
hep-th/0412014.

\bibitem{Witten:1992fb}
E.~Witten, {\em Chern-Simons gauge theory as a string theory,}
Prog.\ Math.\  {\bf 133} (1995) 637 [hep-th/9207094].

\bibitem{LeBrun:1984}
C.~R.~LeBrun, {\em Twistor CR manifolds and three-dimensional
conformal geometry,} Trans.\ Amer.\ Math.\ Soc.\ {\bf 284} (1984)
601.

\bibitem{Rawnsley} J.~H.~Rawnsley, {\em Flat partial connections and
holomorphic structures in smooth vector bundles,}  Proc.\ Amer.\
Math.\ Soc. {\bf 73} (1979) 391.

\bibitem{Atiyah:wi}
M.~F.~Atiyah, N.~J.~Hitchin and I.~M.~Singer, {\em Self-duality in
four-dimensional Riemannian geometry,} Proc.\ Roy.\ Soc.\ Lond.\ A
{\bf 362} (1978) 425.

\bibitem{Penrose:in}
R.~Penrose, {\em The twistor program,} Rept.\ Math.\ Phys.\  {\bf
12} (1977) 65.

\bibitem{Kodaira}
K.~Kodaira, {\em A theorem of completeness of characteristic
systems for analytic families of compact submanifolds of complex
manifolds,} Ann.\ Math.\ {\bf 75} (1962) 146.

\bibitem{Penrose:ca}
R.~Penrose and W.~Rindler, ``Spinors and space-time. Vols. 1 \&
2,'' Cambridge University Press, Cambridge, 1984 \& 1985.

\bibitem{Manin:ds}
Yu.~I.~Manin, ``Gauge field theory and complex geometry,''
Springer, Berlin, 1988 [Russian: Nauka, Moscow, 1984].

\bibitem{Ward:vs}
R.~S.~Ward and R.~O.~Wells, ``Twistor geometry and field theory,''
Cambridge University Press, Cambridge, 1990.

\bibitem{Mason:rf}
L.~J.~Mason and N.~M.~J.~Woodhouse, ``Integrability, self-duality,
and twistor theory,'' Clarendon Press, Oxford, 1996.

\bibitem{Bogomolny:1975de}
E.~B.~Bogomolny, {\em Stability of classical solutions,} Sov.\ J.\
Nucl.\ Phys.\  {\bf 24} (1976) 449.

\bibitem{Prasad:1975kr}
M.~K.~Prasad and C.~M.~Sommerfield, {\em An exact classical
solution for the 't Hooft monopole and the Julia-Zee dyon,} Phys.\
Rev.\ Lett.\  {\bf 35} (1975) 760.

\bibitem{Manton:1977ht}
N.~S.~Manton, {\em Complex structure of monopoles,} Nucl.\ Phys.\
B {\bf 135} (1978) 319.

\bibitem{Atiyah:1988jp}
M.~F.~Atiyah and N.~J.~Hitchin, ``The geometry and dynamics of
magnetic monopoles,'' M.B. Porter lectures, Princeton University
Press, 1988.

\bibitem{Howe:1995md} P.~S.~Howe and G.~G.~Hartwell, {\em A superspace survey,} Class.\ Quant.\ Grav.\  {\bf 12} (1995)
1823.

\bibitem{Nirenberg}
L.~Nirenberg, {\em A complex Frobenius theorem,} Proc.\ Conf.\
Analytic Functions, I. (Princeton University Press, N.J., 1957),
p.\ 172.

\bibitem{Harnad:1984vk}
J.~P.~Harnad, J.~Hurtubise, M.~Legar{\'e} and S.~Shnider, {\em
Constraint equations and field equations in supersymmetric $\CN=3$
Yang-Mills theory,} Nucl.\ Phys.\ B {\bf 256} (1985) 609.

\bibitem{wuyang}
T.~T.~Wu and C.~N.~Yang, in: ``Properties of matter under unusual
conditions,'' edited by H.~Mark and S.~Fernbach (Interscience, New
York, 1969), p. 349.

\bibitem{Popov:2004rt}
A.~D.~Popov, {\em On explicit point multi-monopoles in SU(2) gauge
theory,} to appear in J.\ Math.\ Phys., hep-th/0412042.

\bibitem{Ward:1981jb}
R.~S.~Ward, {\em A Yang-Mills Higgs monopole of charge 2,}
Commun.\ Math.\ Phys.\  {\bf 79} (1981) 317;
R.~S.~Ward, {\em Ans\"{a}tze for self-dual Yang-Mills fields,}
Commun.\ Math.\ Phys.\  {\bf 80} (1981) 563.

\bibitem{Pohlmeyer:1979ya}
         K.~Pohlmeyer,
         {\it On the Lagrangian theory of anti-(self)-dual fields in
         four-dimensional Euclidean space},
         Commun.\ Math.\ Phys.\  {\bf 72} (1980) 37.

\bibitem{Chau:1981gi}
         L.~L.~Chau, M.~L.~Ge and Y.~S.~Wu,
         {\it The Kac-Moody algebra in the self-dual Yang-Mills equation},
         Phys.\ Rev.\ D {\bf 25} (1982) 1086;
         L.~L.~Chau and Wu Yong-Shi,
         {\it More about hidden symmetry algebra for the self-dual Yang-Mills
         system},
         Phys.\ Rev.\ D {\bf 26} (1982) 3581;
         L.~L.~Chau, M.~L.~Ge, A.~Sinha and Y.~S.~Wu,
         {\it Hidden symmetry algebra for the self-dual Yang-Mills equation},
         Phys.\ Lett.\ B {\bf 121} (1983) 391.


\bibitem{Ueno:1982dy}
  K.~Ueno and Y.~Nakamura,
         {\it Transformation theory for anti-(self)-dual equations and the
         Riemann-Hilbert problem},
         Phys.\ Lett.\ B {\bf 109} (1982) 273;
         L.~Dolan,
         {\it A new symmetry group of real self-dual Yang-Mills},
         Phys.\ Lett.\ B {\bf 113} (1982) 387;
         L.~Crane,
         {\it Action of the loop group on the self-dual Yang-Mills equation},
         Commun.\ Math.\ Phys.\  {\bf 110} (1987) 391.

\bibitem{Popov:1995qb}
A.~D.~Popov and C.~R.~Preitschopf, {\em Conformal symmetries of
the self-dual Yang-Mills equations,} Phys.\ Lett.\ B {\bf 374}
(1996) 71 [hep-th/9512130];
A.~D.~Popov,
         {\it Self-dual Yang-Mills: Symmetries and moduli space},
         Rev.\ Math.\ Phys.\  {\bf 11} (1999) 1091
         [hep-th/9803183];
         T.~A.~Ivanova,
         {\it On current algebra of symmetries of the self-dual Yang-Mills
         equations},
         J.\ Math.\ Phys.\  {\bf 39} (1998) 79
         [hep-th/9702144].

\bibitem{Popov:1998fb}
A.~D.~Popov, {\em Holomorphic Chern-Simons-Witten theory: From 2D
to 4D conformal field theories,} Nucl.\ Phys.\ B {\bf 550} (1999)
585 [hep-th/9806239].

\bibitem{Witten:1978xx}
E.~Witten, {\em An interpretation of classical Yang-Mills theory,}
Phys.\ Lett.\ B {\bf 77} (1978) 394.

\bibitem{Isenberg:kk}
J.~Isenberg, P.~B.~Yasskin and P.~S.~Green, {\em Non-self-dual
gauge fields,} Phys.\ Lett.\ B {\bf 78} (1978) 462.

\bibitem{Woodhouse}
N.~M.~J.~Woodhouse, {\em Contour integrals for the ultrahyperbolic
wave equation,} Proc.\ R.\ Soc.\ Lond.\ A {\bf 438} (1992) 197.

\bibitem{Lerner:1992ag}
D.~E.~Lerner, {\em The Linear system for selfdual gauge fields in
a space-time of signature 0,} J.\ Geom.\ Phys.\  {\bf 8} (1992)
211.
\end{thebibliography}
\end{document}